\newcommand{\DoPrePrint}{0} 
\newcommand{\DoFiguresAtEnd}{0}
\ifnum\DoPrePrint=1

  \RequirePackage{lineno}
  \documentclass[aps,prl,preprint,showpacs,superscriptaddress,nofootinbib,linenumbers,floatfix]{revtex4}
  \usepackage{lineno}
\else
  \documentclass[aps,prd,twocolumn,showpacs,superscriptaddress,nofootinbib]{revtex4}

\fi

\usepackage{wrapfig}
\usepackage{amsmath}
\usepackage{graphicx}
\usepackage{amsfonts}
\usepackage{braket}

\usepackage{pdflscape}
\usepackage{array}
\RequirePackage{lineno} 
 \usepackage{lineno}

\usepackage{morefloats}

\usepackage{amsmath}   % need for subequations
\usepackage{graphicx}  % for figures
\usepackage{xspace}
\usepackage{units}

\usepackage{hyperref}
\ifnum\DoFiguresAtEnd=1
\usepackage[printfigures]{figcaps}
\else
\fi

\newcommand{\minerva}{MINERvA\xspace}
\newcommand{\minos}{MINOS\xspace}

\newcommand{\nubar}{\ensuremath{\bar{\nu}}}

\newcommand{\sizecheck}{0} % 0 to do nothing; 1 to check size
\newif\ifpdf
\ifx\pdfoutput\undefined
   \pdffalse
\else
   \pdfoutput=1
   \pdftrue
\fi
\ifpdf
   \usepackage{graphicx}
   \usepackage{epstopdf}
\else
   \usepackage{graphicx}
\fi

\def\enu{$E_\nu^{QE}$~}
\def\enutrue{$E_\nu^{true}$~}
\def\qsq{$Q^2_{QE}$~}
\def\pt{$p_T$~}
\def\pz{$p_\parallel$~}

\def\enunospace{$E_\nu^{QE}$}

\def\qsqnospace{$Q^2_{QE}$}
\def\ptnospace{$p_T$}
\def\pznospace{$p_\parallel$}

\begin{document}
%\linenumbers
\ifnum\DoPrePrint=1
\linenumbers
\fi

\title{Measurement of the muon anti-neutrino double-differential cross
  section for quasi-elastic-like scattering on hydrocarbon
  at~$E_\nu \sim 3.5$~GeV	}	%Lines break automatically or can be forced with \\

%% MANUAL PARTS OF AUTHOR LIST

%% AUTOMATIC LIST (EDITED AS ABOVE)
\newcommand{\Rutgers}{Rutgers, The State University of New Jersey, Piscataway, New Jersey 08854, USA}
\newcommand{\Hampton}{Hampton University, Dept. of Physics, Hampton, VA 23668, USA}
\newcommand{\Dortmund}{Institute of Physics, Dortmund University, 44221, Germany }
\newcommand{\Otterbein}{Department of Physics, Otterbein University, 1 South Grove Street, Westerville, OH, 43081 USA}
\newcommand{\JMU}{James Madison University, Harrisonburg, Virginia 22807, USA}
\newcommand{\Florida}{University of Florida, Department of Physics, Gainesville, FL 32611}
\newcommand{\UCIrvine}{Department of Physics and Astronomy, University of California, Irvine, Irvine, California 92697-4575, USA}
\newcommand{\CBPF}{Centro Brasileiro de Pesquisas F\'{i}sicas, Rua Dr. Xavier Sigaud 150, Urca, Rio de Janeiro, Rio de Janeiro, 22290-180, Brazil}
\newcommand{\PUCP}{Secci\'{o}n F\'{i}sica, Departamento de Ciencias, Pontificia Universidad Cat\'{o}lica del Per\'{u}, Apartado 1761, Lima, Per\'{u}}
\newcommand{\INRM}{Institute for Nuclear Research of the Russian Academy of Sciences, 117312 Moscow, Russia}
\newcommand{\Jlab}{Jefferson Lab, 12000 Jefferson Avenue, Newport News, VA 23606, USA}
\newcommand{\Pittsburgh}{Department of Physics and Astronomy, University of Pittsburgh, Pittsburgh, Pennsylvania 15260, USA}
\newcommand{\Guanajuato}{Campus Le\'{o}n y Campus Guanajuato, Universidad de Guanajuato, Lascurain de Retana No. 5, Colonia Centro, Guanajuato 36000, Guanajuato M\'{e}xico.}
\newcommand{\Athens}{Department of Physics, University of Athens, GR-15771 Athens, Greece}
\newcommand{\Tufts}{Physics Department, Tufts University, Medford, Massachusetts 02155, USA}
\newcommand{\WM}{Department of Physics, College of William \& Mary, Williamsburg, Virginia 23187, USA}
\newcommand{\FNAL}{Fermi National Accelerator Laboratory, Batavia, Illinois 60510, USA}
\newcommand{\Purdue}{Department of Chemistry and Physics, Purdue University Calumet, Hammond, Indiana 46323, USA}
\newcommand{\MCLA}{Massachusetts College of Liberal Arts, 375 Church Street, North Adams, MA 01247}
\newcommand{\UMD}{Department of Physics, University of Minnesota -- Duluth, Duluth, Minnesota 55812, USA}
\newcommand{\Northwestern}{Northwestern University, Evanston, Illinois 60208}
\newcommand{\UNI}{Universidad Nacional de Ingenier\'{i}a, Apartado 31139, Lima, Per\'{u}}
\newcommand{\Rochester}{University of Rochester, Rochester, New York 14627 USA}
\newcommand{\Austin}{Department of Physics, University of Texas, 1 University Station, Austin, Texas 78712, USA}
\newcommand{\USM}{Departamento de F\'{i}sica, Universidad T\'{e}cnica Federico Santa Mar\'{i}a, Avenida Espa\~{n}a 1680 Casilla 110-V, Valpara\'{i}so, Chile}
\newcommand{\Geneva}{University of Geneva, 1211 Geneva 4, Switzerland}
\newcommand{\Chicago}{Enrico Fermi Institute, University of Chicago, Chicago, IL 60637 USA}
\newcommand{\hired}{}
\newcommand{\OregonState}{Department of Physics, Oregon State University, Corvallis, Oregon 97331, USA}
\newcommand{\oxford}{}
\newcommand{\umiss}{University of Mississippi, Oxford, Mississippi 38677, USA}
\newcommand{\upenn}{209 S. 33rd St. Philadelphia, PA 19104}
\newcommand{\AMU}{AMU Campus, Aligarh, Uttar Pradesh 202001, India}
\newcommand{\wroclaw}{University of Wroclaw, plac Uniwersytecki 1, 50-137 Wrocław, Poland}
\newcommand{\Mohali}{Knowledge city, Sector 81, SAS Nagar, Manauli PO 140306}
\newcommand{\chrismarshallThanks}{Now at Lawrence Berkeley National Laboratory, Berkeley, CA 94720, USA}
\newcommand{\joelmousseauThanks}{Now at University of Michigan, Ann Arbor, MI 48109, USA}
\newcommand{\cpatrickThanks}{Now at University College London, London WC1E 6BT, UK}
\newcommand{\jwolcottThanks}{Now at Tufts University, Medford, MA 02155, USA}

% 63+1 total signatories.

\author{C.E.~Patrick}\thanks{\cpatrickThanks}  \affiliation{\Northwestern}
\author{L.~Aliaga}                        \affiliation{\WM}  \affiliation{\PUCP}
\author{A.~Bashyal}                       \affiliation{\OregonState}   % added after original submission
\author{L.~Bellantoni}                    \affiliation{\FNAL}
\author{A.~Bercellie}                     \affiliation{\Rochester}
\author{M.~Betancourt}                    \affiliation{\FNAL}
\author{A.~Bodek}                         \affiliation{\Rochester}
\author{A.~Bravar}                        \affiliation{\Geneva}
\author{H.~Budd}                          \affiliation{\Rochester}
\author{G.~F.~R.~Caceres~V.}              \affiliation{\CBPF}
\author{M.F.~Carneiro}                    \affiliation{\OregonState}
\author{E.~Chavarria}                     \affiliation{\UNI}
\author{H.~da~Motta}                      \affiliation{\CBPF}
\author{S.A.~Dytman}                      \affiliation{\Pittsburgh}
\author{G.A.~D\'{i}az~}                   \affiliation{\Rochester}  \affiliation{\PUCP}
\author{J.~Felix}                         \affiliation{\Guanajuato}
\author{L.~Fields}                        \affiliation{\FNAL}  \affiliation{\Northwestern}
\author{R.~Fine}                          \affiliation{\Rochester}
\author{A.M.~Gago}                        \affiliation{\PUCP}
\author{R.Galindo}                        \affiliation{\USM}
\author{H.~Gallagher}                     \affiliation{\Tufts}
\author{A.~Ghosh}                         \affiliation{\USM}  \affiliation{\CBPF}
\author{R.~Gran}                          \affiliation{\UMD}
\author{J.Y.~Han}                         \affiliation{\Pittsburgh}
\author{D.A.~Harris}                      \affiliation{\FNAL}
\author{S.~Henry}                         \affiliation{\Rochester}
\author{K.~Hurtado}                       \affiliation{\CBPF}  \affiliation{\UNI}
\author{D.Jena}                           \affiliation{\FNAL}
\author{J.~Kleykamp}                      \affiliation{\Rochester}
\author{M.~Kordosky}                      \affiliation{\WM}
\author{T.~Le}                            \affiliation{\Tufts}  \affiliation{\Rutgers}
\author{X.-G.~Lu}                         \affiliation{\oxford}
\author{E.~Maher}                         \affiliation{\MCLA}
\author{S.~Manly}                         \affiliation{\Rochester}
\author{W.A.~Mann}                        \affiliation{\Tufts}
\author{C.M.~Marshall}\thanks{\chrismarshallThanks}  \affiliation{\Rochester}
\author{K.S.~McFarland}                   \affiliation{\Rochester}  \affiliation{\FNAL}
\author{A.M.~McGowan}                     \affiliation{\Rochester}
\author{B.~Messerly}                      \affiliation{\Pittsburgh}
\author{J.~Miller}                        \affiliation{\USM}
\author{A.~Mislivec}                      \affiliation{\Rochester}
\author{J.G.~Morf\'{i}n}                  \affiliation{\FNAL}
\author{J.~Mousseau}\thanks{\joelmousseauThanks}  \affiliation{\Florida}
\author{D.~Naples}                        \affiliation{\Pittsburgh}
\author{J.K.~Nelson}                      \affiliation{\WM}
\author{A.~Norrick}                       \affiliation{\WM}
\author{G.M.~Nowak}                       \affiliation{\OregonState}
\author{Nuruzzaman}                       \affiliation{\Rutgers}  \affiliation{\USM}
\author{V.~Paolone}                       \affiliation{\Pittsburgh}
\author{G.N.~Perdue}                      \affiliation{\FNAL}  \affiliation{\Rochester}
\author{E.~Peters}                        \affiliation{\OregonState}
\author{M.A.~Ram\'{i}rez}                 \affiliation{\Guanajuato}
\author{R.D.~Ransome}                     \affiliation{\Rutgers}
\author{H.~Ray}                           \affiliation{\Florida}
\author{L.~Ren}                           \affiliation{\Pittsburgh}
\author{P.A.~Rodrigues}                   \affiliation{\umiss}  \affiliation{\Rochester}
\author{D.~Ruterbories}                   \affiliation{\Rochester}
\author{H.~Schellman}                     \affiliation{\OregonState}  \affiliation{\Northwestern}
\author{C.J.~Solano~Salinas}              \affiliation{\UNI}
\author{M.~Sultana}                       \affiliation{\Rochester}
\author{S.~S\'{a}nchez~Falero}            \affiliation{\PUCP}
\author{A.M.~Teklu}                       \affiliation{\OregonState}
\author{E.~Valencia}                      \affiliation{\WM}  \affiliation{\Guanajuato}
\author{J.~Wolcott}\thanks{\jwolcottThanks}  \affiliation{\Rochester}
\author{M.Wospakrik}                      \affiliation{\Florida}
\author{B.~Yaeggy}                        \affiliation{\USM}
\author{D.~Zhang}                         \affiliation{\WM}

\collaboration{The MINER$\nu$A Collaboration}\ \noaffiliation
\date{\today}

\begin{abstract}
We present double-differential measurements of anti-neutrino quasi-elastic
scattering in the MINERvA detector.  This study improves on a
previous single differential measurement by using updated reconstruction algorithms and interaction
models,  and provides a complete description of observed muon
kinematics in the form of a
double-differential cross section with respect to muon transverse and
longitudinal momentum.  We include in our signal definition
zero-meson final states arising from multi-nucleon interactions and
from resonant pion production followed by pion absorption in the
primary nucleus.  We find that model agreement is considerably
improved by a model tuned to MINERvA inclusive neutrino scattering
data  that incorporates nuclear effects such as weak nuclear screening
and two-particle, two-hole enhancements.  
\end{abstract}

\ifnum\sizecheck=0  
\maketitle
\fi

\setcounter{secnumdepth}{1}  % voodoo to get section numbers to work.

\section{Introduction}

Although quasi-elastic neutrino interactions are a key signal process for accelerator-based oscillation experiments, models of these 
interactions on nuclei have large ($\sim30\%$) 
uncertainties~\cite{geniemanual, Abe:2015awa}.  These arise from 
several sources, including nucleon form factors and final state
interactions wherein the produced particles interact further before exiting the primary nucleus.  Final 
state interactions can also cause other processes such as resonant
pion production to have a zero-meson final state that will appear with
a quasi-elastic topology in a detector. Interactions with  multi-nucleon states can similarly produce zero-meson final states.  

These and similar sources of uncertainty on other processes dominate the systematic uncertainty budgets of the current 
oscillation measurements such as T2K~\cite{Abe:2017uxa} and Nova~\cite{Adamson:2017gxd} and will limit the 
reach of oscillation experiments such as DUNE~\cite{dunecdr} if not 
further reduced.  Because any one measurement of quasi-elastic 
scattering necessarily measures a superposition of these effects,
lowering model uncertainties on individual parameters will require 
many different measurements to untangle the many unknowns. 

 In this 
article, we present a critical ingredient in this process: a double 
differential measurement of the anti-neutrino quasi-elastic cross 
section as a function of the transverse and longitudinal momentum of 
the final state muon.  We include in our measurement events consistent
with zero-meson final states arising from resonant pion production
followed by pion absorption in the nucleus and from interactions on
multi-nucleon states (frequently referred to as two-particle,
two-hole or 2p2h).  This ensemble of signal processes, which is defined precisely in Section~\ref{sec:signal}, is referred to hereafter as ``QE-like''.    In addition to this primary result, we also present a number 
of auxiliary measurements including double-differential cross sections
as a function of alternate variables, single differential projections,
and comparisons of reconstructed energy near the event vertex to
various models.  The neutrino energy range of 1.5-15 GeV covered by
this measurement is well matched to that of present and future
neutrino oscillations experiments with baselines on the 1000 km scale,
including MINOS\cite{minos}, NOvA\cite{NOvA}, and DUNE\cite{DUNE}.

The measurement described here extends a previous measurement of anti-neutrino quasi-elastic scattering by MINERvA~\cite{laura}  and is a companion to similar studies of neutrino scattering~\cite{arturo}.  The measurement complements other~\minerva QE-like studies that look in detail at the hadronic component of the final state~\cite{tammy} and that study neutrino interaction cross sections as a function of nuclear mass~\cite{Tice:2014pgu,Mousseau:2016snl}.  

This article is organized as follows:  Section~\ref{sec:theory} reviews 
the current status of neutrino-nucleus QE-like scattering models.  Sections~\ref{sec:expt} and~\ref{sec:simulation} review the~\minerva experiment and 
simulation.  Event reconstruction and selection are discussed in
Sections~\ref{sec:reco} and ~\ref{sec:selection}.  The cross section
extraction procedure and systematic uncertainties are detailed in
Sections~\ref{sec:crosssection} and~\ref{sec:systematics}.  The
results and comparisons with models are presented in
Section~\ref{sec:results}, and the article is summarized in
Section~\ref{sec:conclusion}. The Appendices present additional
results, including cross sections with alternate signal definitions;
those planning to use the data are encouraged to use
the supplementary materials, which provide higher numerical precision.

\section{Theory of QE-like Interactions} \label{sec:theory}

In charged-current quasi-elastic scattering on free nucleons, an incoming muon anti-neutrino interacts with a target proton, exchanging a $W^{\pm}$ boson to knock out a neutron and leave a positively charged muon in the final state:

\begin{equation}
\label{eq:antinuccqe}
\bar{\nu}_\mu + p \to \mu^+ + n
\end{equation} 
In this case, it is possible to reconstruct certain characteristics of
the interaction using only the kinematics of the outgoing charged
lepton assuming the initial-state nucleon is at rest.  For a nucleon
bound within a nucleus, the incoming neutrino energy and the four-momentum transfer, $Q^2$, can be estimated as:

\begin{equation}
E_\nu^{QE}=\frac{m_n^2 - (m_p-E_b)^2-m_\mu^2+2(m_p-E_b)E_\mu}{2(m_p-E_b-E_\mu+p_\mu\cos\theta_\mu)}
\label{eq:enu}
\end{equation}

\begin{equation}
Q^2_{QE}=2E_\nu^{QE}(E_\mu-p_\mu\cos\theta_\mu)-m_\mu^2
\label{eq:qsq}
\end{equation}
where $E_b$ is the initial state nucleon's binding energy, taken to be
30 MeV, as described in~\cite{laura}, $E_\nu$ and $E_\mu$ are the
neutrino and muon energy, $p_\mu$ and $\theta_\mu$  are the muon's
momentum and angle with respect to the neutrino, and $m_\mu$, $m_n$,
$m_p$ are the mass of the muon, neutron and proton.  The ${QE}$
subscript and superscript here and throughout the remainder of this
article denotes quantities computed under an assumption of a
quasi-elastic hypothesis with the initial state nucleon at rest.  

In the case of a bound nucleon, Fermi motion and nucleon correlations
mean that the initial state nucleon is not at rest, making the $QE$
kinematic variables  only estimates of the true values.   The final
state interpretation can also be affected, as  an ejected nucleon may
interact with other nucleons while escaping the nucleus. Other
interactions such as resonant pion production can be modified by final
state nuclear effects to have no pions in the final state, thus
appearing QE-like.  Similarly, interactions with correlated pairs of nucleons can also produce final states that appear quasi-elastic.   All of these nuclear effects can cause quasi-elastic neutrino interactions on heavy nuclei to differ substantially from those  on free nucleons. In this section, we discuss the quasi-elastic scattering from a free nucleon, and several contemporary theories that attempt to model the impact of the nuclear environment.  More detail can be found in~\cite{cherylthesis}.

\subsection*{Quasi-elastic Anti-neutrino Scattering on Free Nucleons} 
\label{sec:ccqe_theory}

Because the internal structure of the initial- and final-state
nucleons is governed by the  non-perturbative regime of QCD, it is not
possible to make a precise {\it ab initio} calculation of
the neutrino-nucleon quasi-elastic cross-section; it may instead be
described by nucleon form factors.   In the 1972 review article of
C. Llewellyn-Smith \cite{LlewellynSmith1972}, the differential
quasi-elastic cross section is expressed as a function of two vector,
one axial-vector, one pseudoscalar and two second-order form factors.
 All but the axial form factor are known from electron-nucleon scattering measurements.  The axial form factor must be taken from neutrino scattering or pion electro-production measurements, and is typically parametrized as a dipole:
\begin{equation}
F_A(Q^2) = \frac{g_A}{(1+\frac{Q^2}{M_A^2})^2}.
\end{equation}
The value of the axial form-factor at $Q^2=0$, has been measured through beta-decay experiments \cite{betadecay,betadecay2},
leaving one free parameter, the axial mass $M_A$. Deuterium and
hydrogen bubble chambers  \cite{Miller:1982qi,Kitagaki:1990vs}
have measured the value of $M_A$ on free or quasi-free nucleons.  An
average value of  $M_A = 1.014 \pm 0.014 \text{\:GeV/}c^2$ was
extracted by Bodek et al. \cite{1742-6596-110-8-082004} in
2008. Modern experiments on heavy nuclei have favored higher values of
$M_A$ ~\cite{minibooneccqe, Adamson:2014pgc, t2kccqe}, and the
discrepancy between these and the deuterium experiments has been
attributed to insufficiencies in the nuclear models used to extract
the axial mass on heavy nuclei.  Alternate parameterizations of the
dipole form factor are also available.  In particular, a more general ``z-expansion'' parameterization~\cite{Bhattacharya:2011ah} has been widely adopted in flavor physics and was recently implemented in neutrino event generators.  

\subsection*{Scattering from Nuclei}\label{sec:models}

\label{sec:rfg}

When simulating quasi-elastic scattering in heavy nuclei, the most commonly used nuclear model is the Relativistic Fermi Gas (RFG) model proposed by Smith and Moniz~\cite{rfg}, in which scattering from a nucleon in a nucleus is treated as if the incoming lepton scatters from an independent nucleon (the ``impulse approximation''). However, in the case of the RFG, the target nucleon is not stationary, but has a momentum consistent with the Fermi distribution. Thus the cross section for scattering off the nucleus is replaced by an incoherent sum of cross sections for scattering off of individual nucleons, with the remaining nucleus (depleted by 1 nucleon) as a spectator.

The Local Fermi Gas (LFG) model is a extension to the RFG model in
which a local density approximation \cite{Negele:1970jv,lfg} is
used, so that instead of using a constant average field for the whole
nucleus, the momentum distribution is dependent on a nucleon's
position within the nucleus.  This gives a Fermi motion distribution
that is not sharply peaked at the momentum limit and is both more
natural and reproduces the measured peak of the distribution.

Spectral functions can be also used to improve the Relativistic Fermi Gas model \cite{PhysRevC.56.276}.  The Hamiltonian for a large nucleus is so complicated that it is impractical to try to solve the many-body Schr\"odinger equation for the entire nucleus. However, if a mean field is used to replace the sum of the individual interactions, a  spectral function can be constructed that represents the probability of finding a nucleon with momentum and removal energy within the nucleus.  There are a number of non-Fermi-gas approaches~\cite{Benhar:2005dj,Jachowicz:2002rr,Pandey:2014tza,Lovato:2015qka,Gonzalez-Jimenez:2014eqa,Megias:2016fjk}.

\subsection*{Multi-Nucleon Correlations}
\label{sec:correlations} 

The models described above, which treat individual nucleons
independently, do not fully take into account the nature of the
nuclear force, which has a short range with a repulsive core
\cite{arrington2012}. Interactions between two (or more) spatially
close nucleons can give the individual nucleons very high momenta, far
above the Fermi momentum. Electron
scattering~\cite{PhysRevLett.96.082501} data indicate that approximately 20\% of the nucleons in carbon atoms are part of correlated pairs.  These experiments have observed ejected nucleons consistent with knocked-out partner nucleons~\cite{PhysRevLett.99.072501,Subedi13062008}, with 
 90\% 
of those pairs found to be proton-neutron pairs, with the remainder being $nn$ and $pp$ pairs.  In the case of $np$ pairs, it is expected that charged current QE-like anti-neutrino scattering on protons within a correlated pair would tend to produce two neutrons -- the expected neutron produced by the QE-like interaction, plus the knocked out neutron partner.    

The impact of multi-nucleon states in the initial state has been modeled in a number of ways.  Bodek and Ritchie's modification to the Relativistic Fermi Gas model \cite{Bodek1981} adds a high-momentum tail to the RFG's momentum distribution, based on the nucleon-nucleon correlation function and fits to data as explained in \cite{Gottfried196329}.  While this method attempts to provide a realistic initial momentum distribution, it does not include any model for the ejection of paired nucleons.

Going a step further, Bodek, Budd and Christie \cite{tem} have
developed a ``transverse enhancement model'' (TEM). They fit
inclusive electron scattering data~\cite{carlson,mamyan}, modifying
the nucleon magnetic form factors to accommodate the enhancement of
the transverse response observed in those data.  The resulting form
factor modifications plus an unmodified axial vector form factor were then used to predict neutrino-nucleus
scattering cross sections, producing results that were consistent with both low-energy data from MiniBooNE~\cite{minibooneccqe} and high-energy results from NOMAD~\cite{nomad}.  The TEM fit does not attempt to model additional knocked-out nucleons in either the electron or neutrino case.   Those empirical fits, when
combined with different expressions~\cite{PhysRevC.6.719} for how the structure functions should be related, and attaching
a two-nucleon knockout are used in the GiBUU generator.

While the approaches described above are empirical models that ascribe
effects observed in electron scattering to multi-nucleon processes and
then attempt to predict the effect they might have on neutrino
scattering, 2p2h models attempt to predict multi-nucleon effects in neutrino scattering from first principles.  These models consider pairs of nucleons connected by the exchange of
virtual pions and rho mesons \cite{mec}.   There has been a recent dramatic
expansion of work on 2p2h models, such as those of
Marteau/Martini~\cite{martini}, IFIC Valencia group~\cite{nieves2011},
the SuSA group~\cite{Gonzalez-Jimenez:2014eqa,Megias:2016fjk}, and the
Gent group~\cite{VanCuyck:2017wfn}.  As of this writing, the IFIC
Valencia model is implemented in GENIE, NuWro, and NEUT for the CC
process.   Empirical versions related to the TEM fits to electron
scattering data are also available in GENIE (without two-nucleon
knockout) and GiBUU (with two-nucleon knockout).  Finally, there is a
version that implements a simple empirical shape in W and Q2 developed
for use in electron scattering codes~\cite{doi:10.1063/1.168298} to
generate events as an option in GENIE~\cite{Katori:2013eoa}.

Long-range correlations between nucleons are typically modeled using
an approach known as the random-phase approximation
(RPA)~\cite{rpa}. It is based on the phenomenon, observed in
$\beta$-decay and muon capture experiments, that the electroweak coupling can be
modified by the presence of strongly-interacting nucleons in the
nucleus, when compared to its free-nucleon coupling strength, similar
to the screening of an electric charge in a dielectric. The RPA
approach affects cross-section predictions at low energy transfers
(and low $Q^2$), where a quenching of the axial current reduces the
cross section compared to the RFG prediction. It also introduces a
small cross section enhancement at intermediate $Q^2$.  Multiple RPA
models are available within generators, including those of Nieves~\cite{Nieves:2004wx}, Martini~\cite{martini}, Graczyk and Sobczyk~\cite{Graczyk:2003ru}, and Singh~\cite{Singh:1992dc}.  There is also discussion of the interplay between RPA with the Fermi gas and beyond-the-Fermi-gas models~\cite{Martini:2016eec}~\cite{Nieves:2017lij}.

\subsubsection*{Final-State interactions}\label{sec:fsi}

Non-quasi-elastic processes that undergo final-state interactions within the nucleus can have QE-like final states.  For example, there are three possible anti-neutrino charged-current resonant pion production processes:
\begin{align}
\bar{\nu}_\mu p &\to \mu^+ p \pi^-\\
\bar{\nu}_\mu p &\to \mu^+ n \pi^0\\
\bar{\nu}_\mu n &\to \mu^+ n \pi^-
\end{align} 
In such events, there is a $\sim$20\% possibility that the pion will be absorbed before it exits the nucleus, leaving a QE-like final state of a single muon and one or more nucleons.  

Nearly all available models of final state interactions are
Intranuclear Cascade (INC) models in which final state hadrons are
individually propagated through the nucleus with some probability of
undergoing interactions such as absorption or inelastic scattering
with the nuclear medium.  The details of the interactions vary
significantly across different models (typically implemented as part
of an event generator such as GENIE~\cite{geniemanual} or
NEUT~\cite{NEUT}), but all are tuned to hadron scattering data.  Some
generators (including GENIE) also provide effective cascade models wherein
the cascade of interactions that particles may undergo as they
traverse a nucleus is modeled as a single interaction. At least one alternative to INC models exists in the form of a semi-classical nuclear transport model implemented as part of GiBUU~\cite{gibuu}.

\section{\texorpdfstring{MINER\MakeLowercase{v}A}\:   Experiment}
\label{sec:expt}

\newcommand{\ignore}[1]{{}}

The MINERvA (Main INjector ExpeRiment $\nu$-A) experiment  is situated
in the NuMI (Neutrinos at the Main Injector) neutrino beam at Fermilab. The detector and beam are described in detail in~\cite{minerva_nim} and~\cite{numi}; this section summarizes their main features, focusing on the components relevant to this study.  

\subsection{The NuMI neutrino beam}\label{sec:beam}
Fermilab's NuMI beam uses 120 GeV/c protons from the Main Injector,
which impinge on a 1 meter long graphite target. The resulting  pions
and kaons are focused by a pair of movable parabolic horns. The horn
current polarity can be set to focus positively or negatively charged
mesons, which decay in a 675m-long decay pipe, producing muons and neutrinos. An absorber removes any remaining hadrons from the beam and 200 meters of rock filter out muons, leaving a beam of primarily neutrinos or anti-neutrinos, depending on the horn current polarity.

For the low energy beam configuration used in this work, the peak beam energy was approximately 3~GeV and the horns were configured to focus negative particles.  We use data recorded between November 5, 2010 and February 24, 2011, corresponding to $1.020\times 10^{20}$ protons on target (POT). The Monte Carlo simulation described in the next section corresponds to $9.247\times 10^{20}$ POT.

\subsection{\minerva Detector}
The~\minerva detector is composed of an inner detector (ID) and an outer detector (OD).  The most upstream portion of the ID consists of active scintillator planes interspersed with passive nuclear targets.  This region is used for studies of  the A-dependence of neutrino interaction cross sections, but is not used in the work described here.  Immediately downstream of the nuclear target region is the central tracker, followed by electromagnetic and hadronic calorimeters (ECAL and HCAL). 

The tracker is composed of 124 active scintillator planes, each consisting of 127 strips of doped polystyrene scintillator with a titanium dioxide coating. The strips have a triangular cross section $17$~mm (height) by $33$ mm (base) and vary in length between 122 and 245 cm depending on position within the plane.    Each scintillator plane is installed in one of three orientations, X, U or V. In the X orientation, the strips are vertical. Strips in the U or V planes are oriented at $\pm60^\circ$ with respect to the strips in the X planes. Each module in the active tracker region consists of two planes of scintillator strips, alternating between UX and VX configurations. A 2mm-thick lead collar covers the outermost 15 cm of each plane, forming the side electromagnetic calorimeter.

The downstream ECAL consists of ten modules that are similar to tracker modules, except that the 2mm-thick lead collar is replaced by a 2mm-thick sheet of lead covering the plane.  The 20 hadronic calorimeter (HCAL) modules, downstream of the ECAL, each contain only one plane of scintillator, followed by a 2.54 cm-thick plane of steel.  

Light produced in the scintillator is collected by a 1.2 mm diameter
wavelength-shifting (WLS) optical fiber inserted in a hole passing along the length of the strip and transmitted by the optical fibers to Hamamatsu H8804MOD-2 photomultiplier tubes (PMTs), as described in \cite{minerva_nim}. The full detector has 507 PMTs, each of which consists of 64 pixels.  The PMTs are read out via a data acquisition system that is described in detail in \cite{daq}. Raw PMT counts are transformed into estimated energy deposited in the strip via the calibration chain also described in~\cite{daq}.  The intrinsic time resolution is $~4$ ns.  %referee request

\subsection{The MINOS Near Detector}

The MINOS Near Detector~\cite{minos} is located 2 meters downstream of MINERvA, and is used to measure the charge and momentum of muons exiting the back of~\minerva.  The 1 kTon MINOS detector is composed of 2.54~cm-thick steel planes, interspersed with 1~cm-thick layers of scintillator. The scintillator planes are formed from 4.1~cm-wide scintillator strips, with orientation of the strips alternating between $+45^\circ$ and  $-45^\circ$ to the vertical in successive planes. The first 120 planes are instrumented for fine sampling; in this region, every fifth steel plane is followed by a fully-instrumented scintillator plane, while all other steel planes are followed by a partially-instrumented scintillator plane. The coarse-sampling region, further downstream, has only the fully-instrumented scintillator every five planes; there are no partial scintillator planes in this region.  The MINOS detector is magnetized by a coil that runs in a loop passing through the detector, generating a toroidal field with an average strength of 1.3~T.

\section{\texorpdfstring{MINER\MakeLowercase{v}A}\:   Simulation}
\label{sec:simulation}

\subsection{Beam Flux Simulation}
\label{sec:fluxsim}
MINERvA's simulation chain begins with G4Numi~\cite{leo}, a \textsc{Geant4}~\cite{Agostinelli2003250} based simulation of the NuMI beamline from primary proton beam to the~\minerva detector.  The FTFP\_BERT
% (FRITIOF Precompound - Bertini cascade) 
inelastic scattering model of Geant version 4.9.2.p03 is used.  This raw simulation is found to disagree with existing hadroproduction data from the NA49~\cite{na49} and other experiments~\cite{barton,mipp}, and is therefore corrected so that both differential and total interaction cross sections in the simulation match these external datasets.    
Version 1 of the PPFX package is used to implement these corrections~\cite{flux}. %We use PPFX version 1. 

\begin{figure}\centering 
  \includegraphics[width=1.0\columnwidth]{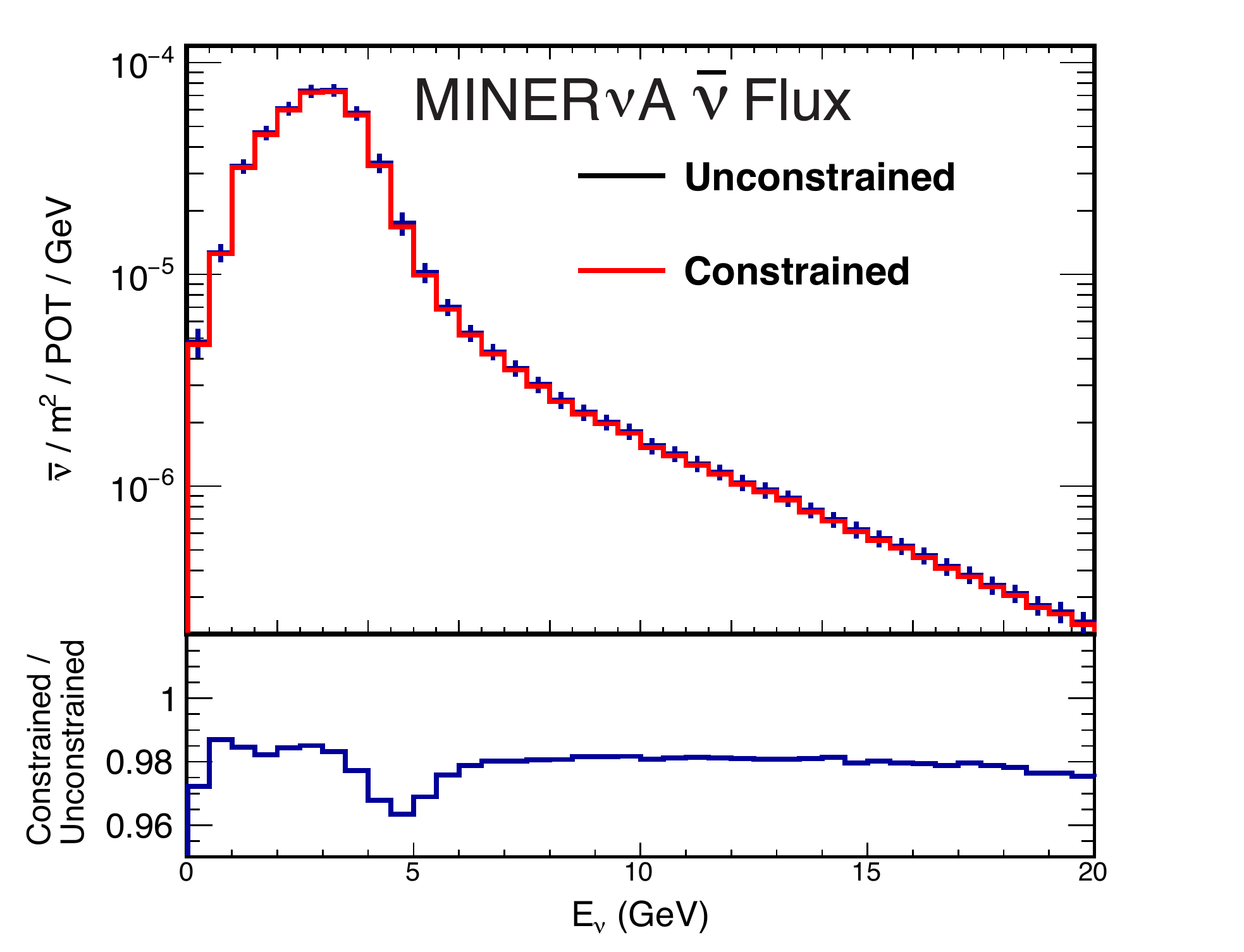}
        	\caption{NuMI flux distributions averaged over the MINERvA fiducial volume for anti-neutrinos as a function of energy, with and without a neutrino-electron scattering constraint.  The top plot shows the constrained (red) and unconstrained (black) distributions, which are separate by less than the line width.  The lower plot shows the ratio of the constrained to unconstrained values. The units are anti-neutrinos/proton on target/m$^2$. 
} 
        \label{fig:flux_constraint_ratio}
\end{figure}

We also use neutrino-electron scattering data collected in the~\minerva detector with the beamline in neutrino mode (focusing positive pions) as an independent constraint on the flux model, as described in~\cite{fluxconstraint}.  This constraint lowers the predicted neutrino flux by 2-4\% depending on neutrino energy.  While an equivalent measurement is not available for the anti-neutrino running mode due to low statistics for the $\nubar-e$ process in that configuration, the known correlations between the neutrino and anti-neutrino fluxes are used to apply this constraint to the anti-neutrino flux distribution. As shown in Fig.~\ref{fig:flux_constraint_ratio}, applying the constraint results in a 1-3\% decrease in the anti-neutrino flux prediction.

\subsection{Neutrino Event Generation}
\label{sec:neutrinosim}
MINERvA uses a modified version of the GENIE neutrino interaction event generator~\cite{genie} version 2.8.4 to model physics processes within the primary interaction nucleus. Simulated event distributions using this generator, with data constraints described in section~\ref{sec:backgrounds},  are used to estimate background levels, resolution effects, acceptance and efficiency.  

 GENIE models the nucleus using the Relativistic Fermi Gas model
 \cite{rfg} incorporating the Bodek-Ritchie high-momentum tail
 \cite{Bodek1981} that simulates short-range correlations. For carbon,
 the maximum momentum for Fermi motion is taken as $k_F=0.221$~GeV/c,
 and Pauli blocking is also included. 
Quasi-elastic cross sections follow Llewellyn-Smith's prescription.  Vector form factors are modeled by default using the BBBA05 model \cite{Bradford2006}.  For the axial vector form factor $f_A$, a dipole form is used, with $f_A(0) =1.2670$ and axial mass $M_A=0.99$~ \text{GeV/}c$^2$~\cite{Kuzmin:2007kr}.

GENIE uses the Rein-Sehgal model \cite{ReinSehgal} to simulate baryon resonance production, which provides cross sections for 16 different resonance states.  The resonant axial mass  $M_A^{RES}$ is taken to be 1.12~GeV/c$^2$.   DIS cross sections are calculated with an effective leading order model with a low-$Q^2$ modification from Bodek and Yang \cite{bodekyang}. Hadronic showering is modeled with the AGKY model \cite{agky}. The Bodek-Yang model also describes other low-energy non-resonant pion production processes.  Re-scattering of nucleons and pions in the nucleus is simulated using the INTRANUKE-hA intra-nucleon hadron cascade package \cite{intranuke}. While the resonant interactions described earlier account for the
majority of pion production, other inelastic processes, as described
by Bodek-Yang \cite{bodekyang} are also possible. In  particular
non-resonant pion production followed by FSI can produce a QE-like
signature. 

\begin{figure}\centering 
                \includegraphics[width=1.0\columnwidth]{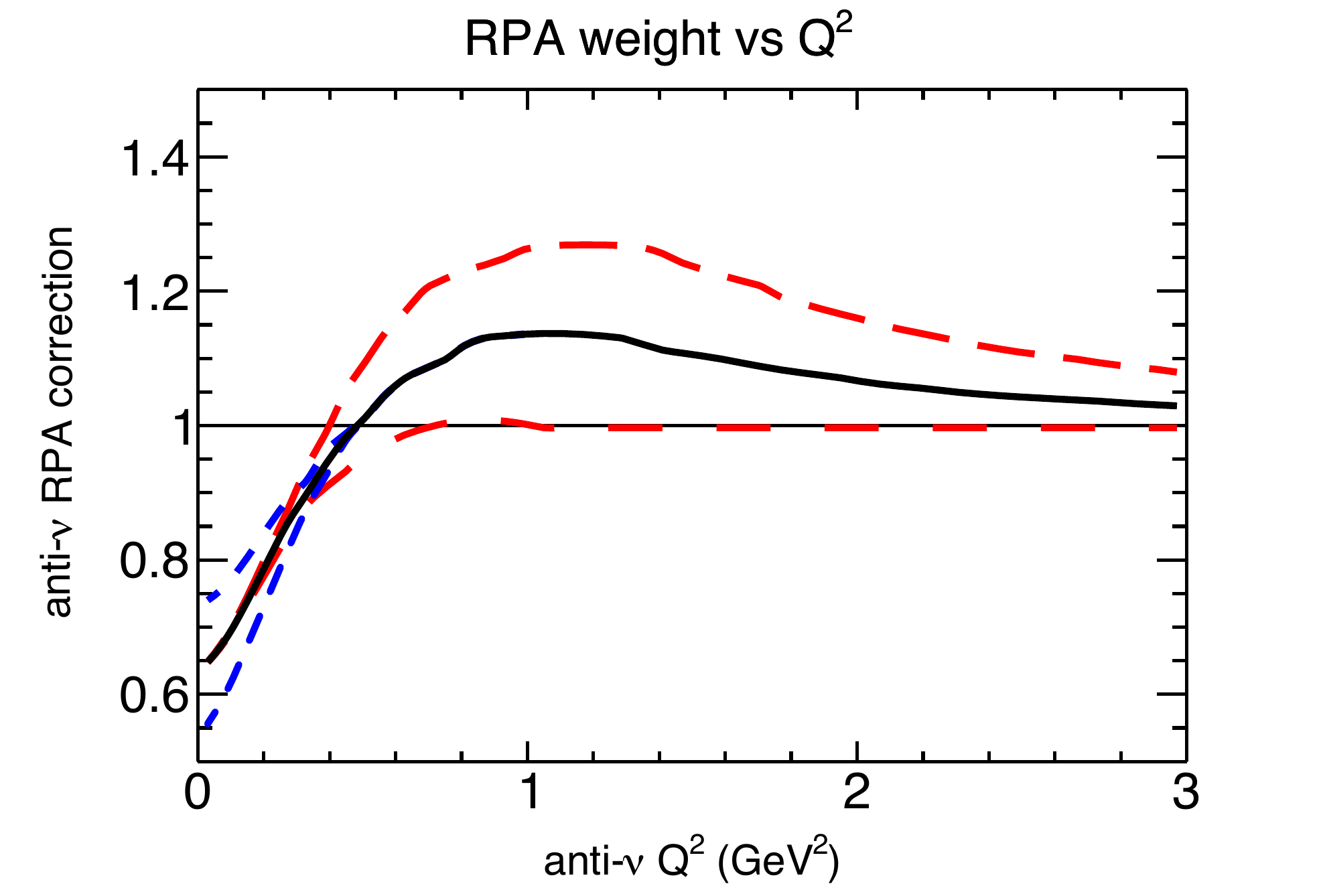}
                 \caption{\label{fig:rpa}Random Phase Approximation correction projected as a function of  generated $Q^2$.  The solid black curve indicates the central value from the relativistic calculation. The short dashed blue  lines indicate uncertainties from low $Q^2$ processes suggested by the application of the model of Nieves et al. to muon capture data, and the long dashed red lines a higher $Q^2$ uncertainty estimated from the difference between the relativistic and non-relativistic calculations. } 
\end{figure}

In addition to the basic processes simulated in GENIE 2.8.4 we also
apply three additional corrections.  First, we reweight quasi-elastic
events as a function of the energy and 3-momentum transfers $q_0$ and
$q_3$ to include   the Random Phase Approximation model as predicted
by the Valencia model of Nieves {\it et al.} \cite{Nieves:2004wx} and
implemented for MINERvA \cite{GranRPA}. Fig.~\ref{fig:rpa} shows the
$Q^2$ dependence of this correction.   Second, QE-like interactions on
multi-nucleon pairs are simulated using the Valencia IFIC model.  We
modify this model to match MINERvA inclusive neutrino scattering data
reported in ~\cite{Rodrigues:2015hik}, which enhances this
contribution by approximately 60\%,  \footnote{The IFIC model has a cutoff of $q_3 < 1.2$ GeV. While this is marginally within the MINERvA kinematic range, the bulk of the distribution is near 0.5$\pm$0.2 GeV and the contribution above 1.2 GeV would be negligible.}

  Finally, the normalization of non-resonant pion production is reduced to
  43\% of the default GENIE 2.8.4 prediction, based on a fit to
  pion-production data on deuterium from bubble-chamber experiments at
  Argonne and Brookhaven National Laboratories \cite{Rvn1pi}.  We
  reduce the uncertainty on the normalization of this process
  to 5\%, based on the same data fit. 
This modified version of GENIE 2.8.4 is hereafter referred to as \minerva-tuned GENIE.  

\subsection{Detector Simulation}
The \textsc{Geant4} toolkit \cite{geant4} v4.9.4p02 with the
QGSP\_BERT physics list is used to simulate  propagation through the
material of the detector and has been validated using a scaled-down
version of the MINERvA detector in a test beam~\cite{testbeam}. The optical and electronics systems are also
simulated, which allows the energy depositions recorded by
\textsc{Geant4} to be converted to a simulated readout that can be
analyzed as if it were MINERvA data.  This simulated data is overlaid
with actual data to include the effects of multiple neutrino
interactions, noise and dead-time, which is a result of the $\sim150$
ns digitization window following activity above threshold, during
which additional deposits will not be recorded.

\section{Event Reconstruction}
\label{sec:reco}

Calibrated energy depositions in the scintillator strips (referred to subsequently as `hits') are reconstructed into anti-neutrino interaction candidates through a series of steps.  First, the ensemble of hits collected over the $10\mu$s long NuMI beam spill are grouped into time slices corresponding to individual neutrino interactions. Hits within the same time slice are then collected into clusters that are adjacent in strip space and contained within the same scintillator plane. The position of the cluster is taken to be the energy-weighted average of the hit (strip) positions; the cluster time is set to the time of the highest-energy hit. 

\subsection{Track Reconstruction}\label{sec:track_reco}

Track reconstruction begins by collecting clusters within a single
time slice into `seeds' containing three clusters in consecutive
planes of the same (X,U or V) orientation that fit to a straight line.
Seeds are merged into track candidates within each view (X, U and V),
and candidates are formed into 3-dimensional tracks, which are fitted
with a Kalman filter
routine~\cite{Fruhwirth1987444,Luchsinger1993263}, in combination with
additional untracked clusters in planes adjacent to the track. This
allows tracks to be extrapolated through areas of high activity (such
as a hadron shower).  This algorithm is then repeated to identify
additional charged particles in the event until no further tracks are identified.

A similar reconstruction algorithm is performed in parallel in the MINOS detector, where time slices are selected by looking at hits clustered in space and time. The hits in a given time slice are then formed into clusters, which are grouped into tracks if their positions are correlated. Each track's path is then estimated using a Kalman filter; unlike in MINERvA, MINOS tracks curve due to the detector's magnetic field.  For tracks stopping within the detector and not entering the coil, the track's momentum is estimated via range; otherwise, the momentum is estimated via curvature through the Kalman fit.   For the data considered here, MINOS's magnet was configured to focus positive muons.  

Once tracks have been formed in both~\minerva and MINOS, they are then
matched between the two detectors.   MINOS tracks are matched to
MINERvA muons when activity is measured in the last five planes of
MINERvA, and a track starts in the first four planes of MINOS within
200ns of the MINERvA track time. The MINERvA track is extrapolated
forwards to the first MINOS plane, and the MINOS track is extrapolated
back to the last plane of MINERvA. The point of closest approach
between the two tracks is required to be less than 40 cm.

The final step of track reconstruction is known as muon ``cleaning''.  MINOS-\minerva matched tracks are deemed to be muons.  Energy beyond the expected deposition of a minimum-ionizing particle is removed from the muon track and added to the ensemble of unmatched clusters considered for further reconstruction.

\subsection{Recoil energy reconstruction}\label{sec:recoil_reco}
We refer to final-state energy not associated with the muon track  as ``recoil energy''.  In this study we consider only energy deposited in the tracker and ECAL portions of the detector, and further require recoil cluster times to be between 20 ns before and 35 ns after the path-length-corrected average time of clusters on the muon track.  We also exclude all clusters likely to be due to PMT cross talk and clusters within 10 cm of the muon vertex from the recoil energy sum, to minimize dependence on simulations of energy near the vertex, which are sensitive to details of final-state and multi-nucleon interactions.  Energy in all remaining clusters is summed and calorimetrically corrected:
\begin{equation}\label{eq:total_recoil}
E_{\text{recoil}}\equiv \sum\limits_{i} C^{sd}_i E_i
\end{equation}
where $C^{sd}_i$ is a calorimetric constant obtained from the simulation for sub-detector $i$ that corrects for the passive material fraction in that sub-detector (1.22 for the tracker and 2.013 for the ECAL).

\section{Event Selection}
\label{sec:selection}\label{sec:signal}

Before identifying selection criteria for isolating signal events, it
is necessary to clearly define what is meant by ``signal''.  For
MINERvA's first studies of quasi-elastic
scattering~\cite{laura,arturo}, we attempted to measure events
in which the underlying neutrino-nucleon interaction was
quasi-elastic, regardless of how those events were modified by final
state interactions.  Several other experiments have recently published
measurements~\cite{miniboone_antinu, minibooneccqe,t2kccqe} of QE-like
events with a final state of an appropriately-charged muon, plus
nucleons.  In this case, resonant pion production events where the
pion is absorbed become part of the signal to be measured.  However in
MINERvA's scintillator tracker, which is able to resolve proton tracks
above a kinetic energy of 120 MeV, and to detect the energy of lower
energy particles, this definition is not ideal.   For this study, we
define our signal to be events that are anti-neutrino charged-current events occurring in the MINERvA tracker fiducial volume, have post-FSI final states without mesons, prompt photons above nuclear de-excitation energies, heavy baryons, or protons above our proton tracking kinetic energy threshold of 120 MeV, and  include a muon emitted at an angle with respect to the beam of less than 20 degrees, $1.5$~GeV$<p_\parallel<15$~GeV and $p_T<1.5$~GeV (matching the region where tracks can be reconstructed in both \minerva and MINOS with well-reconstructed momentum).
This is similar to the QE-like (often called CC0Pi) definitions used
by other experiments~\cite{minibooneccqe,t2k_ccqe2016}, modified
slightly to correspond to MINERvA's acceptance, which is poor for events with high angle muons, very low or very high momentum muons, but able to reject high momentum protons.  We also report alternate results where the signal definition consists of interactions that were initially generated in GENIE as quasi-elastic (that is, no resonant or deep inelastic scatters, but including scatters from nucleons in correlated pairs with zero-meson final states), regardless of the final-state particles produced.  

We begin the event selection by identifying time slices containing at
least one track reconstructed in the~\minerva detector and matched to
a track in the MINOS detector as described in
Section~\ref{sec:track_reco}.  This provides a high purity sample of
charged-current events.  To isolate anti-neutrino event candidates, we
further require that the charge-momentum ratio (q/p) returned by the
MINOS Kalman fit be positive. Because we also require no visible
proton in the final state, the remaining neutrino contamination in our
samples is quite low -  0.6\% in simulation - and is accounted for in
the acceptance calculation.    Because MINERvA experiences some dead
time  after an event has been recorded, we further require that no
more than one strip immediately upstream of the track vertex
(projected along the track direction) or immediately adjacent to these
strips be dead at the time of the neutrino event.  This reduces
candidates arising from through-going muons generated upstream of the
detector to less than 0.1\% .   We require the reconstructed interaction vertex to be within the fiducial volume of our detector; the vertex must be within a hexagon of apothem 850~mm and fall within modules 27 to 80, inclusive, corresponding to 108 tracking planes.  We also require our reconstructed muon longitudinal momentum to be less than 15~GeV. This removes very energetic, forward-going muons that have poor energy reconstruction in MINOS.  

To reduce backgrounds from non-QE-like events, we require that no
tracks other than the muon track be reconstructed between 20 ns before
and 35 ns after the muon track (the same time window used for recoil
energy reconstruction).  This reduces backgrounds from events with
charged pions, particularly at high \qsq where the recoil cut
described below is very loose, while the narrow time window minimizes the likelihood that signal events are rejected due to overlapping neutrino interactions.

\newcommand{\colorkey}{ In the simulation, signal events include both quasi-elastic events (purple), 2p2h scatters (green) and resonant or DIS events (pink and red) with a QE-like signature. The backgrounds consist of quasi-elastic and 2p2h events with non-QE-like signature (hatched purple and green), and non-QE events (resonant and DIS) without a QE-like signature (hatched pink and red).  }

\begin{figure}[ht]
\centering
\includegraphics[width=1.0\columnwidth]{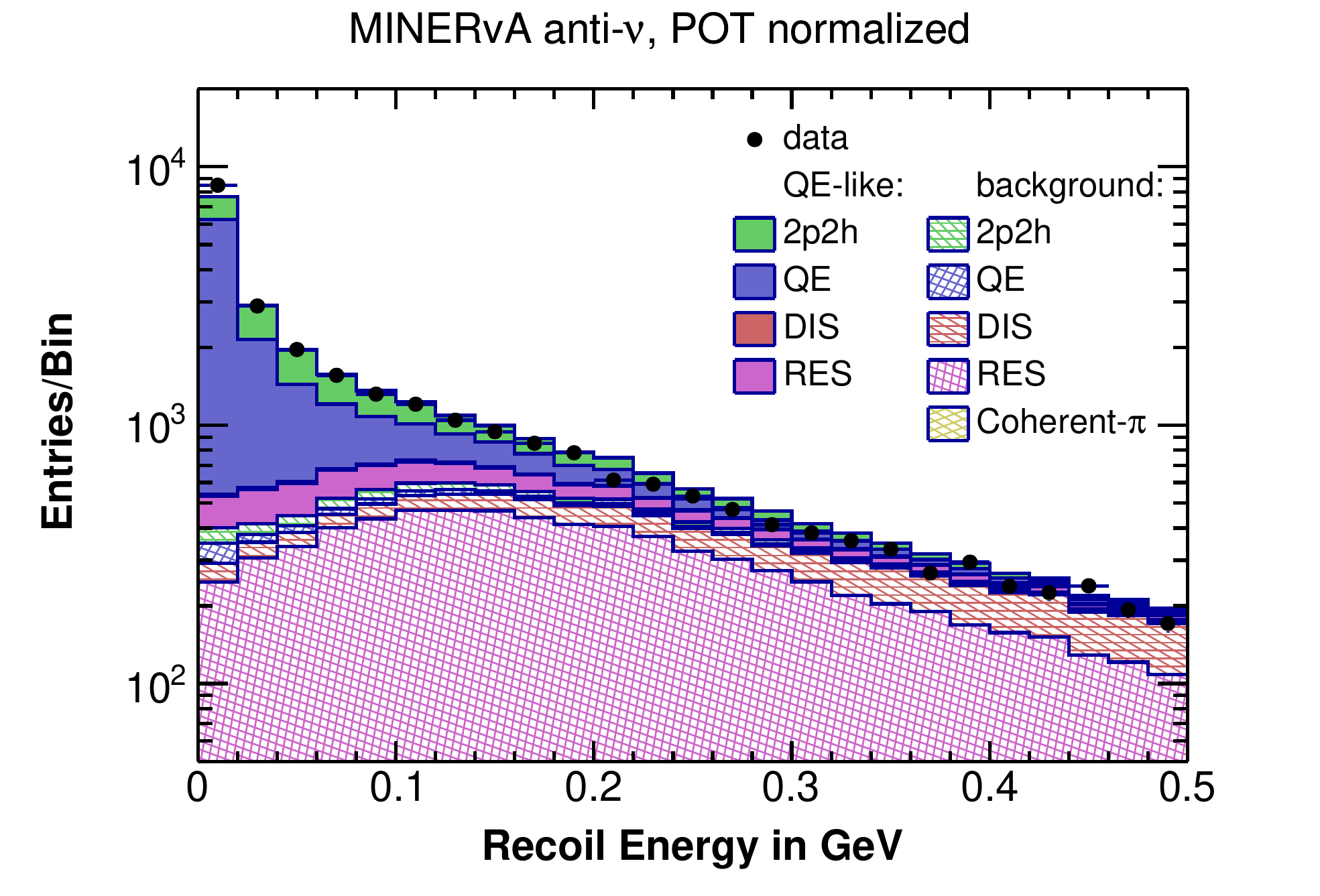}
\caption{ \small Recoil energy for data (points) and simulation (colors, POT-normalized to data).\colorkey  All cuts except the recoil cut are applied.}
\label{fig:recoil_energy}
\end{figure}

\begin{figure}[ht]
\centering
\includegraphics[width=1.0\columnwidth]{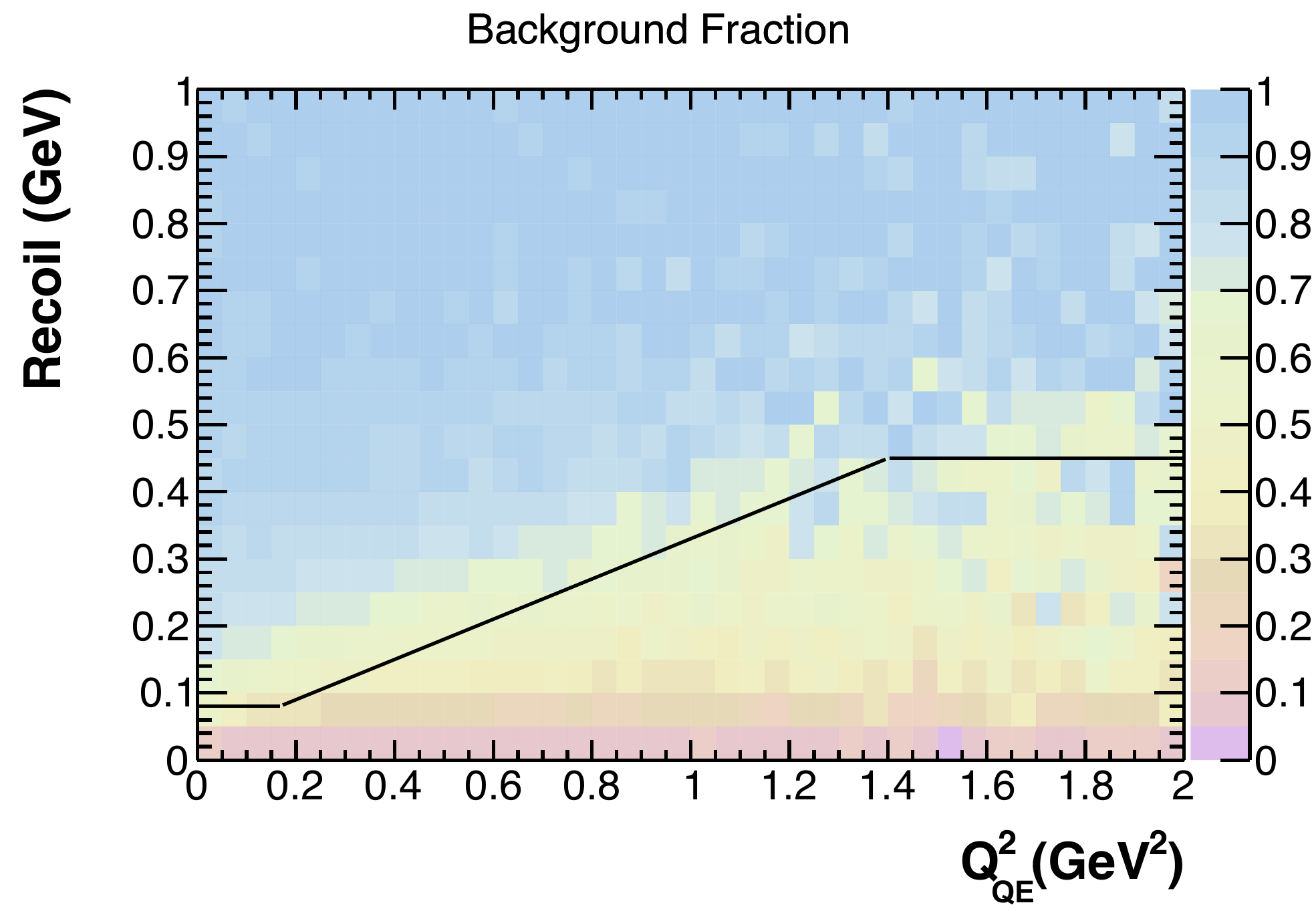}
\caption{ \small Simulated QE-like background fraction before the recoil cut, in bins of $Q^2$ and recoil energy.  The line shows the $Q^2$ dependent recoil energy selection used to optimize efficiency and purity.}
\label{fig:recoil_energy2}
\end{figure}

Charged pions and high-energy protons do not always leave reconstructable tracks; they do, however, deposit clusters of energy in the detector. We therefore consider recoil energy, reconstructed as described in Section~\ref{sec:recoil_reco} and shown in Fig.~\ref{fig:recoil_energy}. We find that the purity  the QE-like sample depends on both the recoil energy and on the \qsq of the interaction, with high \qsq interactions having larger recoil (see Fig.~\ref{fig:recoil_energy2}). We therefore apply a \qsq dependent cut on the recoil energy:
\begin{eqnarray*}
E_{recoil} &<&  \max(0.08, 0.03 + 0.3\times Q^2_{QE}) \mathrm{\:GeV\:and} \\
E_{recoil} &<& 0.450 \mathrm{\:GeV} 
\end{eqnarray*}
where \qsq is in units of GeV$^2$.

\section{Cross Section Extraction}
\label{sec:crosssection}

The double-differential cross section versus variables $x$ and $y$ in bin $(i,j)$ is constructed using:

\begin{equation}\label{eq:xsec}
{\left(\frac{\mathop{d^2\sigma}}{\mathop{dx} \mathop{dy}}\right)}_{ij} = \frac{ U\left(N_{\alpha\beta} - N^{bkgd}_{\alpha\beta}\right)}{\epsilon_{ij}(\Phi T ) (\Delta x_i) (\Delta y_j)}
\end{equation}

Where  $N_{\alpha\beta} $ is the number of data events reconstructed
in bin $(\alpha,\beta)$, $N^{bkgd}_{\alpha\beta} $ is the estimated
number of background events reconstructed in bin $(\alpha,\beta)$, $U$
is an unfolding operation transforming reconstructed bin
$(\alpha,\beta)$ to true bin $(i,j)$, $\epsilon_{ij}$ is the product
of reconstruction efficiency and detector acceptance for events in
true bin $(i,j)$, $\Phi$ is the flux of incoming anti-neutrinos
(either integrated or for the given bin, as described later), $T$ is
the number of scattering targets (here, the number of nucleons), and 
$\Delta x_i$ ($\Delta y_j$)is the width of bin $i$ ($j$).

We report our primary cross section measurement in bins of muon transverse (\ptnospace) and longitudinal momentum (\pznospace) with respect to the neutrino beam direction.   We choose these as our primary results as they are quantities that we have directly measured.  For comparison with other experiments, we also report auxiliary measurements vs. \qsq and \enunospace, both reconstructed in the quasi-elastic hypothesis from the muon kinematics (see Eqs. \ref{eq:enu} and \ref{eq:qsq}). The bin boundaries are shown in Table \ref{tab:binboundaries}.

%\onecolumngrid
\begin{table}[t]
\begin{tabular}{l l}
\pt (GeV/c) & $0,0.15,0.25,0.4,0.7,1.0,1.5$ \\
\pz (GeV/c) & $1.5,2.0,2.5,3.0,3.5,4.0,5.0,6.0,8.0,10.0,15.0$ \\
\qsq (GeV$^2$) & $0.0,0.025,0.05,0.1,0.2,0.4,0.8,1.2,2.0$  \\
\enu (GeV) & $1.5,2.0,2.5,3.0,3.5,4.0,5.0,6.0,7.0,8.0,10.0$  \\
\end{tabular}
\caption{Bin boundaries.\label{tab:binboundaries}}
\end{table}
%\twocolumngrid 

\begin{figure}[ht]
\centering
\includegraphics[width=1.0\columnwidth]{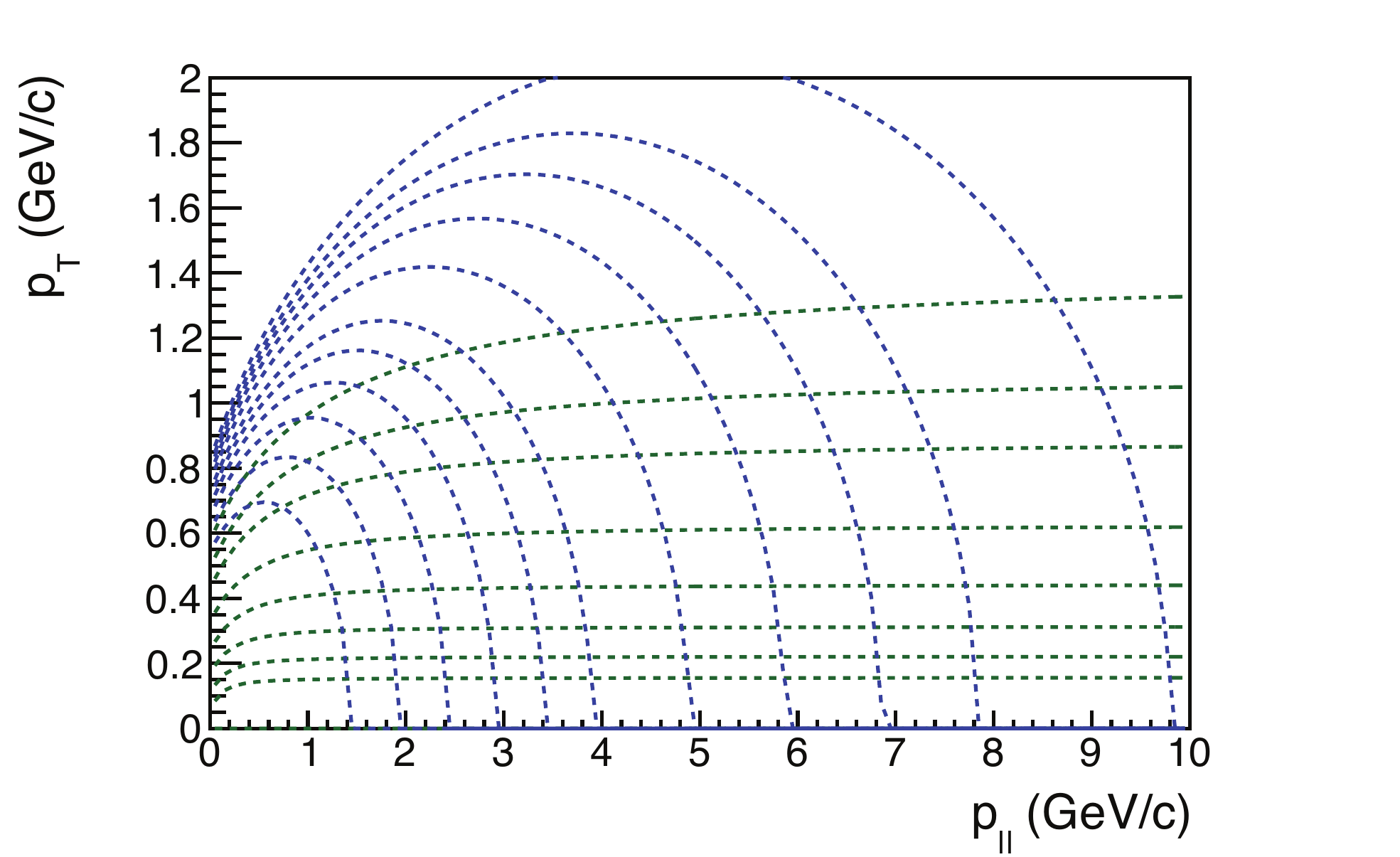}
\caption{ \small Relationship between \enu and \qsq in the
  quasi-elastic hypothesis, and muon kinematic variables \pt and
  \pznospace. The dashed lines show constant values of \enunospace
  (blue)  and \qsq (green) corresponding to our \enu and \qsq bin
  boundaries, which are given in Table~\ref{tab:binboundaries}.}
\label{fig:enu_qsq_lines}
\end{figure}

  Two bins at highest \pt and lowest \pz and four bins at highest \qsq
  and lowest \enu are not reported due to poor acceptance in those
  regions.  Note that, as \qsq and \enu are reconstructed from the
  muon kinematics, they are both functions of both \pz and \ptnospace. Figure \ref{fig:enu_qsq_lines} shows lines of constant \qsq and \enunospace, projected onto the \pznospace/\pt phase space. For most of the region considered by this analysis, \enu correlates fairly well with \pznospace, and \qsq with \pt. This simplification breaks down at high \pt and low \pznospace.  For both versions of the double-differential cross sections, we also report projections onto each axis, resulting in one-dimensional distributions of \ptnospace, \pznospace, \qsqnospace, and \enunospace.  

For the single differential cross section versus \enunospace, we report a flux-weighted cross section, where each bin has been divided by the flux integrated over the energy range of that bin only, rather than the entire anti-neutrino flux integrated over all energies.  Note that care must be taken in interpreting this quantity, as \enu does not correspond exactly to true anti-neutrino energy.

\begin{figure*}

         \includegraphics[width=0.9\textwidth]{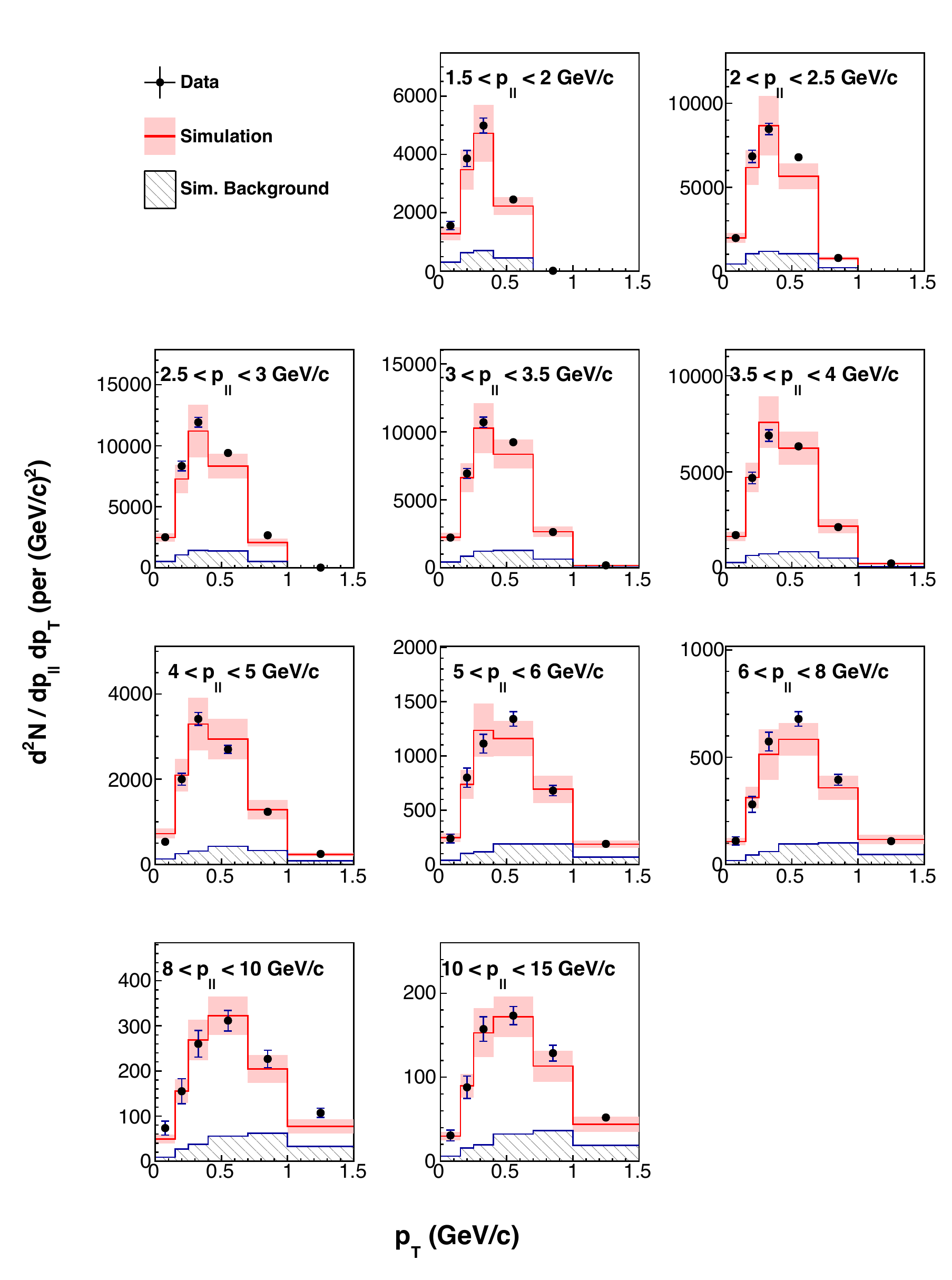}
         \caption{Reconstructed event counts  vs. muon transverse momentum, in bins of muon longitudinal momentum.  Uncertainties  on the data are indicated by error bars; uncertainty on the Monte Carlo is indicated by a pink shaded bar.  The data uncertainty is statistical; the Monte Carlo simulation includes all sources of systematic uncertainty, including uncertainties on the GENIE signal model.  The estimated background  contribution is shown by the hatched area.}      
\label{fig:raw_pt_binned}

 \end{figure*}

A total of 17,621 interactions pass our reconstruction cuts for data. Distributions of these events versus muon \ptnospace, in bins of \pz are shown in Fig.~\ref{fig:raw_pt_binned}. 

\subsection{Background subtraction}
\label{sec:backgrounds}

\begin{figure*}[ht]
\centering
\includegraphics[width=0.49\textwidth]{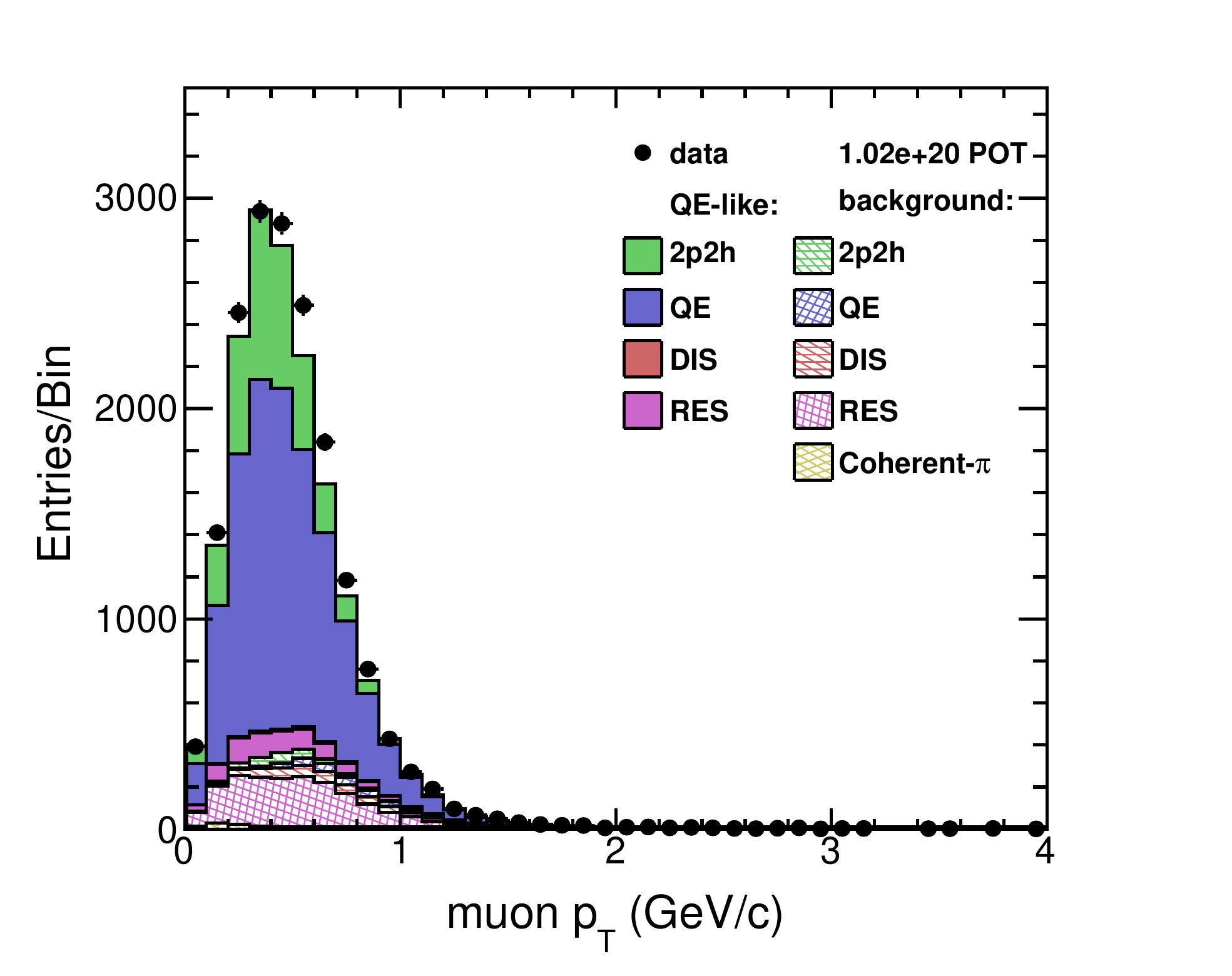}
\includegraphics[width=0.49\textwidth]{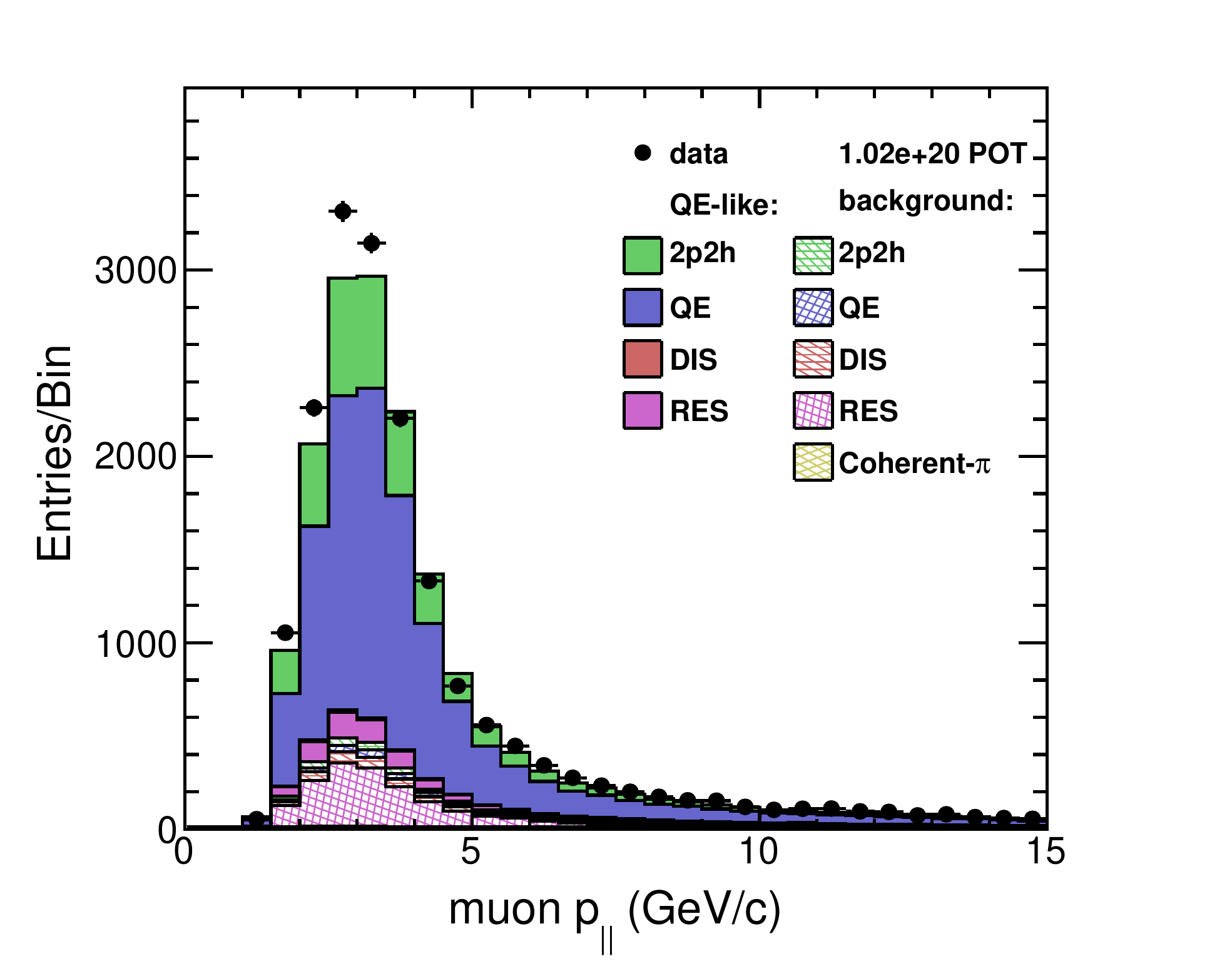}
\includegraphics[width=0.49\textwidth]{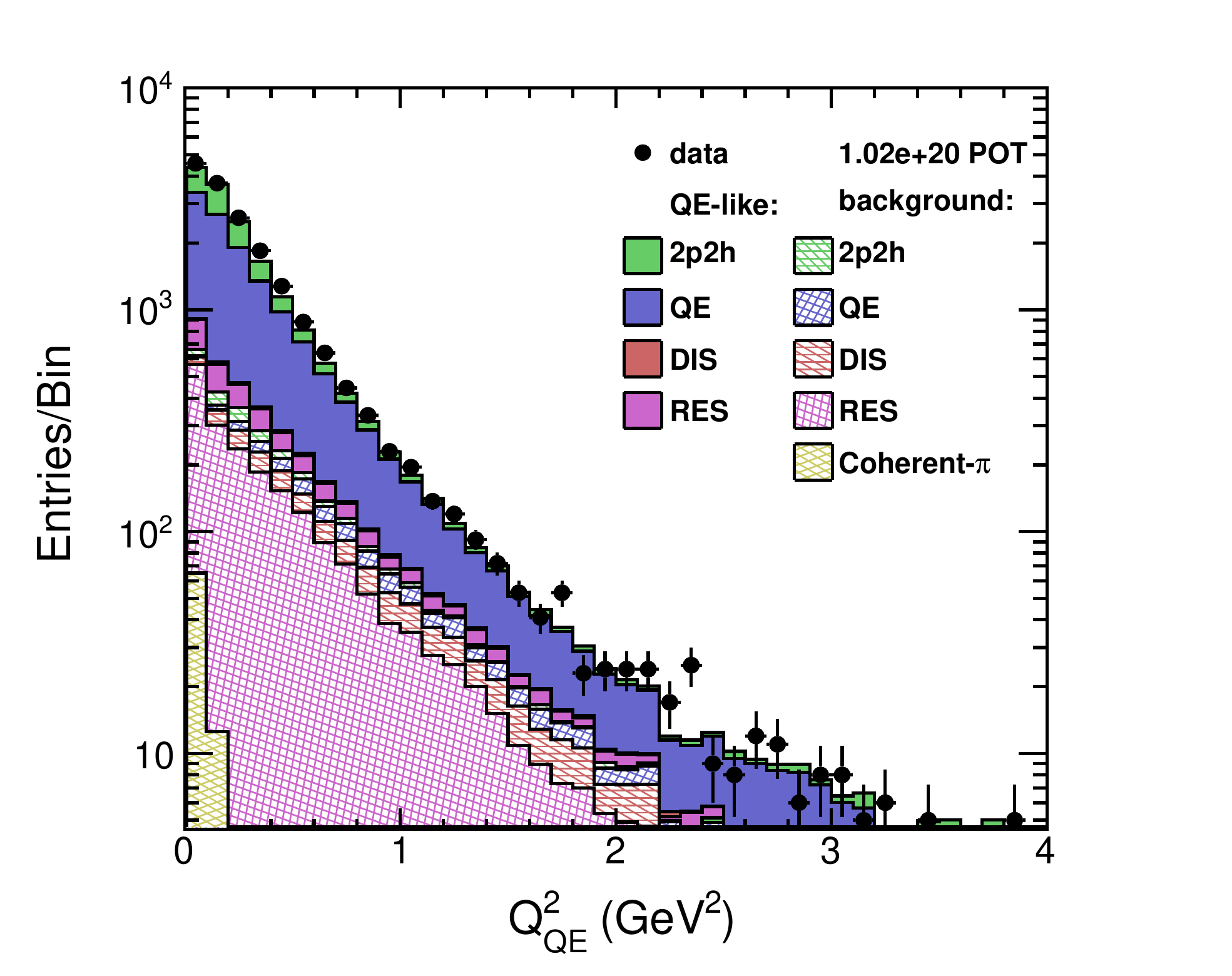}
\includegraphics[width=0.49\textwidth]{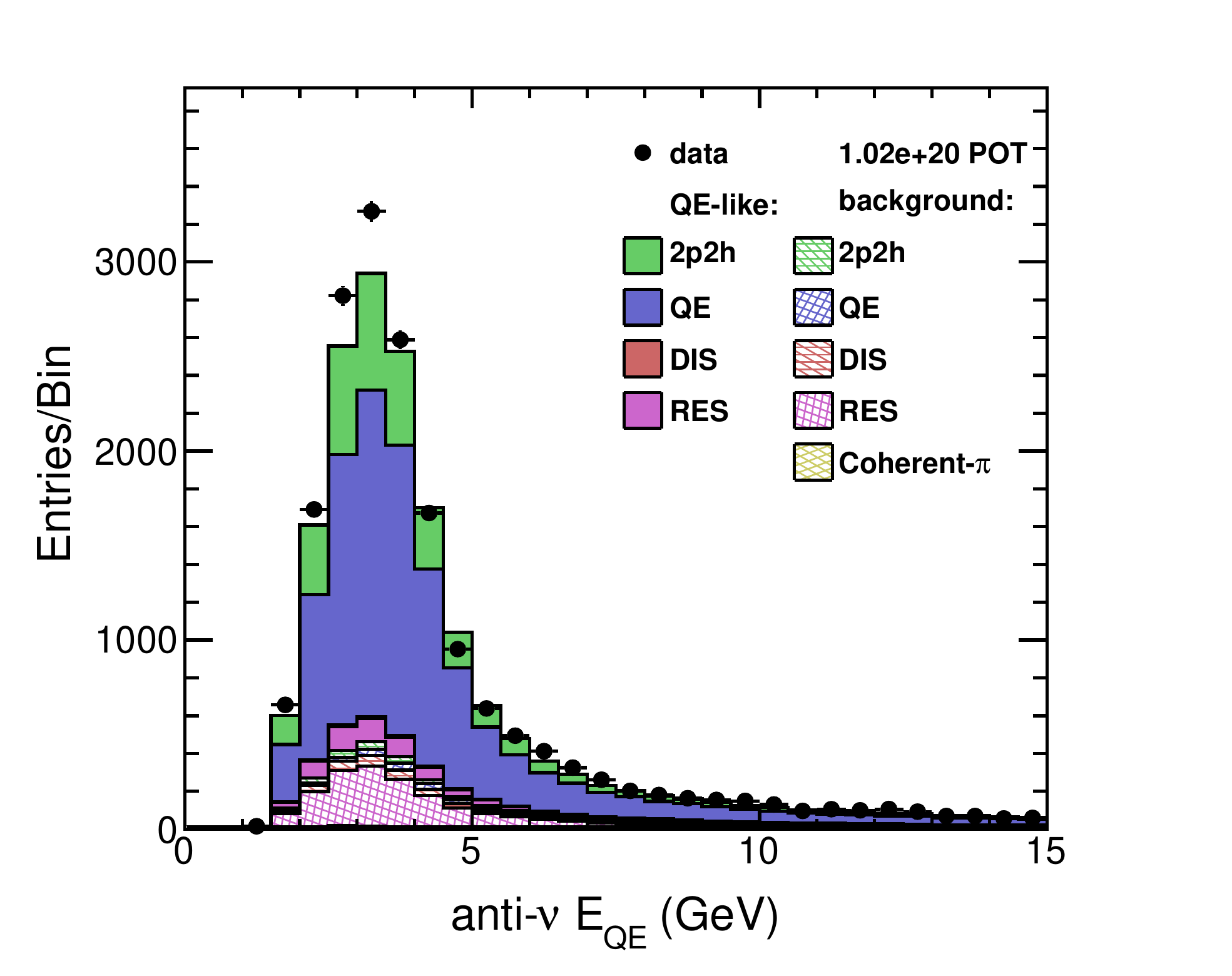}
\caption{Distributions of signal and background events vs. muon
  transverse and longitudinal momentum, \qsq and \enu. Here the
  \minerva-tuned GENIE simulation is absolutely normalized to the POT of the data sample, but the background corrections described in this section have not yet been applied.  The \qsq distribution is shown on a log scale to highlight the high \qsq region. \colorkey}
\label{fig:sig_bkgd_ptpz}
\end{figure*}

The term $N^{bkgd}_{\alpha\beta} $ in Eq. \ref{eq:xsec} refers to the estimated number of reconstructed data events that correspond to background processes.   Recall that our QE-like signal, explained in section \ref{sec:selection}, is defined as having a final state containing a $\mu^+$, any number of neutrons, any number of protons with less than 120~MeV kinetic energy, and no pions, other hadrons, or prompt photons. Thus, background events in our sample could, for example, correspond to resonant events with pions that did not make a track, and that generated recoil distributions that fell within our cuts. Figure~\ref{fig:sig_bkgd_ptpz} shows \pt, \pz, \enu, and \qsq distributions in the data and simulation, with the latter subdivided into signal and background.

%\onecolumngrid

%\twocolumngrid

Backgrounds in this analysis arise primarily from events involving charged pions.   MINERvA's charged pion production analysis \cite{McGivern:2016bwh} suggests that GENIE 2.8.4 over-predicts the rate of resonant pion production. We therefore use a data-driven fitting procedure to constrain the backgrounds predicted by GENIE.  Since the constraint can in principle be different in each $p_T/p_\parallel$ bin, the fit would ideally be done separately in each bin.  However, the limited statistics of our data sample caused attempts to fit each bin separately to fail.  The fits are instead performed separately for five regions of the $p_T/p_\parallel$ phase spaces, chosen by combining $p_T/p_\parallel$ bins with similar background shapes. 

For each of the five regions, the recoil energy, after all other cuts, is compared for data, and for signal and background Monte Carlo.  
 The TFractionFitter tool, part of the ROOT framework \cite{root}, is used to perform a fractional fit of the simulation to data, where the relative normalization of the signal and background distributions is allowed to vary.  The shapes of the distributions are not varied.   

\begin{figure}[h]
\centering
%\begin{subfigure}[b]{0.48\textwidth}
\includegraphics[width=1.0\columnwidth]{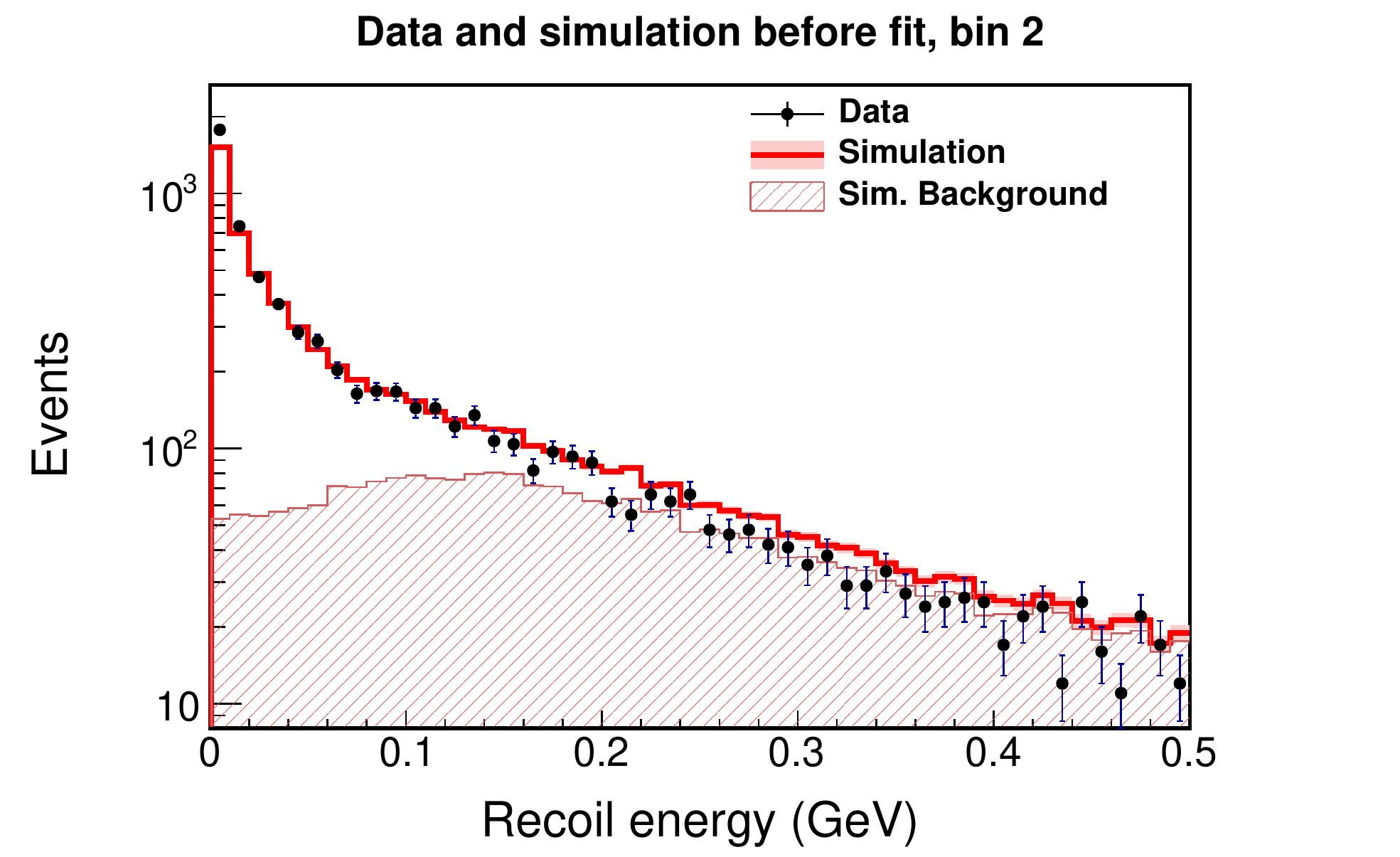}
%\end{subfigure}
\includegraphics[width=1.0\columnwidth]{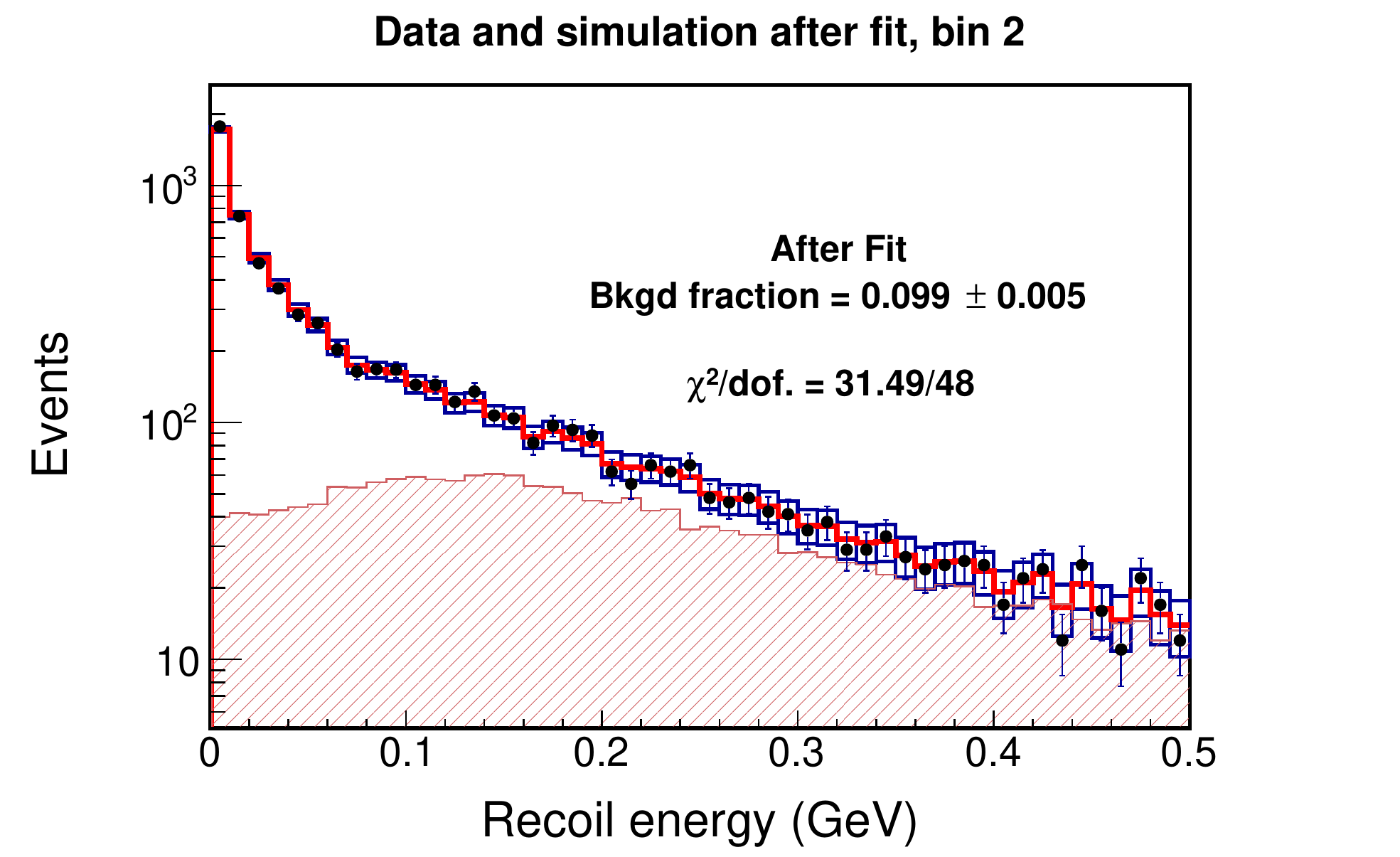}
\caption{Area normalized recoil distributions before (above) and after (below) background tuning for the bin corresponding to $0.25< p_T < 0.4$~GeV. The blue boxes indicate the uncertainty  in the simulation (red curve) estimated by TFractionFitter. }
\label{fig:bkgd_ptpz_bin_2}
\end{figure}

Figure \ref{fig:bkgd_ptpz_bin_2} shows the recoil distributions in data and (area-normalized) simulation for one of the five regions of  $p_T/p_\parallel$, before and after tuning the signal and background fractions. In each bin, a scale is extracted corresponding to the factor by which the background fraction was rescaled relative to the nominal simulation to give the best fit. The estimated background fraction in each bin of the data distribution corresponds to the background fraction of the Monte Carlo in that bin, multiplied by this scale factor.

The scales for the $p_T$ vs. $p_\parallel$ regions are shown in Table \ref{tab:bkgd_scales}. In most cases, as suggested by \cite{PhysRevD.92.092008}, the simulation is found to predict too high a fraction of  background events. 

\

\begin{table*}[ht]
\centering
\begin{tabular}
%{| p{5.8cm}  | p{1.8cm}| p{2.8cm}  | p{1.8cm} | p{1.8cm}|}
{ l    r  r  r  r  r }
\hline\hline
Bin & \pt range  & \pz range & background  & background &  $\chi^2$/dof. \\
&(GeV/c) & (GeV/c)& rescale factor&fraction &\\
\hline

0 & 0.00 - 0.15  & 1.5 - 15 &0.609$\pm$0.060&0.130$\pm$0.013&0.68\\%BACKFITTEX
1 & 0.15 - 0.25 & 1.5 - 15 &0.680$\pm$0.046&0.110$\pm$0.008&0.70\\%BACKFITTEX
2 & 0.25 - 0.40  & 1.5 - 15 &0.750$\pm$0.034&0.099$\pm$0.005&0.64\\%BACKFITTEX
3 & 0.40 - 1.50 & 1.5 - 4.0&0.840$\pm$0.033&0.17$\pm$0.007&0.78\\%BACKFITTEX
4 & 0.40 - 1.50  & 4.0 - 15&1.00$\pm$0.046&0.25$\pm$0.011&0.50\\%BACKFITTEX
\hline
\hline

\end{tabular}
\caption{ Summary of the fits to determine the background fraction in data. The scale applied to the background to match the data, the resulting background fraction in the signal region  and  the $\chi^2$/DOF of the fit are shown. } % TFractionFitter introduces additional degrees of freedom to the fit to account for statistical fluctuation in the simulation template shapes, causing the $\chi^2$/DOF after the fit to be  }
\label{tab:bkgd_scales}
\end{table*}

\begin{figure*}[t]
\centering
\includegraphics[width=1.0\columnwidth]{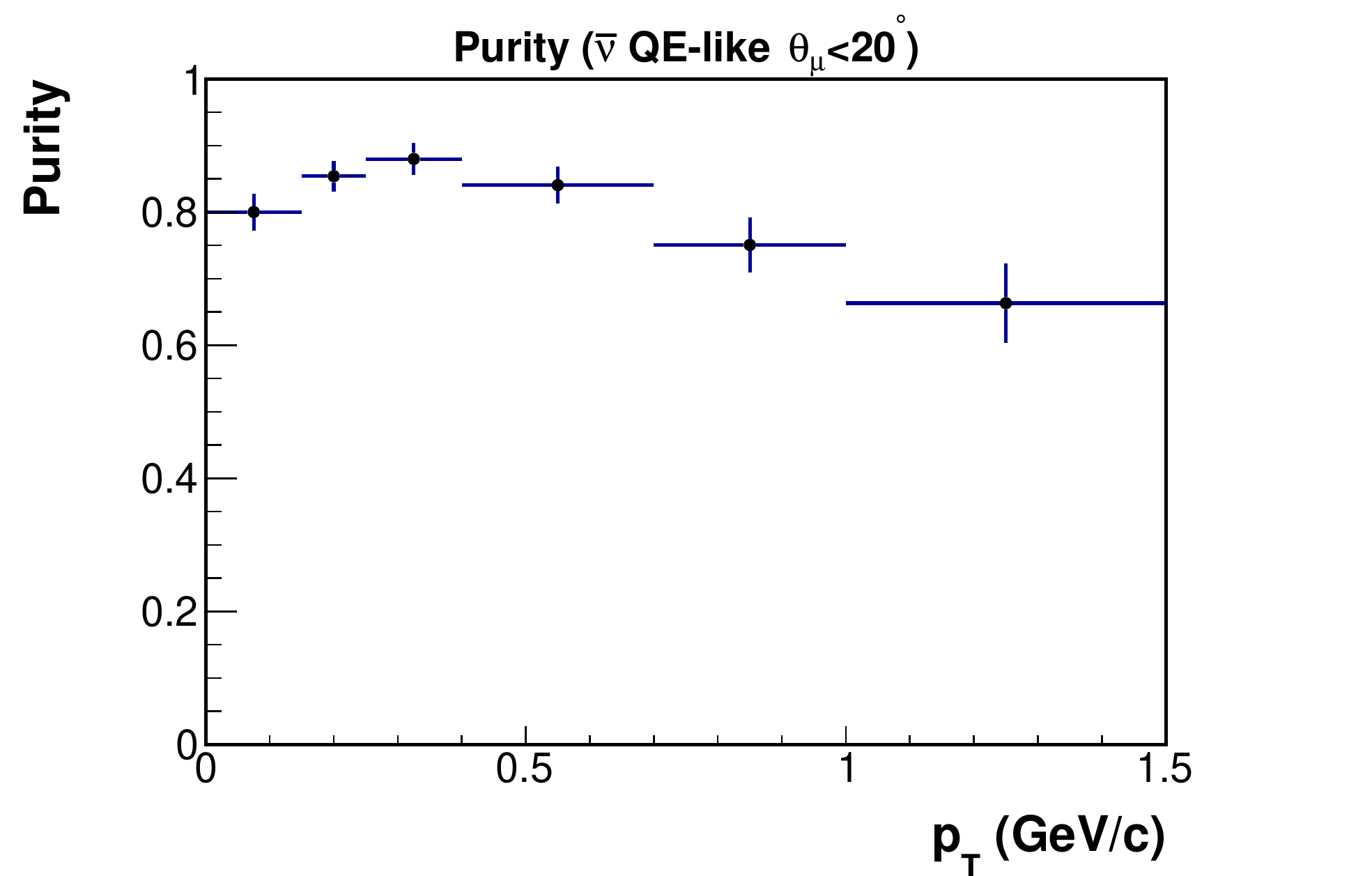}
\includegraphics[width=1.0\columnwidth]{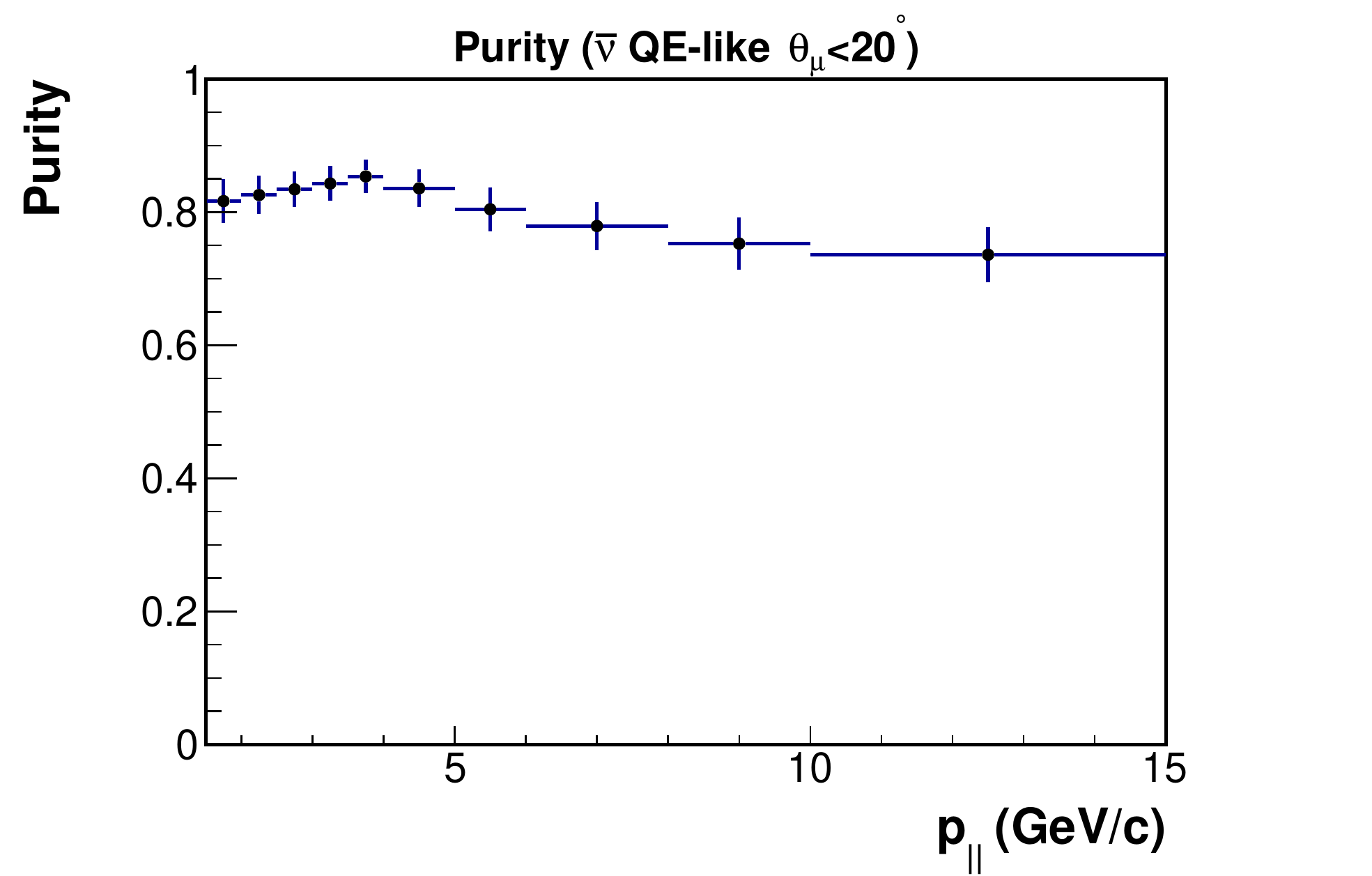}
\caption{ Purity (signal/total) as a function of kinematic variables.}
\label{fig:purity}
\end{figure*}

Figure \ref{fig:purity} shows the signal fraction as a function of the
muon kinematic variables.   After background subtraction, the signal
data sample has estimated 14,839 events.

%\onecolumngrid

\subsection{Unfolding}

\begin {figure}\centering
\includegraphics[width=1.0\columnwidth]{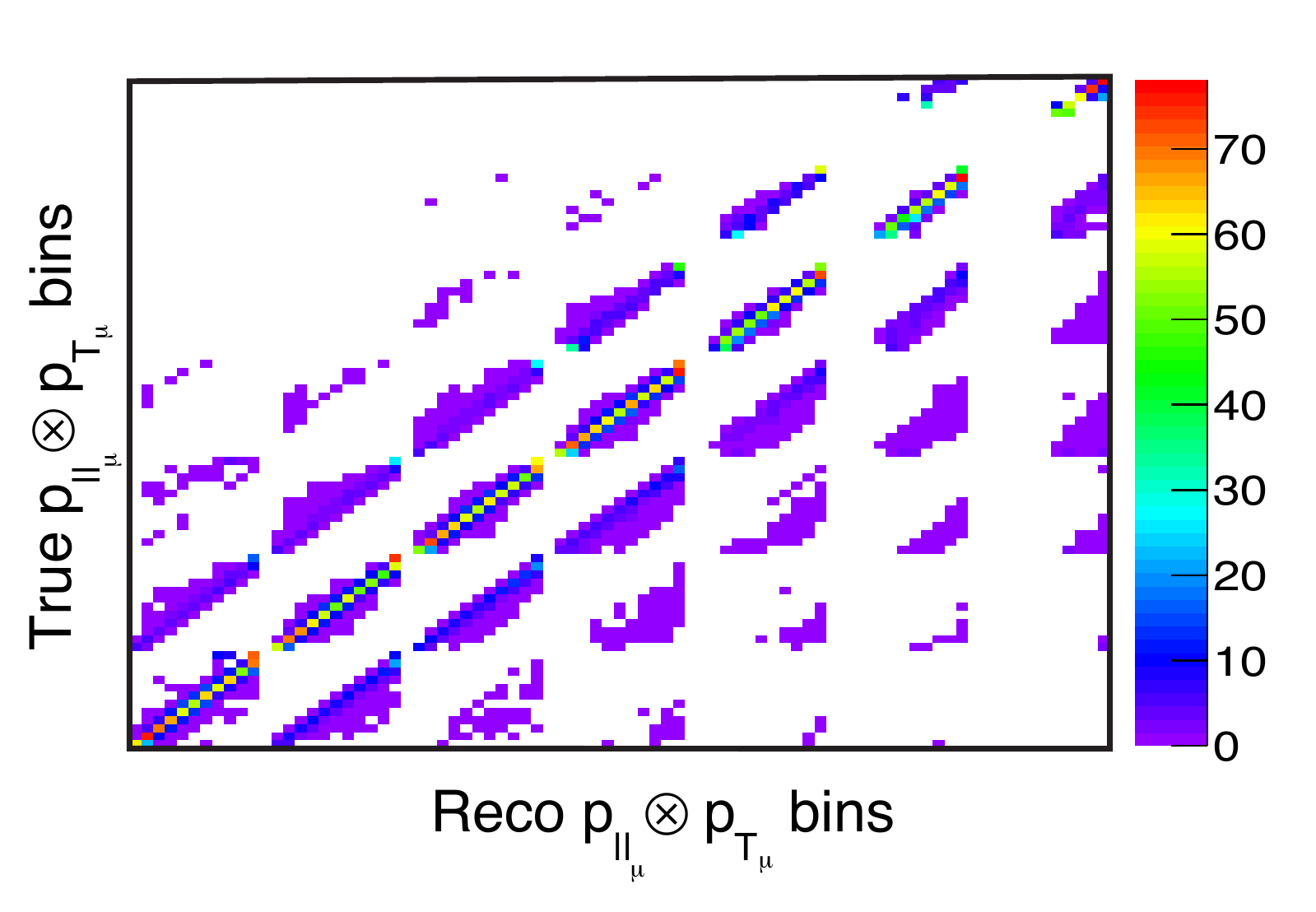}
        	\caption{ Migration matrix for the \pz$\otimes$\pt distribution. The $x$ axis corresponds to reconstructed bins, the $y$ to true. The large cells correspond to $p_T$ bins while the small cells are the \pz bins within a \pt bin.  The high \pt overflow bin is included.} 
\label{fig:migration}
\end{figure}

Detector smearing is corrected using a migration matrix that describes the relationship between true and reconstructed bins of \pt and \pznospace.  The migration matrix for our simulated reconstructed QE-like signal distribution is shown in Fig.~\ref{fig:migration}. The $x$ axis indicates bins in the reconstructed variables, where the bins of \pz are repeated for each bin of \pt. The $y$ axis indicates bins in the true variables, arranged in the same way. Thus any events on the diagonal were reconstructed in the correct bin of both \pz and \pt. An event reconstructed in the wrong bin of  \pz  (but the right \pt bin) will be  displayed in another bin in the same subplot; one reconstructed in the wrong \pt bin will appear in a different subplot.

We use the iterative method of D'Agostini~\cite{bayesian}, as
implemented in the ROOT package RooUnfold \cite{roounfold}, with four
iterations.  The unfolding procedure was validated using an ensemble
test, in which ten data-sized subsamples of the simulation were
selected and warped by an adjustment of the quasi-elastic axial mass
by $\pm25$\%. These samples were then unfolded using the migration
matrix generated from the full un-warped simulation; the warped
simulation was recovered within four iterations.   

\subsection{Efficiency and acceptance correction}

The unfolded distributions are then corrected for detector acceptance
and reconstruction efficiency.  The most significant effect on
acceptance is from the requirement that final-state muons are matched
in MINOS, limiting the muon's angle with respect to the beam line to a
maximum of $20^\circ$.  The MINOS-match requirement also limits our
ability to accept muons with low longitudinal momentum $\lesssim 1.5$
GeV/c which will stop in MINERvA or not produce enough activity to be
analyzed in the MINOS spectrometer.  The largest source of inefficiency is due to the \qsq-dependent $E_{\text{recoil}}$ cut.

We estimate the product of acceptance and efficiency using the full
\minerva-tuned GENIE+\textsc{Geant4} simulation:
\begin{equation}\label{eq:acceptance}
\epsilon_{ij} = \frac {N_{ij}^{\text{generated and reconstructed}}} {N_{ij}^{\text{generated}}},
\end{equation}
where ${N_{ij}^{\text{generated and reconstructed}}}$ is the number of simulated events generated in \pt bin i and \pz bin j that also pass  all reconstruction cuts (except the fiducial cuts on position and muon angle), and $ {N_{ij}^{\text{generated}}}$ is the total number of events generated in \pt bin $i$ and \pz bin $j$. 

\begin{figure}[t]
\centering
\includegraphics[width=1.0\columnwidth]{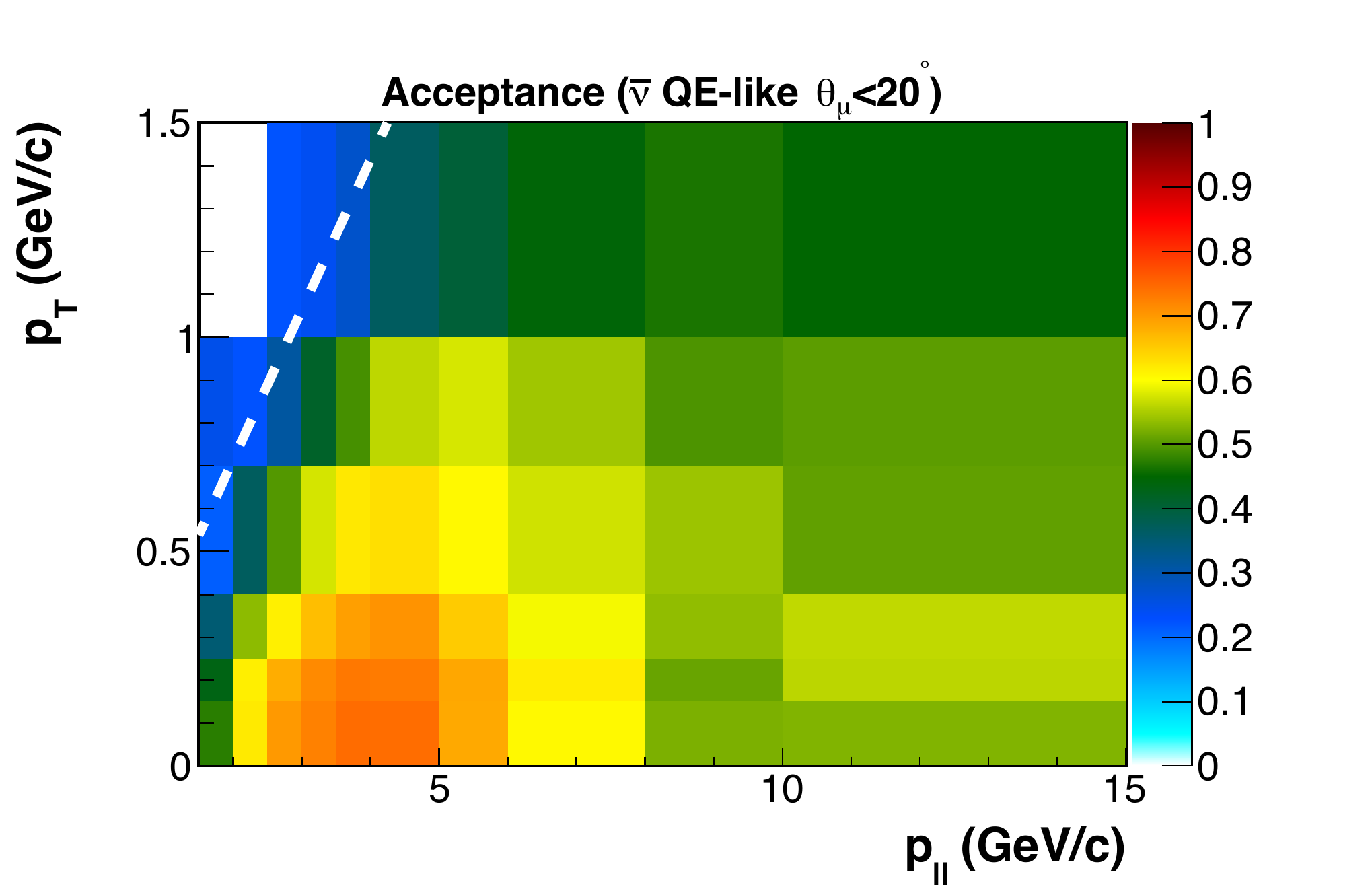}
\caption{Efficiency $\times$ acceptance versus generated \pt and  \pz, for QE-like events in the simulation. The dashed white line illustrates the 20 degree limit imposed on reconstructed muon angles.}
\label{fig:efficiency}
\end{figure}

Figure \ref{fig:efficiency} shows the product of efficiency and acceptance vs. \pz and \pt. The low acceptance at high \pt and low \pz is due to the MINOS match requirement and angle cut. The efficiency also decreases at higher energies, where interactions are more likely to include large amounts of recoil energy and may be vetoed by our \qsq-dependent  $E_{\text{recoil}}$ cut.  The overall efficiency $\times$ acceptance  of the sample is 52.5\%.

\subsection{Flux and Target Number Correction}
\label{sec:flux}

To convert an acceptance-corrected distribution to a cross section, we divide by the number of nucleons, the total number of protons on target (POT) producing the neutrino beam, and the estimated anti-neutrino flux per POT. These are summarized in Table \ref{tab:pot_etc}.

\begin{table}[h]\begin{center}
\begin{tabular}
{c  c }
\hline\hline
Quantity & Value\\
\hline
Protons on target (data) & $1.020\times 10^{20}$ \\
Protons on target (simulation) & $9.247\times 10^{20}$  \\
Number of targets & $3.23478\times 10^{30}$ nucleons \\
Integrated flux &\ $2.340\times10^{-8}\; \bar{\nu}_\mu / \text{cm}^{2}$ / POT\\ 
\hline\hline
\end{tabular}
 \caption{{Normalization factors used in the cross section calculations. The flux used in the quoted cross sections is integrated from 0-100 GeV.}}        
        \label{tab:pot_etc}
        \end{center}
\end{table}

The NuMI beam's flux prediction is explained in detail in \cite{leo}, and is summarized in section \ref{sec:fluxsim}.  For distributions in  the \pt/\pz phase space, we report flux-integrated cross sections. We do the same for the single-differential cross section $d\sigma/dQ^2_{QE}$. We integrate over the entire available flux range of 0-100~GeV, to get a total integrated flux of $2.295\times10^{-8}$~cm$^{-2}$ per proton on target.

\begin{figure}\centering
\includegraphics[width=1.0\columnwidth]{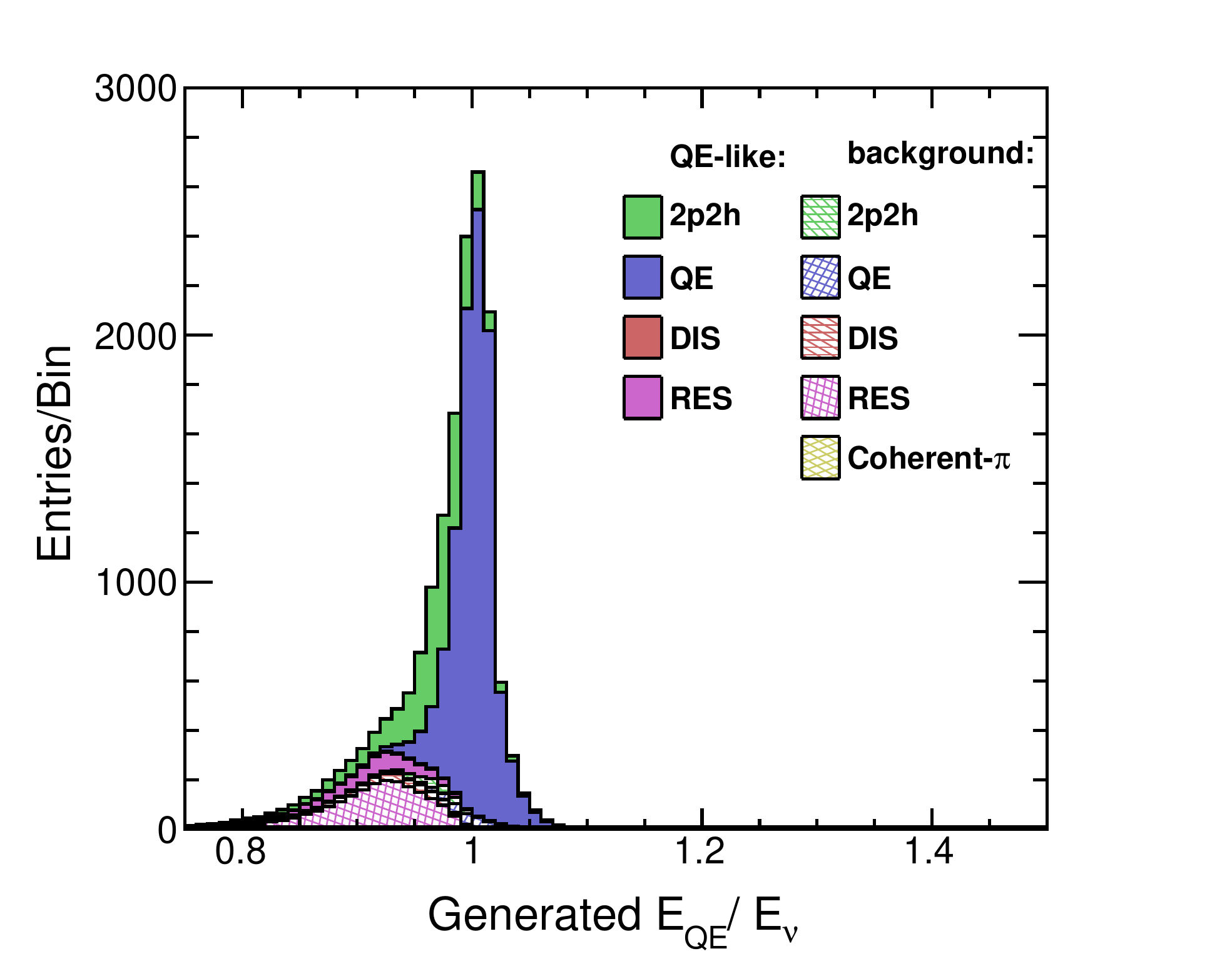}
\caption{Ratio between the neutrino energy reconstructed from true muon kinematics and the true neutrino beam energy in the simulation. True quasi-elastic events show dispersion due to Fermi motion while 2p2h and resonance events tend to underestimate the true neutrino energy.}
\label{fig:enu_true_reco}
\end{figure}

For the cross section as a function of \enu / \qsq, one can create an approximate flux-weighted cross section, where the number of events in each  $E_\nu^{QE}$ bin is normalized by the flux in the corresponding $E_\nu$ bin.  This is not a true total cross section, since $E_\nu^{QE}$ is not the true neutrino energy, except in the case of quasi-elastic scatters off of hydrogen.  However \enu is closely correlated to \enutrue (see Fig.~\ref{fig:enu_true_reco}), making the flux-weighted cross section a close approximation of the total cross section versus energy.

The target for an anti-neutrino quasi-elastic scatter (Eq.~\ref{eq:antinuccqe}) is a proton.  For QE-like scattering, it is possible
that a scattering process could originate on a neutron
(e.g. $\bar{\nu}_\mu n \to \mu^+ \Delta^-$) where the resonance decays
$\Delta^- \to n \pi^-$ and the pion is absorbed, or on a nucleon pair.
We use the total number of nucleons in the fiducial volume as the
target number normalization.  The fiducial volume is made up of a
combination of polystyrene, doping agents, epoxy and light-tight
coating.  The predominant material is polystyrene, which is composed
of equal parts carbon and hydrogen.  A full summary of the composition
of the MINERvA tracker is available in~\cite{minerva_nim}. We estimate
that the fiducial volume used in this analysis contains $3.23\times
10^{30}$ nucleons, of which $1.76\times 10^{30}$  are protons.  

\section{Systematic Uncertainties}
\label{sec:systematics}

Systematic uncertainties on the cross section measurements arise from
many sources.  To assess these systematic uncertainties, we vary
parameters in the simulation within their uncertainties and
recalculate the cross-sections using new estimates of efficiency,
backgrounds, unsmearing, flux and target number corrections.   The
difference between this new cross section and the original result is
taken to be the systematic uncertainty on the cross section due to
that source.  The sources of systematic uncertainty are discussed
below.  Systematic uncertainties in each category, and total
systematic uncertainties, are available in Appendix~\ref{app:systematics}.

\subsection{Flux Uncertainties}
The simulation of the NuMI flux and its uncertainties are described in
detail in \cite{flux}. Uncertainties in the anti-neutrino flux arise
primarily from uncertainties in hadron production rates and in
parameters that control the alignment of the NuMI focusing system,
such as the position of the focusing horns.  These uncertainties are
constrained with both external data and with a \minerva measurement of
elastic neutrino scattering on electrons~\cite{fluxconstraint}.
The total
uncertainty in the focusing peak is approximately 8\%, and rises to
11\% at the falling edge of the focusing peak, where beam focusing uncertainties are large.

\begin{table*}[ht]
\begin{tabular}
%{| p{7cm} |  p{2cm} | p{3cm} |}
{ l  c  r}\hline\hline
Parameter  & Variation & \% effect\\
\hline
Quasi-elastic axial mass (fixed normalization) & $\pm 15\%$  & $<2\%$\\ %\hline%\hline
Quasi-elastic normalization  & $+20\%-15\%$ & 2-4\%\\ %\hline
Vector form factor model  & BBBA05 $\rightarrow$ Dipole & $<1\%$ \\ %\hline
Pauli suppression  &  $30\%$ & $<2\%$\\ %\hline
NC Axial mass & $\pm 25\%$ &$<0.5\%$\\ %\hline
Strange axial form factor for NC &$\pm 30\%$&$<0.5\%$\\ %\hline
NC resonance production rate  & $\pm 20\%$ &$<0.5\%$\\ %\hline
Axial mass for resonance production &$\pm 20\%$ & 3-6\%\\ %\hline
Vector mass for resonance production &$\pm 3\%$ & $<1\%$\\ %\hline
%Deep inelastic scattering normalization & & $<0.5\%$\\ %\hline
Non-resonant 1-pion production rate ($\nu:n$ or $\bar{\nu}:p$) &  $\pm 5\%$ & $<0.5\%$\\ %\hline
Non-resonant 2-pion production rate ($\nu:n$ or $\bar{\nu}:p$) & $\pm 50\%$ &$<0.5\%$\\ %\hline
Non-resonant 1-pion production rate ($\nu:p$ or $\bar{\nu}:n$) &  $\pm 50\%$& $<0.5\%$\\ %\hline
Non-resonant 2-pion production rate ($\nu:p$ or $\bar{\nu}:n$) &  $\pm 50\%$ &$<0.5\%$\\ %\hline
%Neutron mean free path & $\pm 20\%$ & $<1\%$\\ %\hline
Neutron mean free path & $\pm 20\%$ & $1-5\%$\\ %\hline
Pion mean free path &  $\pm 20\%$ &  $<1\%$\\ %\hline
Nucleon elastic scattering cross section & $\pm 30\%$ &  $<1\%$\\ %\hline
Pion elastic scattering cross section & $\pm 10\%$ &  $<1\%$\\ %\hline
Nucleon inelastic scattering cross section & $\pm 40\%$&  $<1\%$\ \\ %\hline
Pion inelastic scattering cross section & $\pm 40\%$& 3-5\%  \\ %\hline
Nucleon charge exchange cross section & $\pm 50\%$ &  $<1\%$\\ %\hline
Pion charge exchange cross section & $\pm 50\%$ &  $<1\%$\\ %\hline
Nucleon absorption cross section & $\pm 20\%$ & $<2\%$ \\ %\hline
Pion absorption cross section & $\pm 20\%$ & 3-5\% \\ %\hline
Nucleon pion production cross section & $\pm 20\%$ &  $<1\%$\\ %\hline
Pion pion production cross section & $\pm 20\%$ &  $<1\%$\\ %\hline
DIS hadronization model adjustment  & $\pm 20\%$ &  $<1\%$\\ %\hline
Pion angle distribution (resonant events) & Isotropic $\to$ Rein-Sehgal &  $<0.5\%$\\ %\hline
Resonant decay photon branching ratio &  $\pm 50\%$ & \\
\hline\hline
\end{tabular}
 \caption{\small{Summary of variable GENIE uncertainties}}        
        \label{tab:genie_uncertainties}
\end{table*}

\begin{table}[hb]
\begin{tabular}
%{| p{7cm} |  p{2cm} | p{3cm} |}
{ l  l  r}\hline\hline
Parameter \ \ \ \ \  & Variation & \% effect\\
\hline
RPA high $Q^2$ & turn off relativistic effects&$<$ 1\%\\
RPA low $Q^2$ & estimate from muon capture&0-2\%\\
2p2h np only & tune to 2p2h np only &$<$ 1\% \\ 
2p2h pp only & tune to 2p2h pp only &0-2\%  \\
1p1h only & turn off 2p2h but tune 1p1h&0-5\%\\
\hline\hline
\end{tabular}
 \caption{\small{Summary of uncertainties in the  cross sections
     extracted using the \minerva-tuned GENIE due to the 2p2h and RPA enhancements of default GENIE 2.8.4 model. Almost all bins have uncertainties of less than 2\% with the largest effects only seen in the highest $p_T$ bins.}}        
        \label{tab:tune_uncertainties}
\end{table}

\subsection{Muon Reconstruction Uncertainties}
There are several uncertainties associated with reconstruction of the
muon track, arising from uncertainties in the muon energy scale,
tracking efficiencies, angular resolution and vertex reconstruction.
The most significant of these is the muon energy scale uncertainty,
which has contributions from several sources, including an 11 MeV
uncertainty in energy loss from MINERvA's material assay, a 30 MeV
uncertainty in the energy deposition rate in MINERvA, $\frac{dE}{dx}$
and a momentum-dependent uncertainty for MINOS muon energy
reconstruction \cite{minos_nim}.  The MINOS uncertainty is 2\% for
muons whose momentum is measured by range, added in quadrature with
either 0.6\% for muons whose momentum is measured by curvature to be
above 1GeV, or 2.5\% for muons whose momentum is measured by curvature to be below 1GeV.

Muon tracking efficiencies in \minerva and \minos are measured by
reconstructing tracks in one detector, extrapolating to the other
detector, and observing the fraction of tracks matched in both
detectors in data and in the simulation.  The simulation is corrected
for small discrepancies between tracking efficiencies, 0.5\% $\pm$
0.25\% for \minerva and 0.5 (2.5)\% $\pm$ 0.25 (1.25)\% for \minos for
muons with momentum greater than (less than) 3.0 GeV.

Potential angular reconstruction biases are estimated both by cutting
tracks in half and comparing the reconstructed angles of both halves,
as well as studies of forward-going events such as neutrino-electron
scattering and low hadronic recoil events.  These studies limit additional angular smearing or bias in the data relative to the simulation to below 1 milliradian.  

Smearing of reconstruction vertices causes some events within the
fiducial volume to be misreconstructed outside the fiducial volume,
and visa versa.  We estimate the uncertainty due to this effect by smearing reconstructed vertices in the simulation by 0.9 mm in x, 1.25 mm in y and 1 cm in z; this results in a negligible change in measured cross section.  

The MINERvA tracker consists of primarily scintillator strips, with smaller portions of epoxy, tape, reflective coating and wavelength shifting fibers.  The total uncertainty on the mass of the tracker is 1.4\%~\cite{minerva_nim}.  

\subsection{Model Uncertainties}

Models used in the simulation include various parameters that carry uncertainties. These include uncertainties in signal, background and final state interaction models.  Most of these are evaluated using the reweighting prescription and parameter uncertainties recommended by the GENIE collaboration~\cite{geniemanual}.  These parameters are listed in Table~\ref{tab:genie_uncertainties}, along with the amount by which they are varied and the approximate effect on the cross sections.

GENIE uncertainties that change particle fates cannot be modeled using the re-weighting method. In this case, we generate an alternative simulated sample in which these parameters, including the effective nuclear radius, formation zone and hadronization model, have been adjusted.  

The RPA correction described in section \ref{sec:neutrinosim} is
applied when calculating the central values.  The correction is varied
within the uncertainties shown in figure \ref{fig:rpa}.  Similarly,
the addition of the 2p2h process is estimated by adding events from correlated pairs as described in section \ref{sec:neutrinosim} and reference \cite{PhysRevD.88.113007}. The uncertainty on this is determined by using several variations of the tuning procedure, including fits that allow interactions on $pp$ pairs, $np$ pairs, or single nucleon interactions to be tuned.  The differences in cross-section obtained using these three variants from the standard simulation are added in quadrature as a systematic error due to the 2p2h model.
Table \ref{tab:tune_uncertainties} summarizes the effects of these variations on the extracted cross sections.

\subsection{Recoil reconstruction uncertainties}\label{sec:recoil_uncertainties}

Several sources of uncertainties can affect the reconstruction of recoil energy, which can in turn change the background estimates and efficiencies used to estimate the cross sections.  Quasi-elastic anti-neutrino events have a hadronic final state consisting of a neutron.  In order for neutrons to deposit recoil energy in the detector, they must undergo an interaction, with resulting charged particles (usually protons) then depositing visible energy.  The most significant source of uncertainty  associated with neutrons is due to the \textsc{Geant4} neutron interaction model.  To evaluate this uncertainty, we vary the mean free path of neutrons in the detector, with the variations spanning discrepancies between \textsc{Geant4} and thin target neutron scattering data on copper, iron and carbon~\cite{Abfalterer:2001gw,Schimmerling:1973bb,Voss:1956,Slypen:1995fm,Franz:1989cf,Tippawan:2008xk,Bevilacqua:2013rfq,Zanelli:1981zz}. 

Energy response of protons has been measured in the \minerva test beam detector~\cite{testbeam}.  To propagate uncertainties on this measurement to the cross sections, we shift simulated recoil energy deposited by protons by uncertainties derived from comparisons of the test beam measurements and \textsc{Geant4}. The variation depends on the proton energy: 4\% below 50~MeV  and 3.5\% above 50~MeV.  
The proton response affects our event rate measurement by less than $1\%$ across our whole phase space, as the track cut removes many protons, and only a small amount of those that remain pass into our selected sample by making this shift.

The pion calorimetric response has also been constrained by test beam studies %\cite{minerva9474}
to an accuracy of 4\%, for pions with a kinetic energy between 400 and 1900 MeV. We thus separate our pions into two categories - ``constrained'' within this energy range, and ``unconstrained'' outside of it. Pions within the constrained range have their energy fraction varied by $\pm4\%$, while others have it varied by $\pm5\%$. The pion response has only a minor effect ($<1\%$ across our whole phase space) on our cross sections.

For the other particles (electromagnetic and kaons), we vary the recoil by $\pm3\%$. This uncertainty was derived by observing the energy response for Michel electrons (electrons from muon decay), which have a well-known energy spectrum. This change mainly affects the $Q^2$ (${p_T}$) shape, and contributes its maximum of around 1\% uncertainty at low $Q^2$.

These uncertainties are dominated by neutron interaction modeling, which
ranges from 2-6\%; the other uncertainties are less than 1\%.

\section{Results}
\label{sec:results}

\begin{figure*}
\centering 
\includegraphics[width=0.9\textwidth]{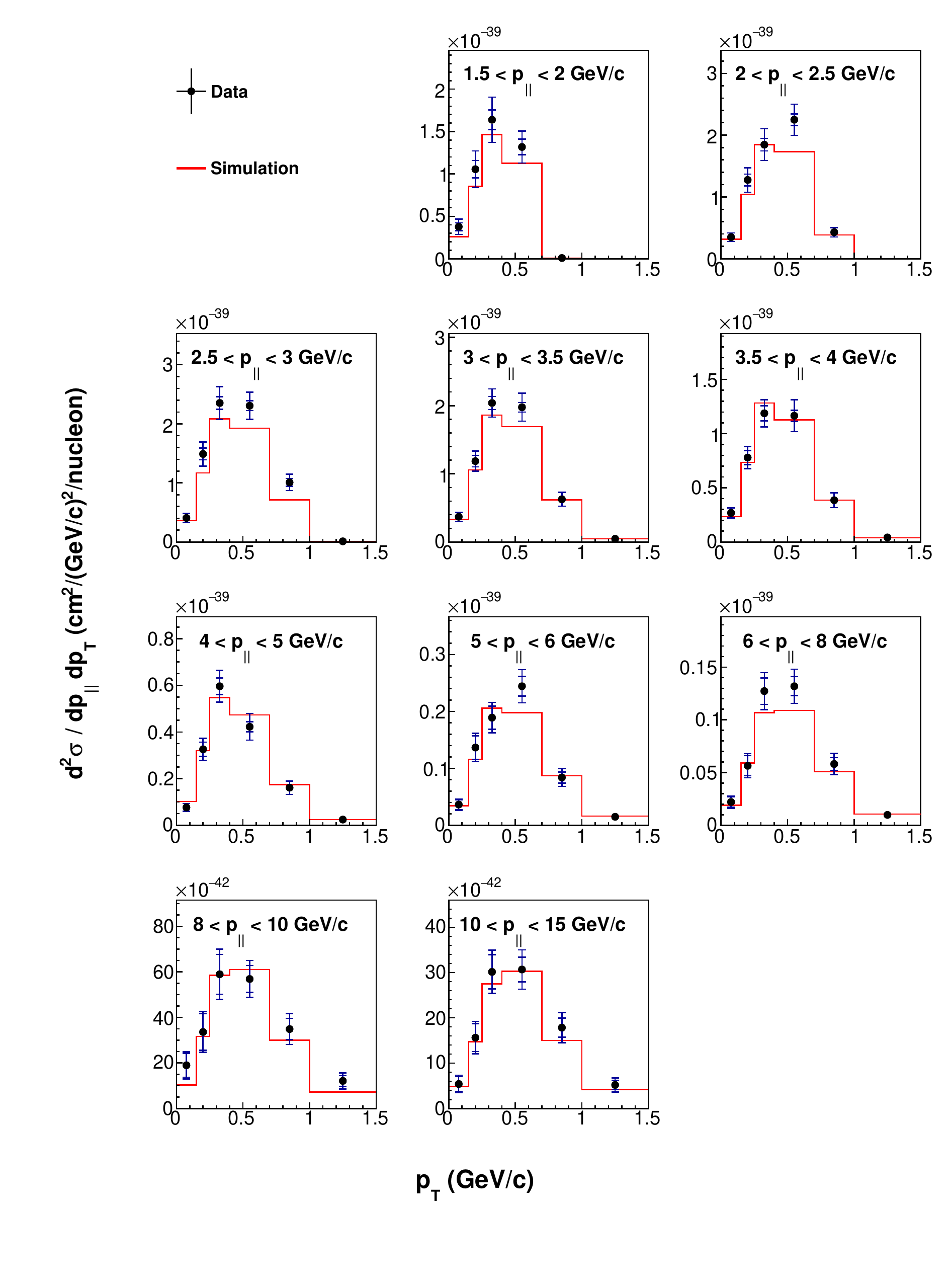}
\caption{\label{fig:cross_sections_pt_pz}Double-differential  QE-like
  cross section vs. muon transverse momentum, in bins of muon
  longitudinal momentum.  Inner error bars show statistical
  uncertainties; outer error bars show total (statistical and
  systematic) uncertainty. The red histogram shows the \minerva-tuned GENIE 2.8.4 model used to estimate smearing and acceptance. These results are tabulated in Tables \ref{tab:xsec_pzpt}--\ref{tab:sys_pzpt}.}      
\end{figure*}

\begin{figure*}
  \centering  \includegraphics[width=0.9\textwidth]{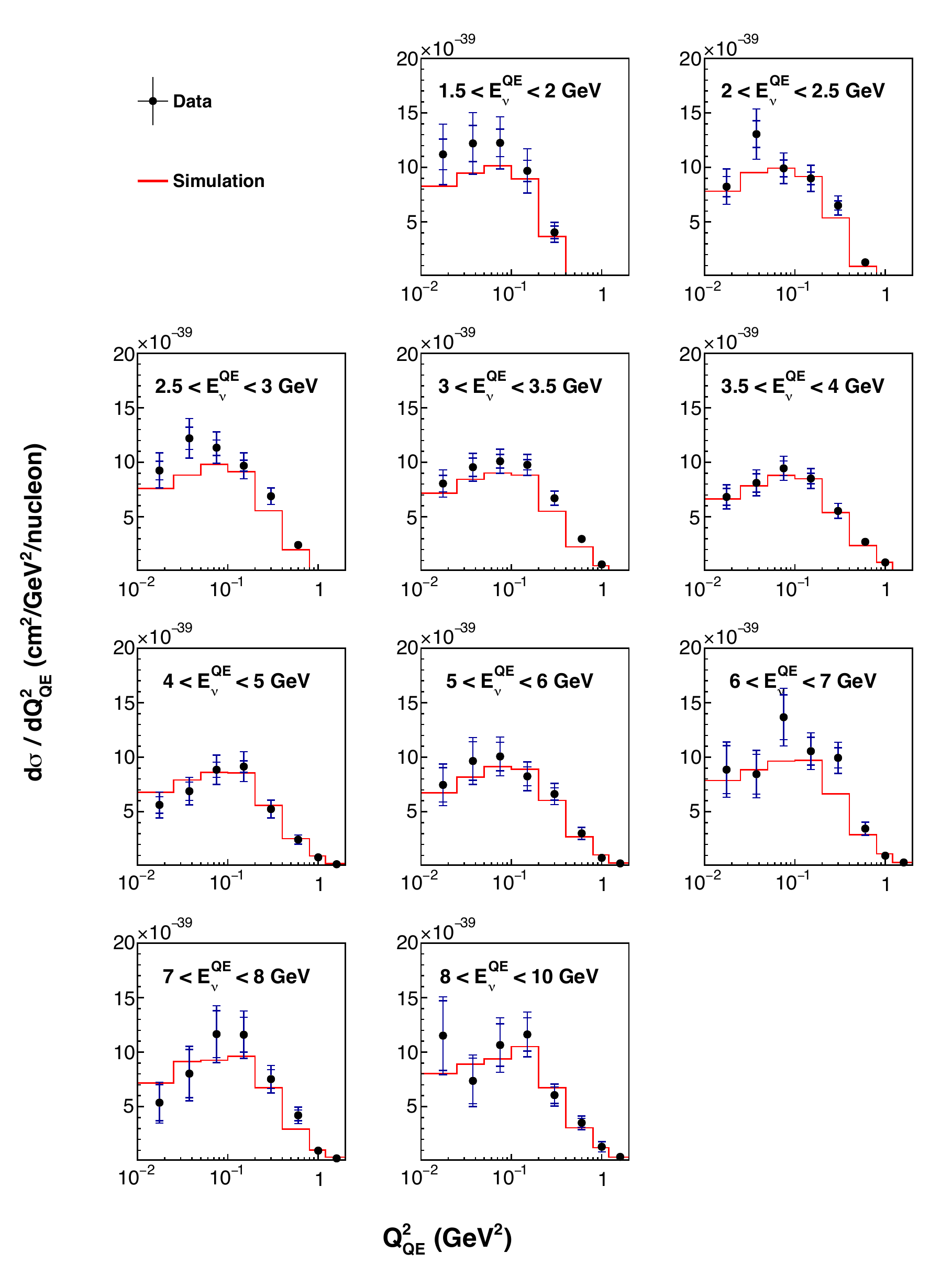}
        \caption{\label{fig:cross_sections_enuqe_qsq} Differential
          QE-like cross section  $d\sigma(E^{QE}_\nu)/dQ^2_{QE}$, in
          bins of $E^{QE}_\nu$.  Inner error bars show statistical
          uncertainties; outer error bars show total (statistical and
          systematic) uncertainty. The red histogram shows the
          \minerva-tuned GENIE model used to estimate smearing and acceptance. These results are tabulated in Tables \ref{tab:xsec_eqe}--\ref{tab:sys_eqe}.}       

\end{figure*}

\begin{figure*}
  \centering 
    \includegraphics[width=0.9\textwidth]{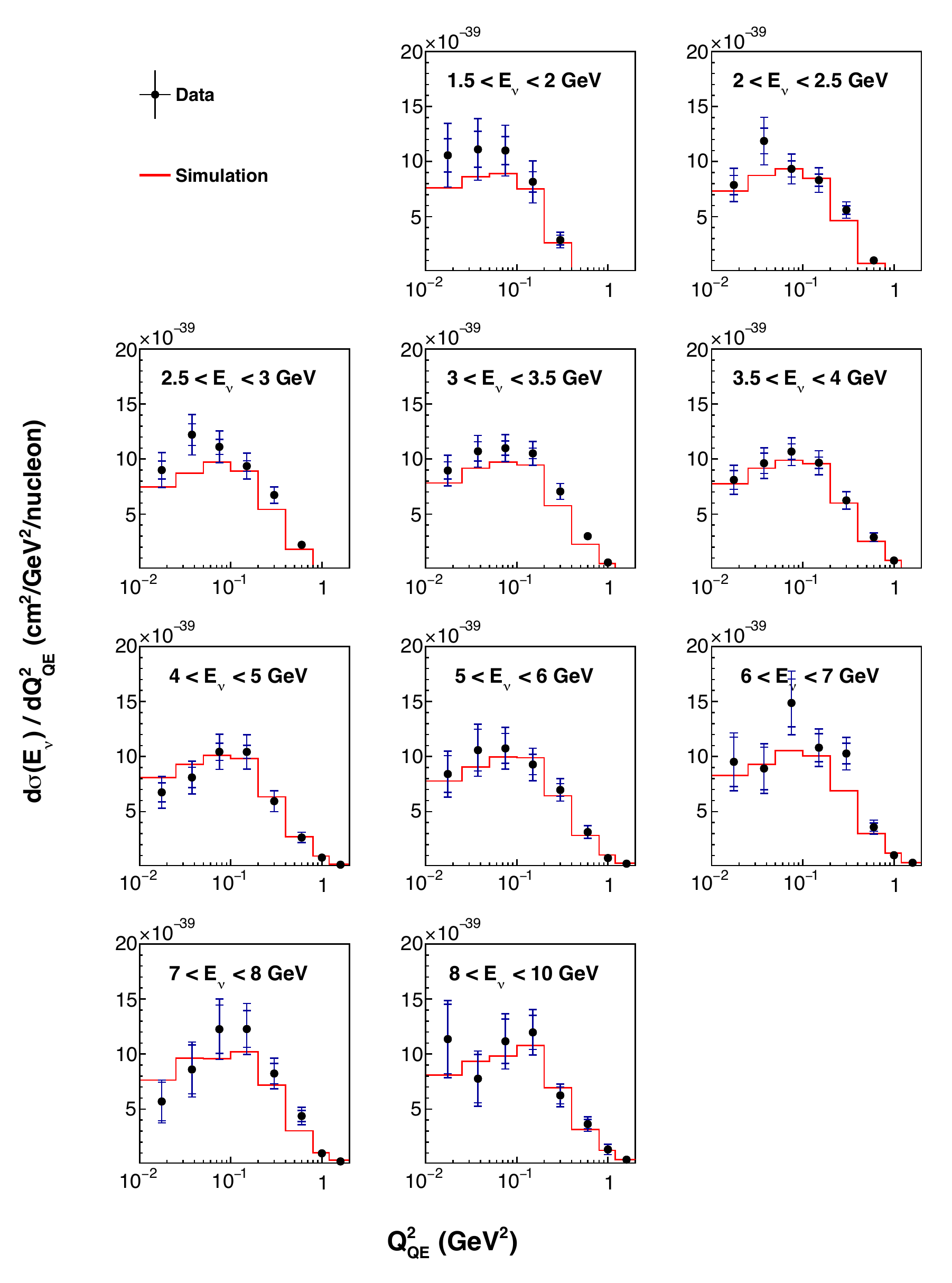}
        \caption{\label{fig:cross_sections_enu_qsq}Differential
          QE-like cross section  $d\sigma(E^{true}_\nu)/dQ^2_{QE}$, in
          bins of $E^{true}_\nu$. The red histogram shows the
          \minerva-tuned GENIE model used to estimate smearing and acceptance. These results are tabulated in Tables \ref{tab:xsec_enu}--\ref{tab:sys_enu}.}       

\end{figure*}

Double-differential cross sections vs. \pt and \pz are shown in
Fig.~\ref{fig:cross_sections_pt_pz}. Double-differential cross
sections vs. \enu (\enutrue) and \qsq are shown in
Fig.~\ref{fig:cross_sections_enuqe_qsq}
(Fig.~\ref{fig:cross_sections_enu_qsq}).  
In each case,
simulated cross sections are also plotted, where the simulation uses
the \minerva-tuned GENIE model described in~\ref{sec:neutrinosim}. Results corrected to
a quasi-elastic, rather than QE-like, signal definition are available in
Appendix~\ref{app:ccqe}.

The \minerva-tuned GENIE model agrees well with MINERvA data, in spite of
the fact that the tune was made to an independent (neutrino rather
than antineutrino) dataset.  In the following sections, we compare
these results with many alternate models and discuss the impact of the
individual components of the \minerva tune.  

\begin{figure*}
\centering 
\includegraphics[trim={1.2cm 2.2cm 0 0 }, width=0.9\textwidth]{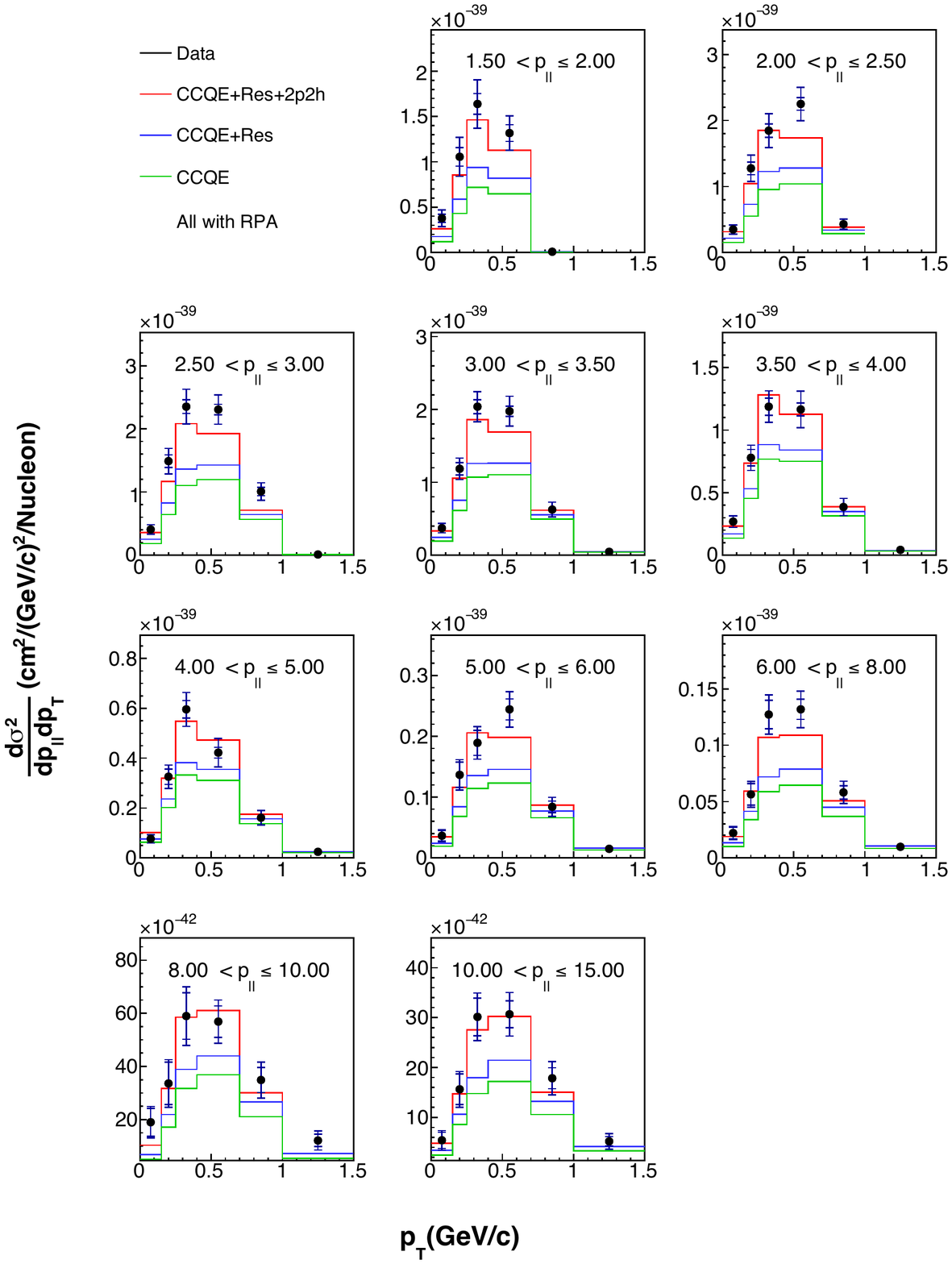}
\caption{\label{fig:components} Double-differential  QE-like
  cross section vs. muon transverse momentum, in bins of muon
  longitudinal momentum.  Inner error bars show statistical
  uncertainties; outer error bars show total (statistical and
  systematic) uncertainty. The red histogram shows the \minerva-tuned GENIE 2.8.4 model used to estimate smearing and acceptance which is CCQE + resonant + 2p2h. The green histogram shows events produced as CCQE, the blue curve shows default GENIE which is CCQE + resonant.  These results are tabulated in Tables \ref{tab:xsec_pzpt}--\ref{tab:sys_pzpt} and the models are included in the supplementary material.}      
\end{figure*}

Figure \ref{fig:components} shows the data and the different components included in the MnvGENIE model.  Events generated as pure 1p1h CCQE are shown in green, the resonance (and a small DIS) component is added to make the blue histogram. The addition of tuned 2p2h yields the red MnvGENIE curve.  RPA corrections are applied to all three.    %added for referee

 \begin{figure*}
 \centering  \includegraphics[trim={1.2cm 1.2cm 0 0 }, width=0.9\textwidth]{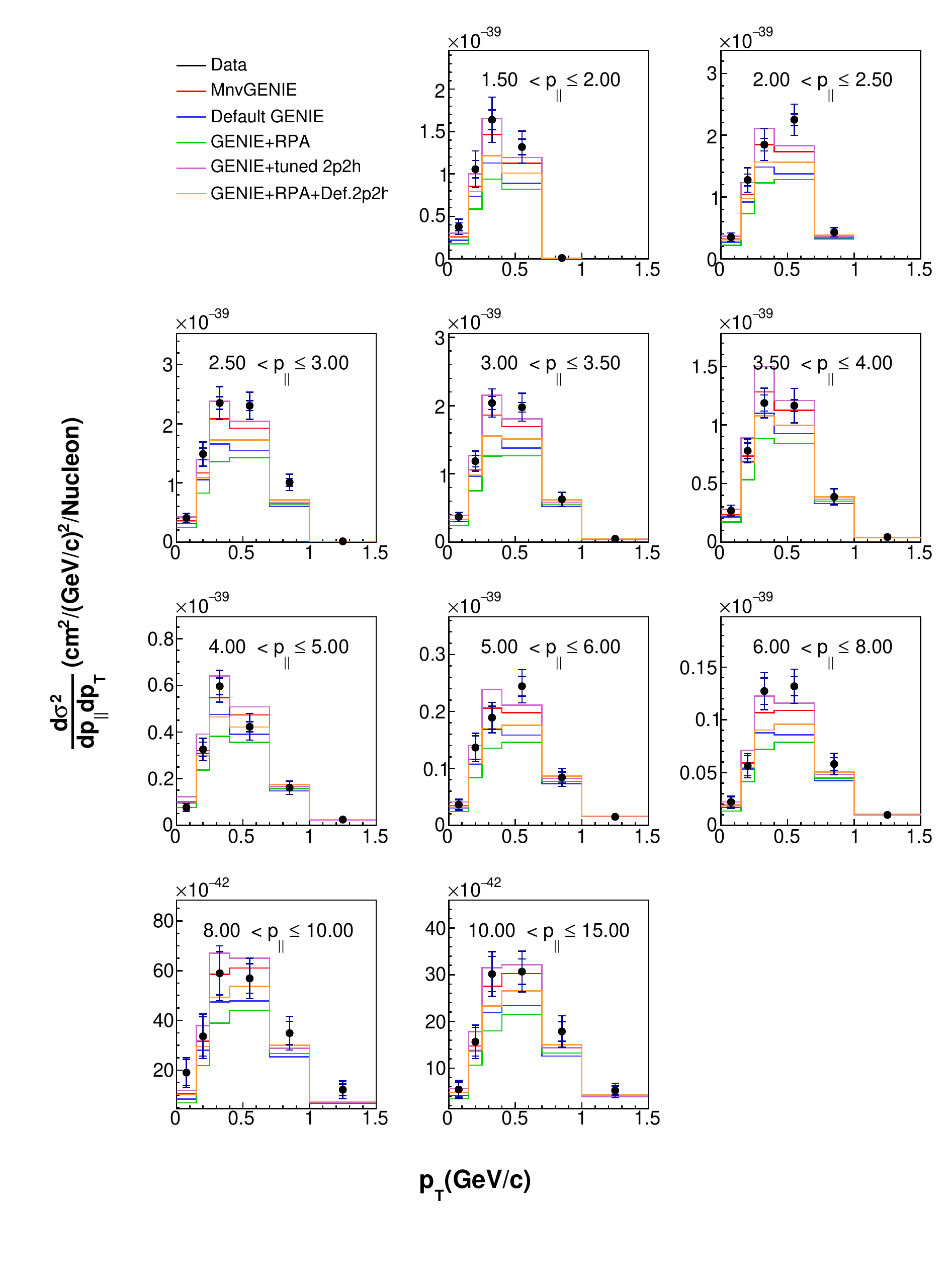}
 \caption{  \label{fig:GENIE_CV}{Double-differential  QE-like cross
     section vs. muon transverse momentum, in bins of muon
     longitudinal momentum (black circles) compared to \minerva-tuned GENIE (red curve, includes RPA and MINERvA-tuned 2p2h), GENIE without any modifications except the single non-resonant pion correction discussed in section \ref{sec:neutrinosim} (blue), GENIE with the RPA weight but no 2p2h component  (green), GENIE with MINERvA-tuned 2p2h but no RPA (violet), and GENIE with RPA and untuned 2p2h  (orange).  Inner error bars show statistical uncertainties; outer error bars show total (statistical and systematic) uncertainty.}}
\end{figure*}

 \begin{figure*}
     \includegraphics[width=0.49\textwidth]{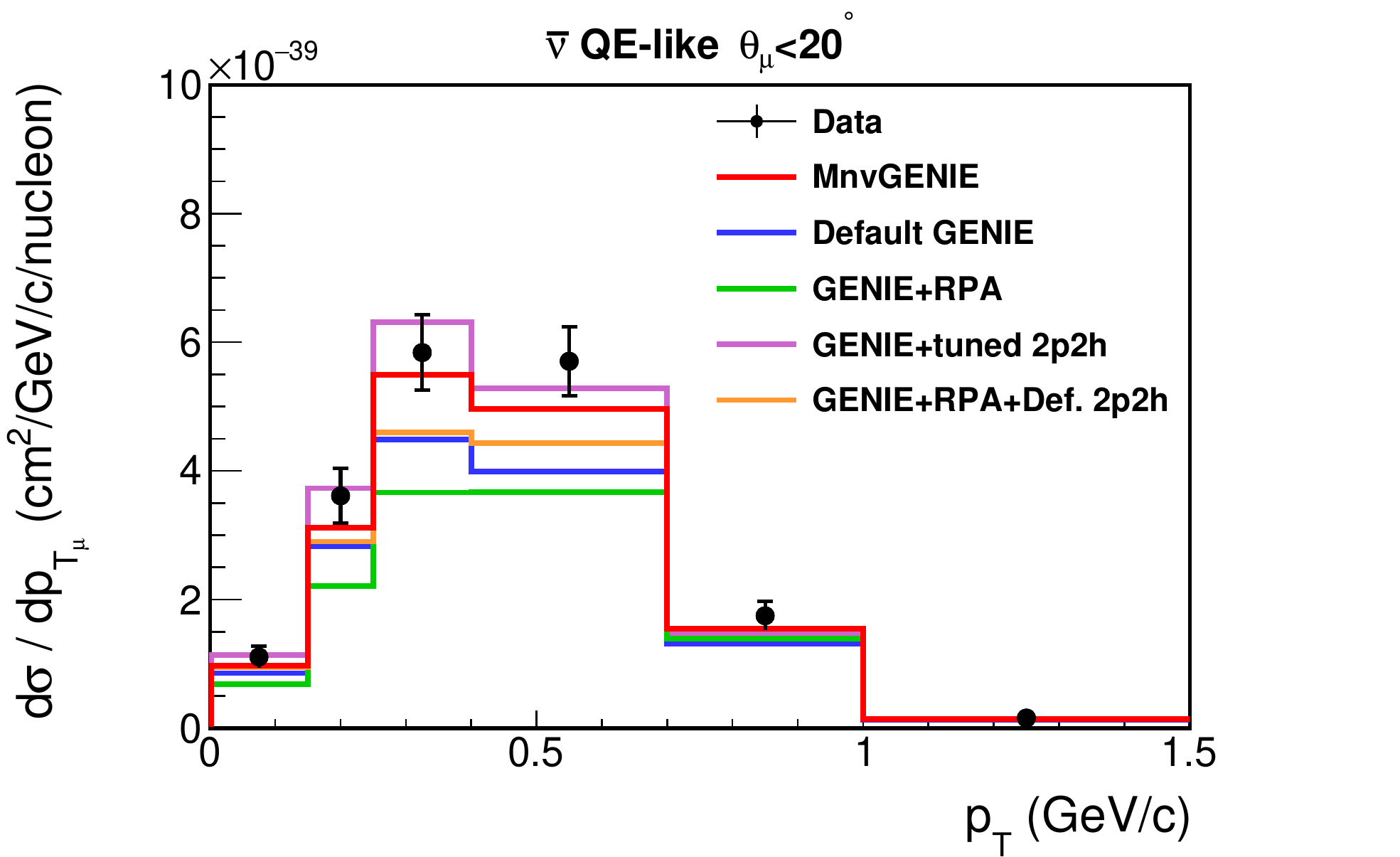}
     \includegraphics[width=0.49\textwidth]{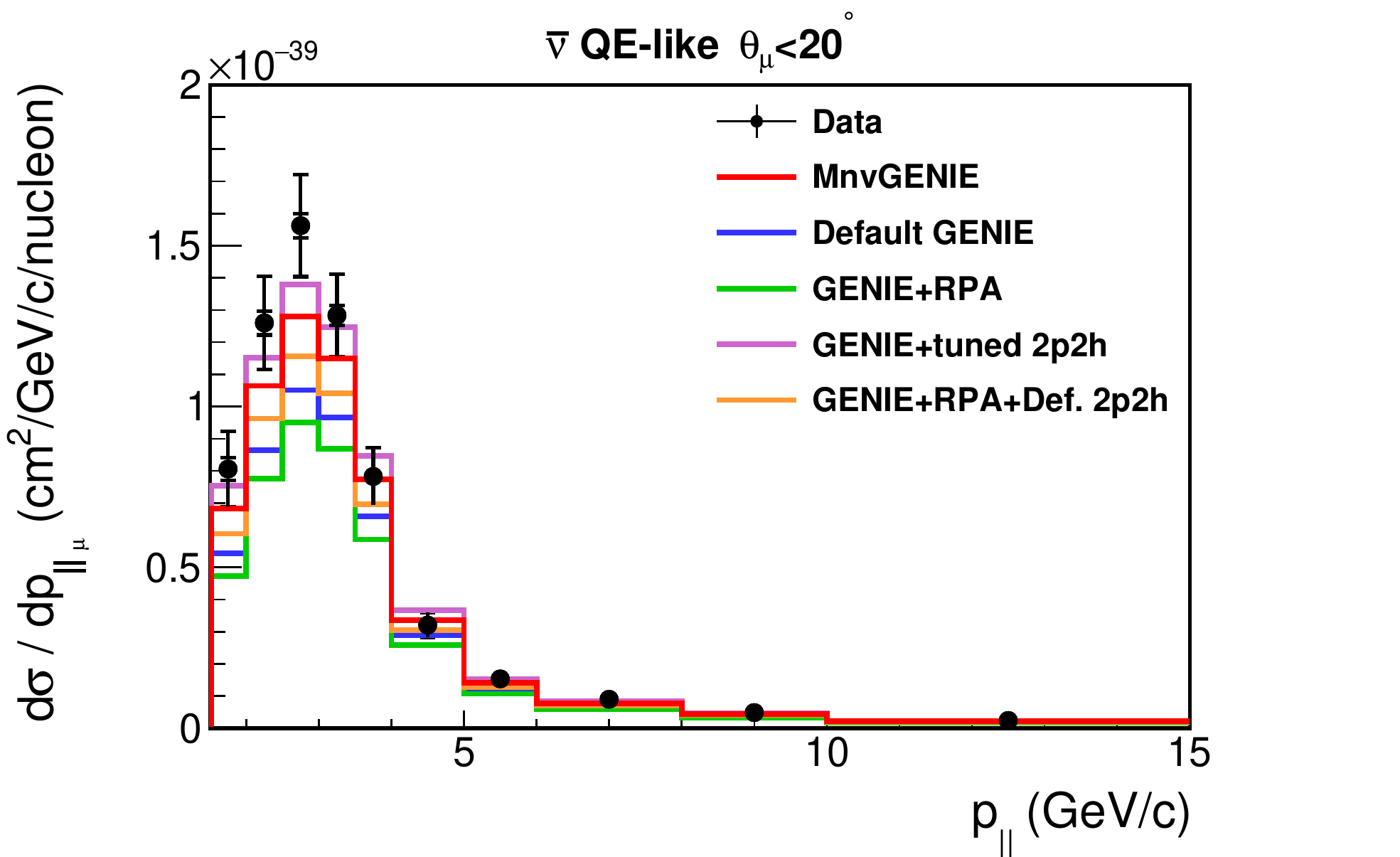}
     \includegraphics[width=0.49\textwidth]{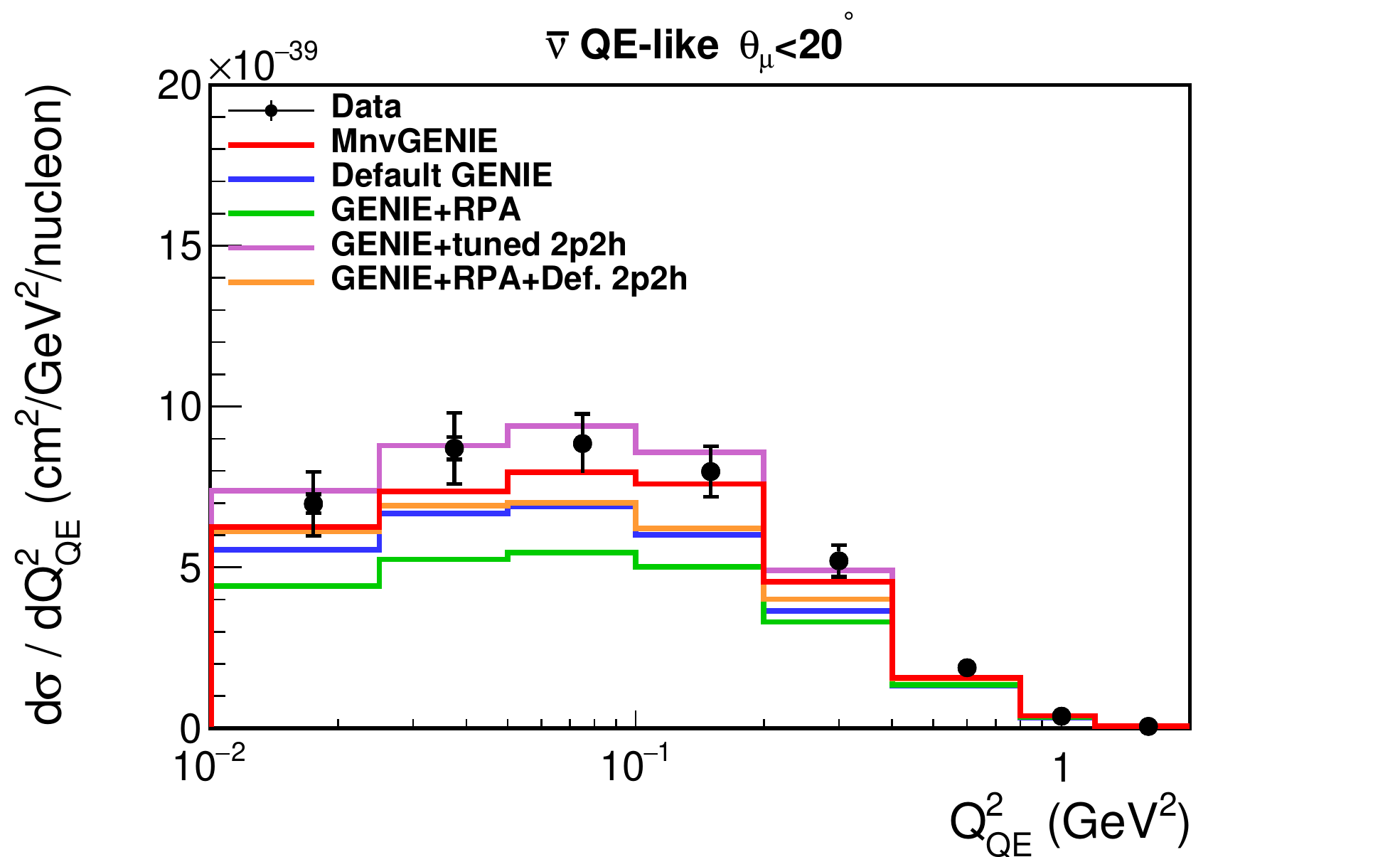}
     \includegraphics[width=0.49\textwidth]{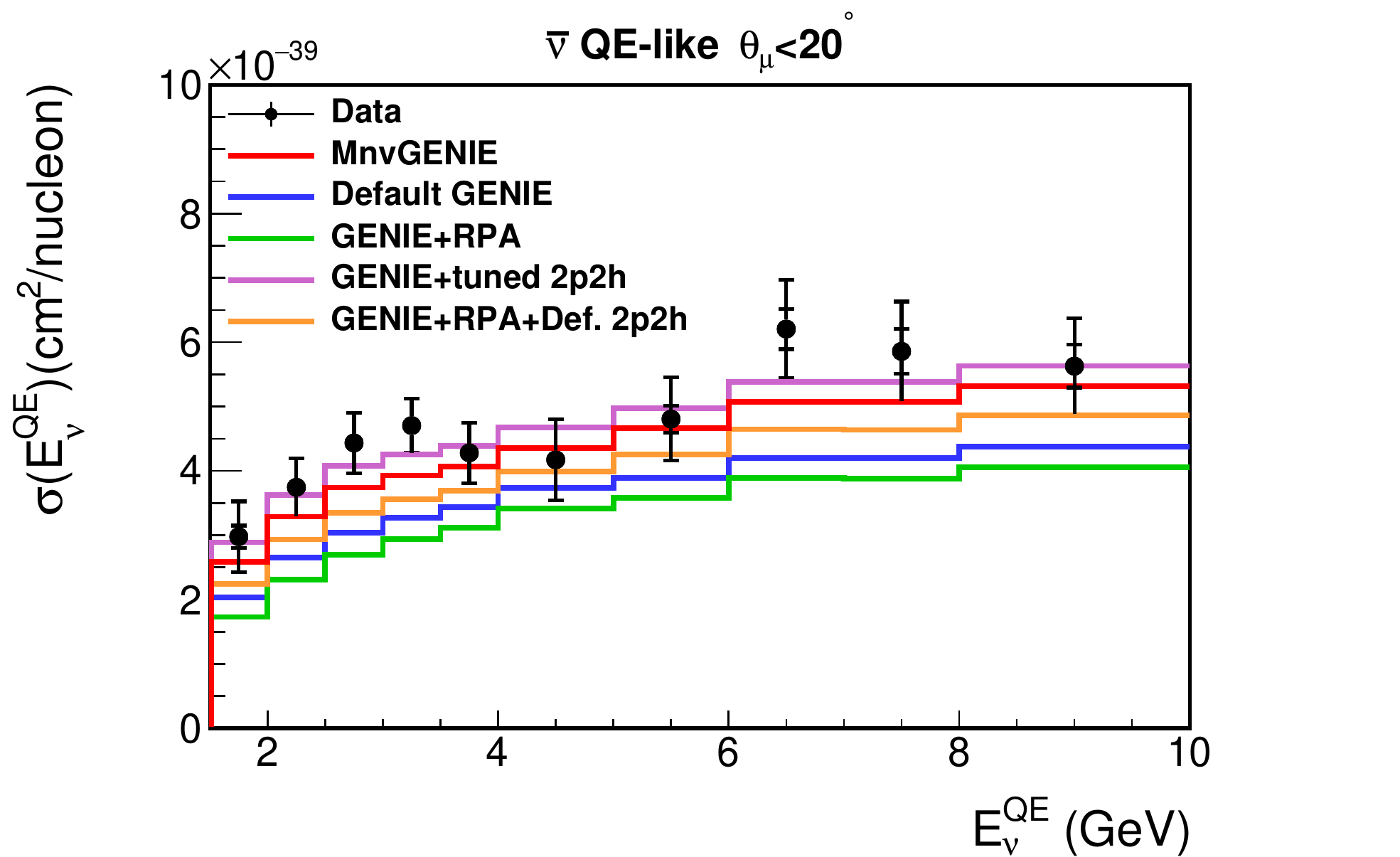}
\caption{Single-differential projections of the double-differential QE-like cross section measurements compared to \minerva-tuned GENIE (red curve, includes RPA and MINERvA-tuned 2p2h), GENIE without any modifications except the single non-resonant pion correction discussed in section \ref{sec:neutrinosim} (blue), GENIE with the RPA weight but no 2p2h component  (green), GENIE with MINERvA-tuned 2p2h but no RPA (violet), and GENIE with RPA and untuned 2p2h  (orange).  Inner error bars show statistical uncertainties; outer error bars show total (statistical and systematic) uncertainty. \label{fig:cross_sections_qelike_models2} These results are tabulated in Tables \ref{tab:xsec_pt}, \ref{tab:xsec_pz}, \ref{tab:xsec_enuqe} and \ref{tab:xsec_qsq_intenuqe}.}
\end{figure*}

 \begin{figure*}
 \centering  \includegraphics[trim={1.2cm 1.2cm 0 0 }, width=0.9\textwidth]{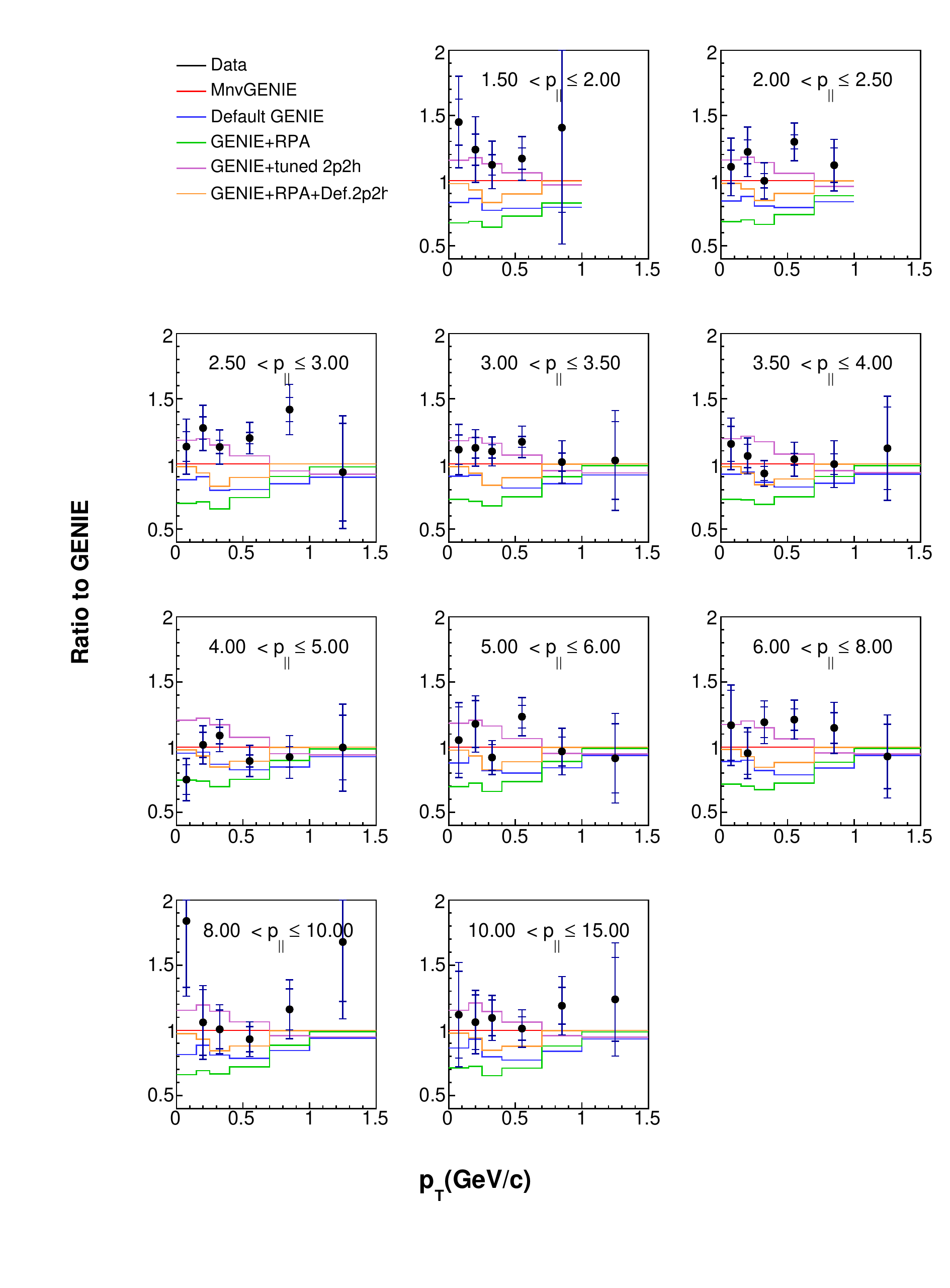}
 \caption{  \label{fig:GENIE_CV_ratio} Ratio of the
   double-differential  QE-like cross sections vs. muon transverse
   momentum shown in the previous figure, in bins of muon longitudinal
   momentum.  In all cases, the denominator for the ratio is
   \minerva-tuned GENIE (includes RPA and MINERvA-tuned 2p2h); the
   numerators are our measured cross section (black circles) ,
   MINERvA-tuned GENIE (red), GENIE without any modifications (blue), GENIE with the RPA weight but no 2p2h component  (green), GENIE with MINERvA-tuned 2p2h but no RPA (violet), and GENIE with RPA and untuned 2p2h  (orange).  Inner error bars show statistical uncertainties; outer error bars show total (statistical and systematic) uncertainty. }
\end{figure*}

\subsection{Comparisons to Alternate GENIE Models}

Figure \ref{fig:GENIE_CV} shows the measured differential cross
sections and several variations on the GENIE model; Single
differential projections versus \ptnospace, \pznospace, \enunospace,
and \qsq are shown in 
Fig.~\ref{fig:cross_sections_qelike_models2} for the same variations.  

In particular, the MINERvA-tuned GENIE (MnvGENIE in figures) model is GENIE
with RPA and \minerva-tuned 2p2h effects added\footnote{Although the default configuration of GENIE 2.8.4 is used here, it is by design very similar to the default configuration of current release GENIE release,
2.12.8.  Application of the
tuning procedure to GENIE 2.12.8 would approximately reproduce
the \minerva-tuned model presented here.}. Other permutations of RPA, 2p2h
and \minerva-tuned 2p2h are also shown.   Figure \ref{fig:GENIE_CV_ratio} shows
the ratio of the measured differential cross section to the
MINERvA-tuned GENIE model.  Table \ref{chisquaremore} shows the
$\chi^2$ for 58 degrees of freedom for the models shown in that figure
and for additional theoretical models described in Section~\ref{sec:othermodels}.

A standard $\chi^2$ comparison for these models relative to the data using the statistical uncertainties derived from the data gives a best agreement (the green curve with RPA but no 2p2h contribution) with the curve that lies furthest away from the data.  This is due to the dominance of multiplicative normalization uncertainties in the covariance matrix, which leads to the well known pathology of Peelle's Pertinent Puzzle\cite{Peelle, BoxCox}.  
This effect is well documented in the nuclear cross section literature
\cite{nuclearfits} and, in the limit  of pure multiplicative
uncertainties, a $\chi^2$ derived from the log of the cross section is
preferred to one derived from the cross section itself, since in the
former case the multiplicative factors are normally distributed. 

 In addition, the $\chi^2$ statistic is known to have biases when
uncertainties estimated from the counting statistics of individual data points are used.  For statistical
 uncertainties estimated to be proportional to $\sqrt{N_i}$, points
 that fluctuate downwards are given smaller estimated uncertainties and hence greater
 weight in the $\chi^2$ calculation.  For normalization uncertainties,
 the effect is even greater with the uncertainty being directly
 proportional to $N_i$.   The normalization uncertainties in these
 data are highly correlated from bin to bin and substantial relative
 to the other uncertainties. For this reason, we report the $\chi^2$
 using both the cross section itself (linear) and the log of the cross
 section (log-normal)  in Table \ref{chisquaremore} in the next
 section. The lowest log-normal $\chi^2$ is for the \minerva-tuned GENIE model with default 2p2h and the RPA correction which appears as the orange curve in Fig. \ref{fig:cross_sections_qelike_models2}, \ref{fig:GENIE_CV} and \ref{fig:GENIE_CV_ratio}.

\subsection{Comparisons to NuWro Models}
\label{sec:othermodels}
\begin{figure*}
  \centering \includegraphics[trim={1.2cm 1.2cm 0 0 }, width=0.9\textwidth]{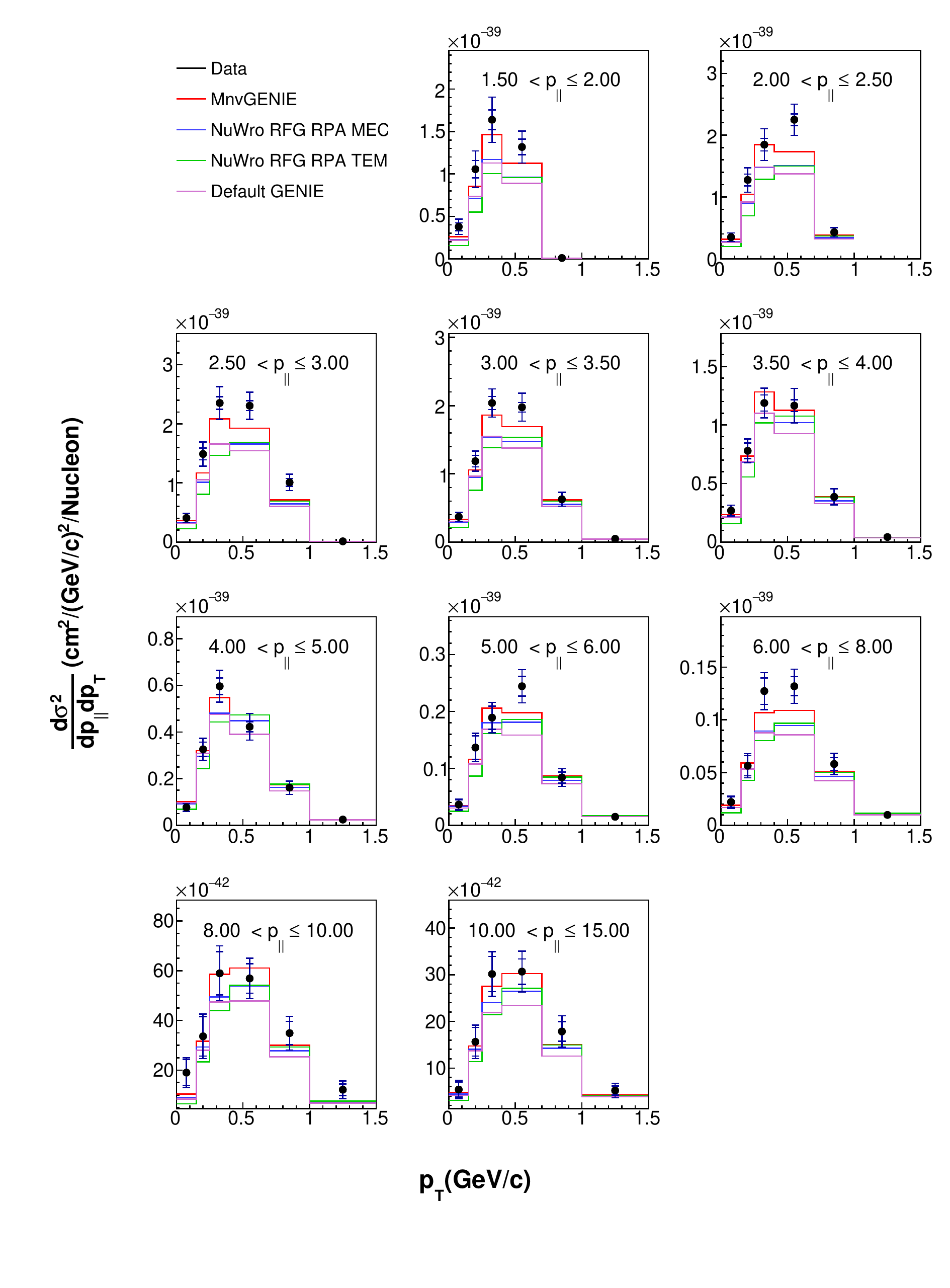}
  \caption{  \label{fig:RFG_RPA}{Double-differential  QE-like cross section vs. muon transverse momentum, in bins of muon longitudinal momentum compared to the MINERvA-tuned GENIE (red curve), the NuWro Nieves RFG model with Random Phase Approximation  and Meson Exchange current (MEC) (blue curve) and the NuWro Relativistic Fermi Gas RFG model with RPA and Transverse Enhancement added (green curve).  Inner error bars show statistical uncertainties; outer error bars show total (statistical and systematic) uncertainty.}}

\end{figure*}

\begin{figure*}
  \centering \includegraphics[trim={1.2cm 1.2cm 0 0 }, width=0.9\textwidth]{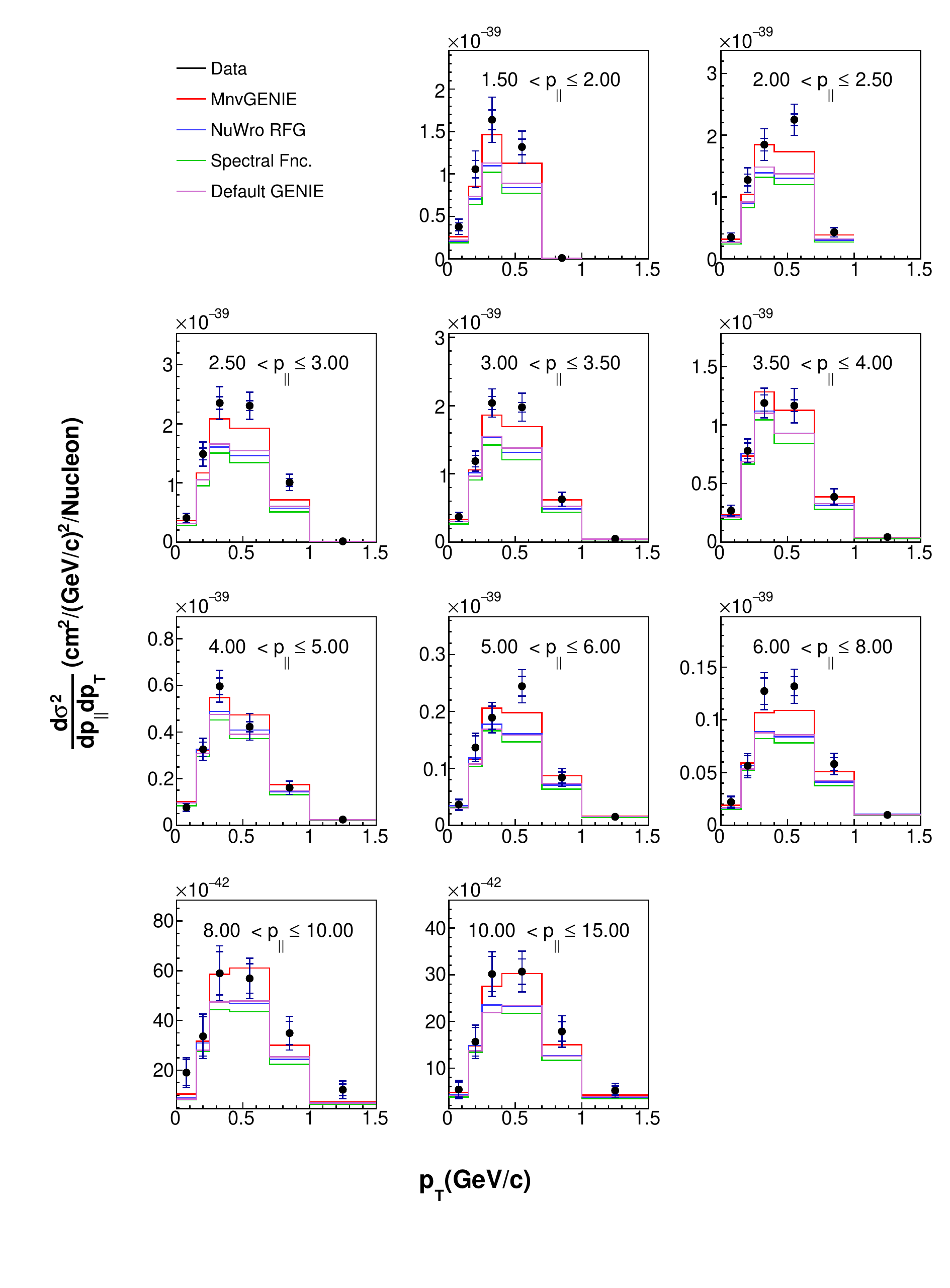}
  \caption{  \label{fig:Spectral_Function}{Double-differential  QE-like cross section vs. muon transverse momentum, in bins of muon longitudinal momentum compared to the MINERvA-tuned GENIE (red curve), NuWro and GENIE RFG implementations (blue and purple)  and the NuWro Spectral Function model (green curve).  Inner error bars show statistical uncertainties; outer error bars show total (statistical and systematic) uncertainty. }}

\end{figure*}

We have also compared the data to several models available in the
NuWro event generator.  Table \ref{chisquaremore} summarizes the
agreement between the data and these models while figures
\ref{fig:RFG_RPA} and \ref{fig:Spectral_Function} show the
comparisons.  NuWro also includes models of 2p2h and RPA, and
additionally includes an implementation of the Transverse Enhancement
Model describe in Section~\ref{sec:correlations}.  The Relativistic Fermi Gas nuclear model is labeled GFG
(Global Fermi Gas) to distinguish it from an alternate LFG (Local
Fermi Gas) nuclear model.  A Spectral Function model is also available
in NuWro and included in the comparisons.   

 All of the NuWro
 models have higher $\chi^2$ values than the \minerva-tuned GENIE model.  Even
 when comparing very similar primary interaction models (e.g. default
 GENIE\footnote{Our 'default' GENIE includes a correction to the single
   non-resonant pion production rate based on bubble chamber inputs discussed in
   section \ref{sec:neutrinosim}.  This has little effect on the
   CCQE-like cross section prediction, since very few CCQE-like events arise from non-resonant
   pion production. } and NuWro GFG without RPA or 2p2h) between the two generators, the agreement with data is quite different.  We believe this is due to the different FSI models used by NuWro and GENIE, which impact the predicted contribution to the cross section from events that include a pion that is absorbed before exiting the primary nucleus.  Of the NuWro models, the preferred model includes RPA and 2p2h contributions, as is also the case with GENIE variants described above.

\begin{table*}
\begin{tabular}{l r r}
\hline\hline
Model	& conventional $\chi^2$&\ \ \ 	Log-Normal $\chi^2$\\
\hline
GENIE+def. 2p2h+RPA&	 \ 70.4&	\  96.5\\ % \MnvGENIE_RPA_MEC_notune
\bf MINERvA tuned GENIE&\ 	 81.2&	 \ 98.4\\ % \MnvGENIE
GENIE+RPA&	\  66.1&	117.9\\ % \MnvGENIE_RPA_noMEC
Untuned GENIE&	\  80.6&	131.0\\ % \Untuned_GENIE
GENIE+2p2h&	149.7&	154.1\\ % \MnvGENIE_noRPA_MEC
\hline
NuWro GFG+2p2h+RPA&	 \ 72.4&	105.0\\ % \gfg_rpa_Nieves_ma099
NuWro LFG&	 \ 85.4&	111.3\\ % \lfg_norpa_tem_ma099
NuWro GFG+TEM&	 \ 86.9&	113.7\\ % \gfg_norpa_tem_ma099
NuWro GFG+TEM+RPA&	\  80.9&	125.7\\ % \gfg_rpa_tem_ma099
NuWro GFG&	108.4&	177.5\\ % \gfg_norpa_nomec_ma099
NuWro Spectral Function&	\  94.8&	184.6\\ % \sf_norpa_nomec_ma099
NuWro GFG+2p2h&	153.7&	185.9\\ % \gfg_norpa_Nieves_ma099

\hline
\hline
\end{tabular}
\caption{\label{chisquaremore} This table summarizes the
  $\chi^2$ values for comparisons of the differential cross section
  $d\sigma^2/dp_Tdp_\parallel$ to  a wide variety of models.  The top
  5 are GENIE variations  illustrated in Fig.~\ref{fig:GENIE_CV} while the bottom 7 are variations of the
  NuWro event generator illustrated in Figs.~\ref{fig:RFG_RPA} and \ref{fig:LFG_noRPA}.  The \minerva-tuned GENIE model includes RPA and
  MINERvA-tuned 2p2h and was used in the extraction of the cross section. 
The $\chi^2$ is for 58 degrees of freedom. }
\end{table*}

\begin{figure*}
  \centering \includegraphics[trim={1.2cm 1.2cm 0 0 }, width=0.9\textwidth]{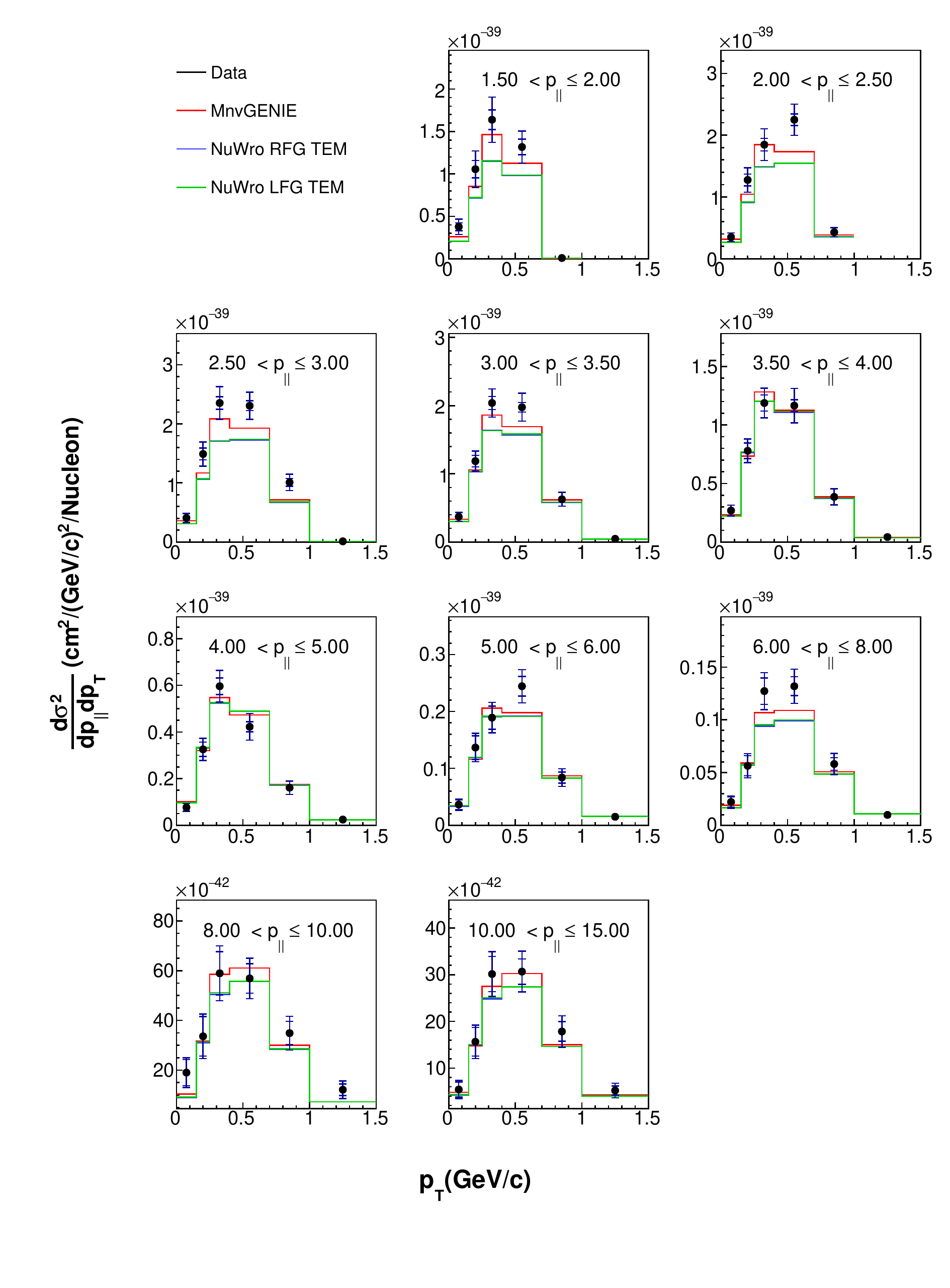}
  \caption{ \label{fig:LFG_noRPA}{Double-differential  QE-like cross
      section vs. muon transverse momentum, in bins of muon
      longitudinal momentum compared to the MINERvA-tuned GENIE (red curve) and the NuWro Relativistic (RFG) Fermi Gas model with Transverse Enhancement (RFG+TEM, blue curve) and the NuWro Local Fermi Gas (LFG) model with Transverse Enhancement (LFG+TEM, green curve).  Inner error bars show statistical uncertainties; outer error bars show total (statistical and systematic) uncertainty. }}

\end{figure*}

\subsection{Comparisons to other experiments}

\begin{figure}
\centering \includegraphics[width=\columnwidth]{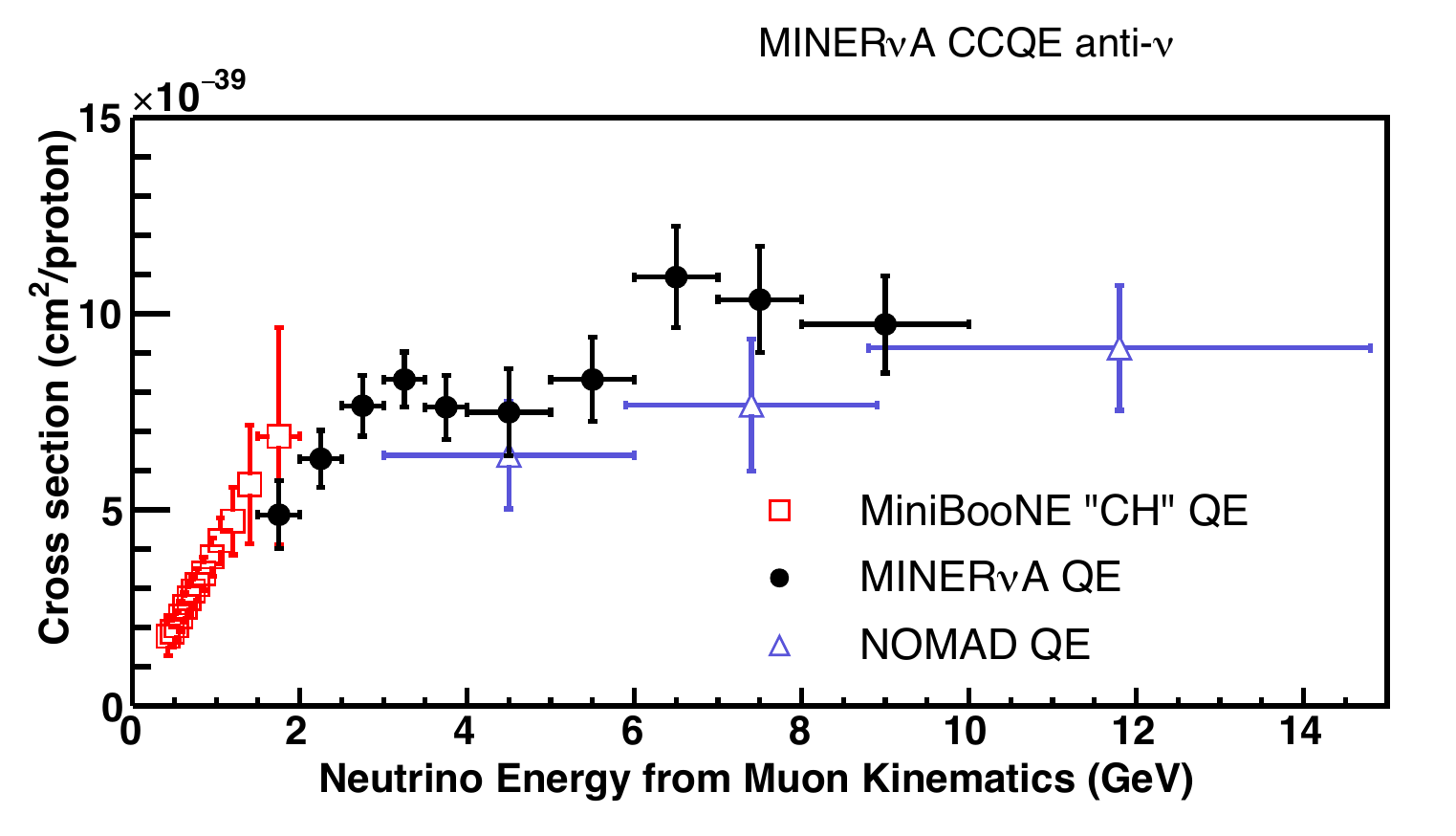}
\caption{\label{fig:sigma_vs_enuQE_true-QE}{MINERvA true CCQE cross section as a function of \enu compared to data from the MiniBooNE and NOMAD experiments.   Error bars show total (statistical and systematic) uncertainty.}}
\end{figure}

\begin{figure}
\centering \includegraphics[width=\columnwidth]{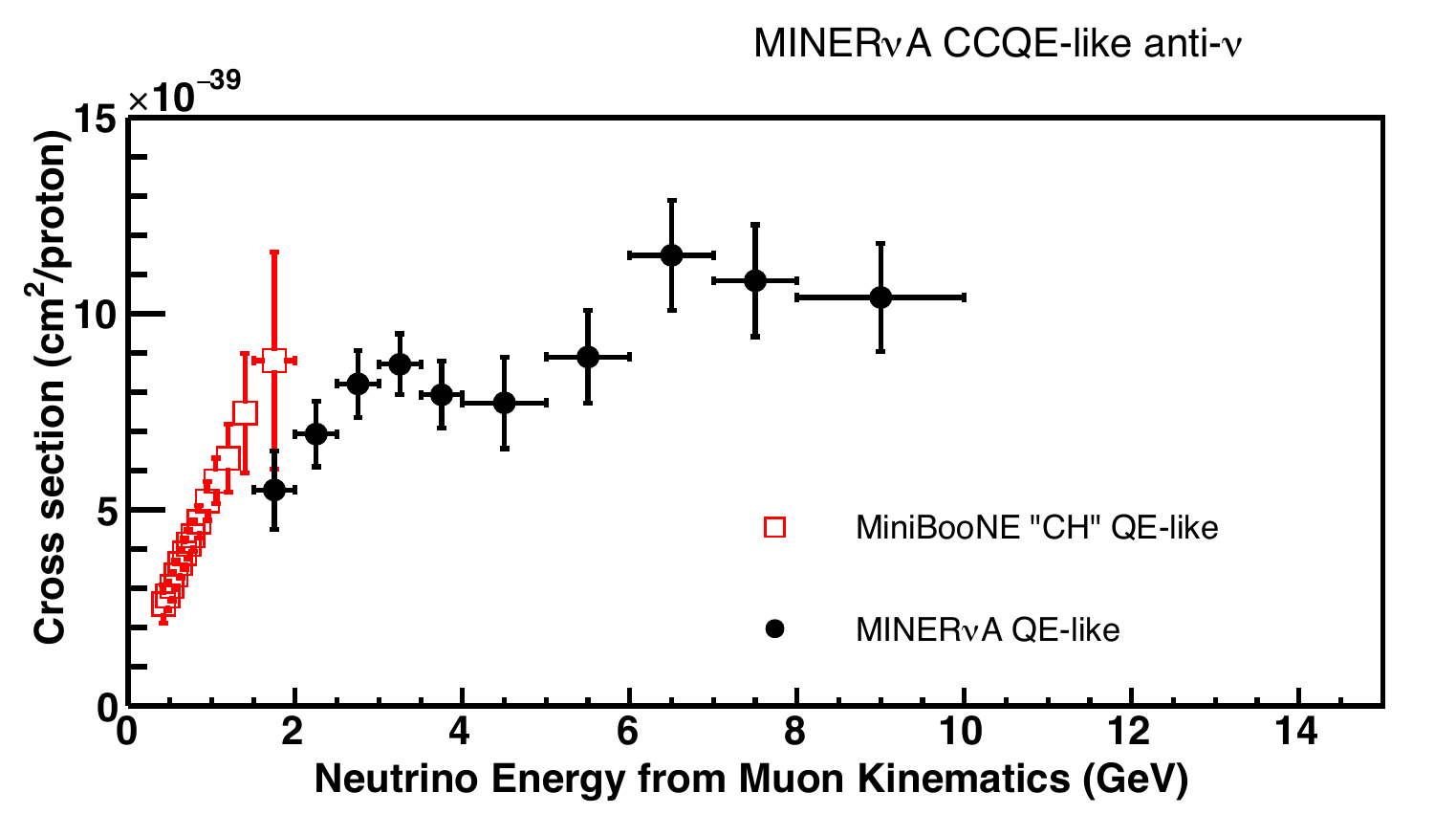}
\caption{\label{fig:sigma_vs_enuQE_qelike}{MINERvA QE-like cross section as a function of \enu compared to data from the MiniBooNE experiment.   The MiniBooNE QE-like definition does not exclude events with proton KE $>$ 120 MeV as MINERvA does so the comparison is not exact. Error bars show total (statistical and systematic) uncertainty.}}
\end{figure}

Figs~\ref{fig:sigma_vs_enuQE_true-QE} and
\ref{fig:sigma_vs_enuQE_qelike} show the cross sections versus \enu,
corrected to cross section/proton, compared to results from MiniBooNE
\cite{miniboone_antinu} and NOMAD \cite{nomad}.  The MiniBooNE cross
sections quoted are the average of their reported cross sections on
mineral oil and their estimated rates on pure carbon, as our
scintillator target lies approximately halfway between those two
compositions.  NOMAD is only shown for the true-QE assumption as they
only quote results for that process. We note that the caveats
discussed in section~\ref{sec:flux} should be taken into account when
comparing these results to other experiments -- namely that this is an
approximation of the energy dependent cross section, and that the
approximations will have different impact on these results than those
measured in beams with different neutrino energy spectra.  Although the MINERvA
data points show a small dip in the $\sim4-6$ GeV region, they are
consistent within uncertainties with models that predict a smoothly
rising cross section in this region (see
Fig.~\ref{fig:cross_sections_qelike_models2}).   The MINERvA neutrino flux~\cite{flux}  changes rapidly in the 4-6 GeV region with uncertainties that are dominated by the
neutrino beam focusing system (see the distribution of uncertainties vs $E_{QE}$ in
Fig.~\ref{fig:xsec_systematics} of
Appendix~\ref{app:systematics}). % response to referee

\subsection{Vertex Energy Distributions}

\begin{figure}
\centering \includegraphics[width=0.49\textwidth]{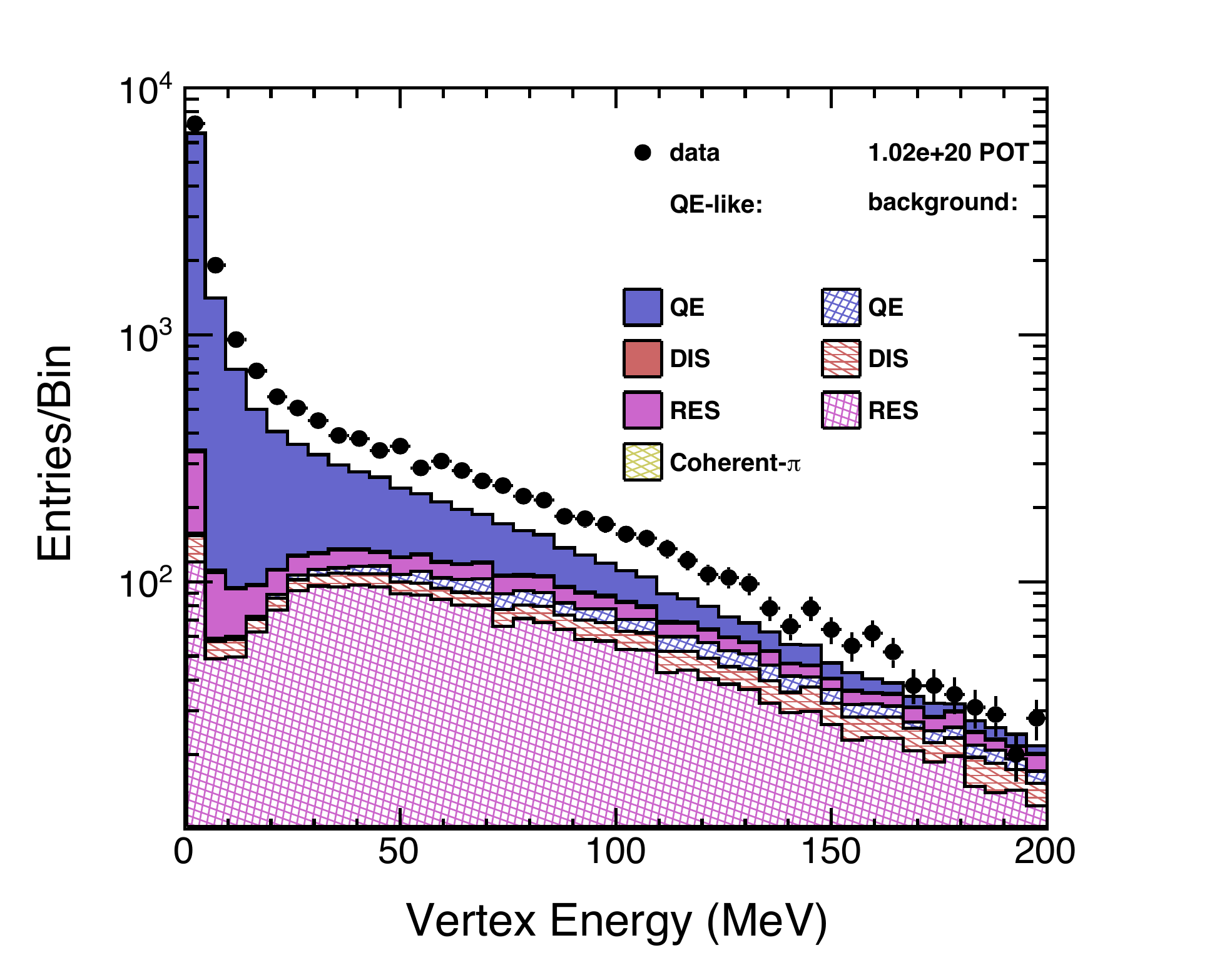}
\centering \includegraphics[width=0.49\textwidth]{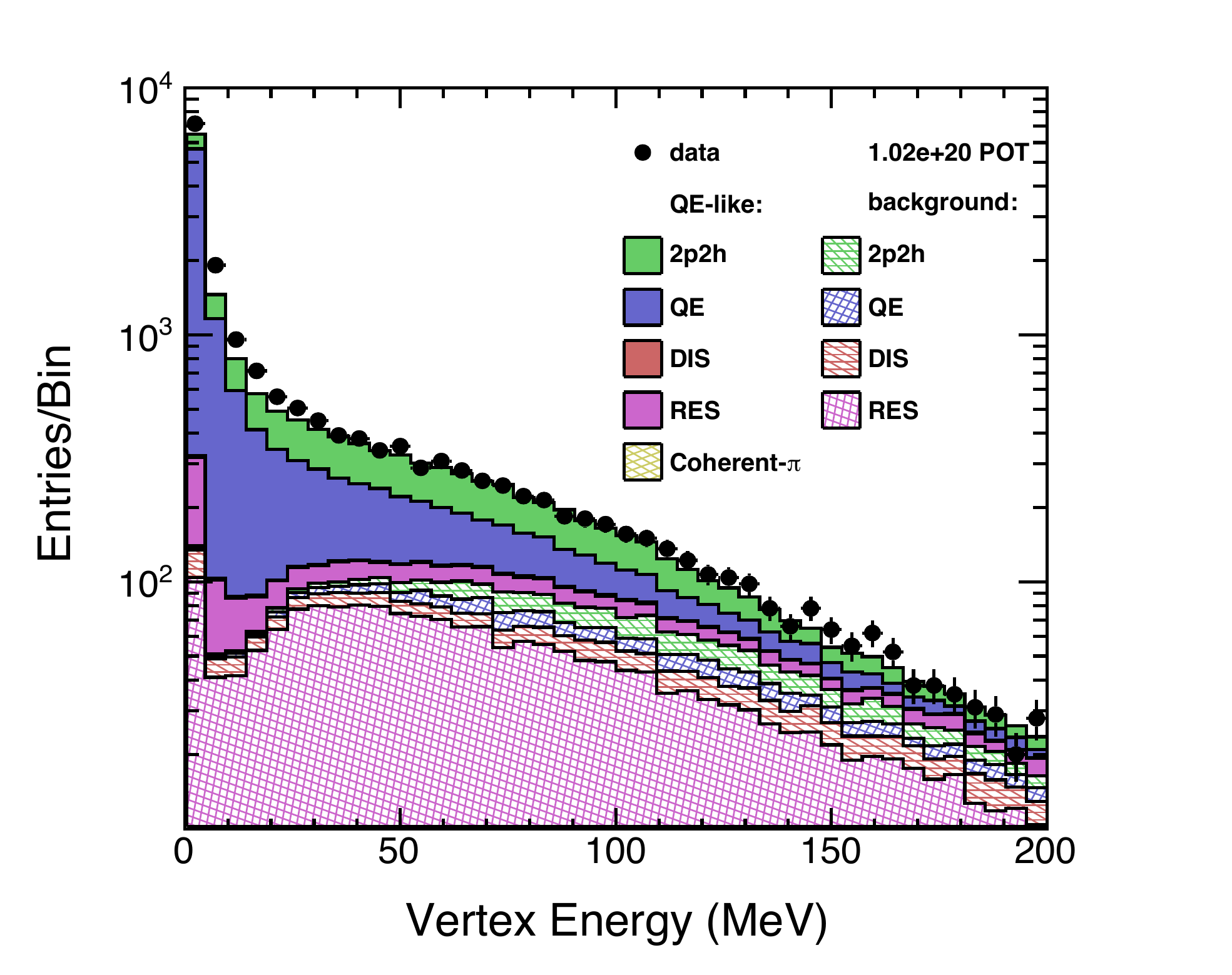}
%\centering \includegraphics[width=0.45\textwidth]{ModelRatios_bkgConst.pdf}
\centering \includegraphics[width=0.49\textwidth]{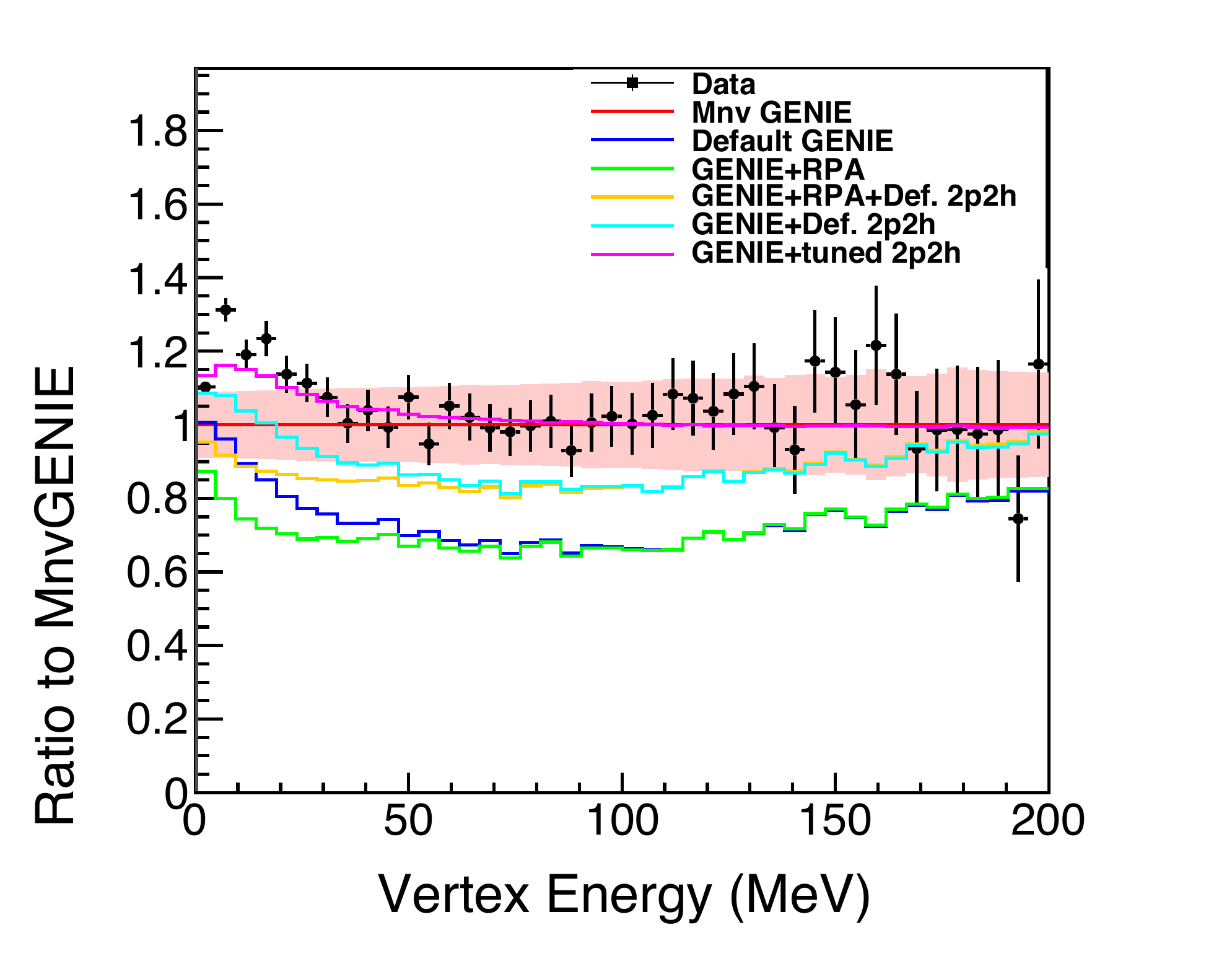}
\caption{\label{fig:vertex}{Reconstructed energy  within 100 mm of the neutrino interaction vertex.  The black points show \minerva data.   In the top plot, the stacked histograms show the predictions of various models using default GENIE 2.8.4.  In the middle plot, the stacked histograms are for the \minerva-tuned GENIE 2.8.4, which includes RPA and \minerva-tuned 2p2h.  The bottom plot shows the ratio of the data and various models to \minerva-tuned GENIE.  Statistical uncertainties only.}}
\end{figure}

Because interactions on multi-nucleon pairs are expected to include additional low-energy nucleons compared to standard QE interactions, reconstructed energy near the interaction vertex is useful for judging the efficacy of 2p2h models.  Figure~\ref{fig:vertex} shows energy reconstructed in scintillator strips that are within 100 mm of the interaction vertex, in the sample used to produce the cross sections discussed earlier, but before background subtraction and efficiency, flux and target number corrections.  Also shown are the expected distributions for default and \minerva-tuned GENIE, and ratios to \minerva-tuned GENIE for the data and several GENIE variants.  Models that omit a 2p2h component have very poor agreement with the data, but the case for RPA suppression is not as strong.  
The model that lies closest to the data is the \minerva-tuned GENIE with the RPA correction omitted.  %response to referee
This is in conflict with similar conclusions drawn from cross-sections versus kinematic distributions, indicating that while \minerva-tuned GENIE is a definite improvement over default GENIE 2.8.4, more improvements are needed to properly simulate the hadronic component of anti-neutrino QE-like interactions.

\section{Conclusion}
\label{sec:conclusion}

We have presented a measurement of a QE-like cross section for anti-neutrino scattering on scintillator.
The signal definition requires no charged pions in the final state and no protons with kinetic energies above 120 MeV. 
This variant of the QE-like definition allows us to include quasi-elastic scatters off of NN pairs in the nucleus in our signal definition and closely matches the actual sensitivity of our detector to low energy protons. 

The main result is presented as a function of muon kinematics \pt and \pznospace. We also present an energy-dependent flux normalized cross section in terms of the neutrino energy and 4-momentum transfer squared as calculated from the muon kinematics using a quasi-elastic assumption.  These data are compared to a large number of models for anti-neutrino QE-like scattering. In particular, we have applied corrections to the default GENIE 2.8.4 scattering model for the Random Phase Approximation and have added 2p2h processes that have been tuned to the observed recoil distributions in an independent MINERvA neutrino scattering sample. This \minerva-tuned model agrees better with our data than default GENIE both visually (Fig.~\ref{fig:GENIE_CV_ratio}) and when a log-normal $\chi^2$ is calculated, as is more appropriate when multiplicative uncertainties are significant.   Moreover, comparisons of reconstructed energy near the interaction vertex between these data and various models indicates poor agreement with models that do not include 2p2h. 

In conclusion, addition of RPA  and 2p2h effects to the simulation substantially improves agreement with the \minerva QE-like data  over  default GENIE.  Addition of either RPA or 2p2h alone is not sufficient. However,  substantial discrepancies between the improved model and data remain, indicating that more model development is needed.  This is the first double-differential measurement of quasi-elastic or QE-like scattering cross sections for anti-neutrinos in this energy range, which is very similar to the expected spectrum of the DUNE experiment, and will be an essential component in the development and tuning of models used in future neutrino oscillation measurements.  
%\clearpage

\ 

\ifnum\sizecheck=0
\begin{acknowledgments}

This document was prepared by members of the MINERvA collaboration using the resources of the Fermi National Accelerator Laboratory (Fermilab), a U.S. Department of Energy, Office of Science, HEP User Facility. Fermilab is managed by Fermi Research Alliance, LLC (FRA), acting under Contract No. DE-AC02-07CH11359.
These resources included support for the \minerva construction project, and support for construction also
was granted by the United States National Science Foundation under
Award PHY-0619727 and by the University of Rochester. Support for
participating scientists was provided by NSF and DOE (USA) by CAPES
and CNPq (Brazil), by CoNaCyT (Mexico), by Proyecto Basal FB 0821, CONICYT PIA ACT1413, Fondecyt 3170845 and 11130133 (Chile), by DGI-PUCP and UDI/VRI-IGI-UNI (Peru),and by the Latin American Center for Physics (CLAF).  We
thank the MINOS Collaboration for use of its near detector data. Finally, we thank the staff of Fermilab for support of the beamline, the detector and computing infrastructure.

\end{acknowledgments}
\fi

\bigskip

\ifnum\sizecheck=0
\bibliographystyle{apsrev4-1}
\bibliography{2DAntineutrinoCCQE}
\fi

\clearpage

\section{Appendix: Summary of Systematic Uncertainties}
\label{app:systematics}
\begin{figure*}\centering
                \includegraphics[width=0.49\textwidth]{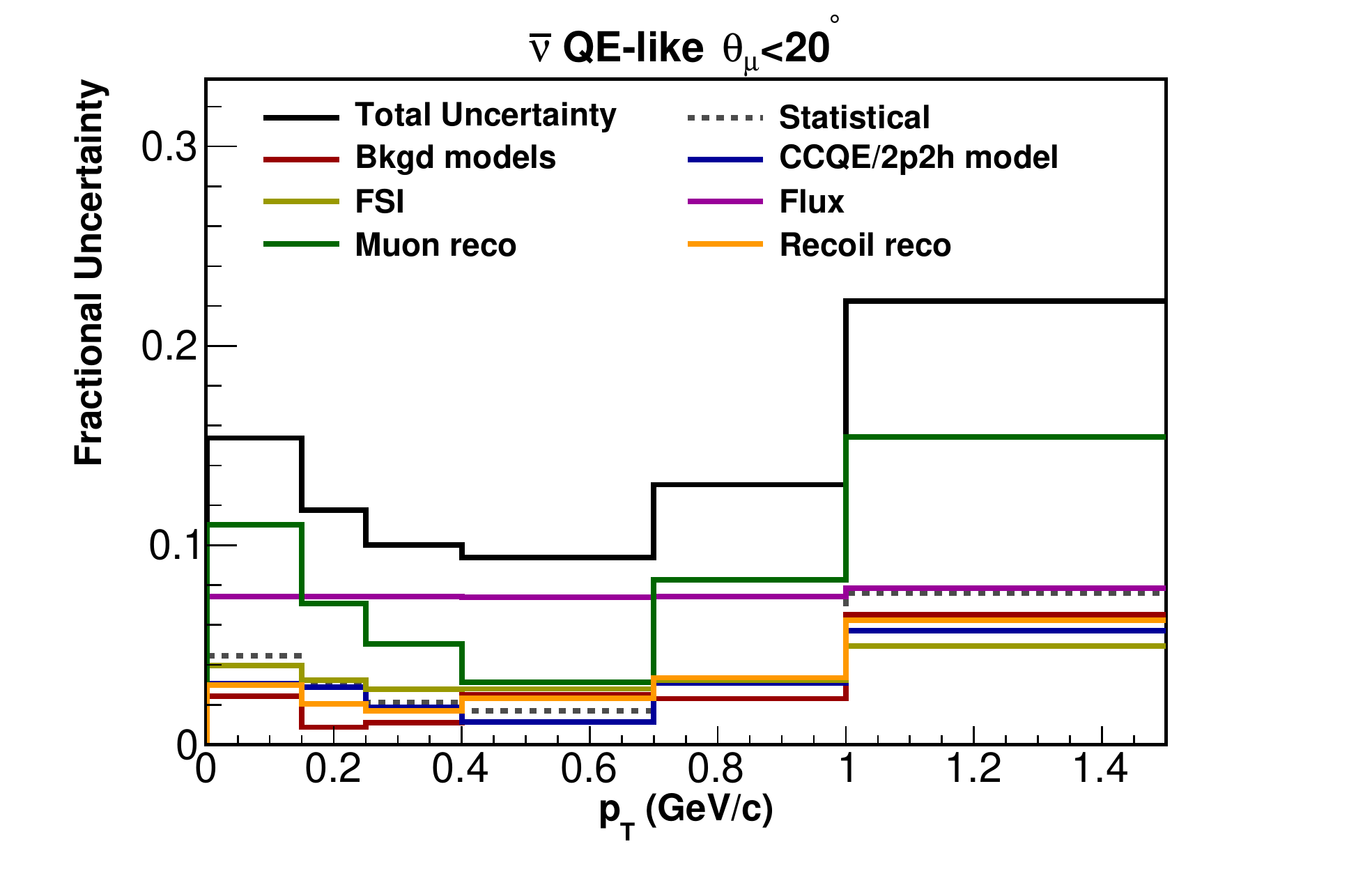}
                \includegraphics[width=0.49\textwidth]{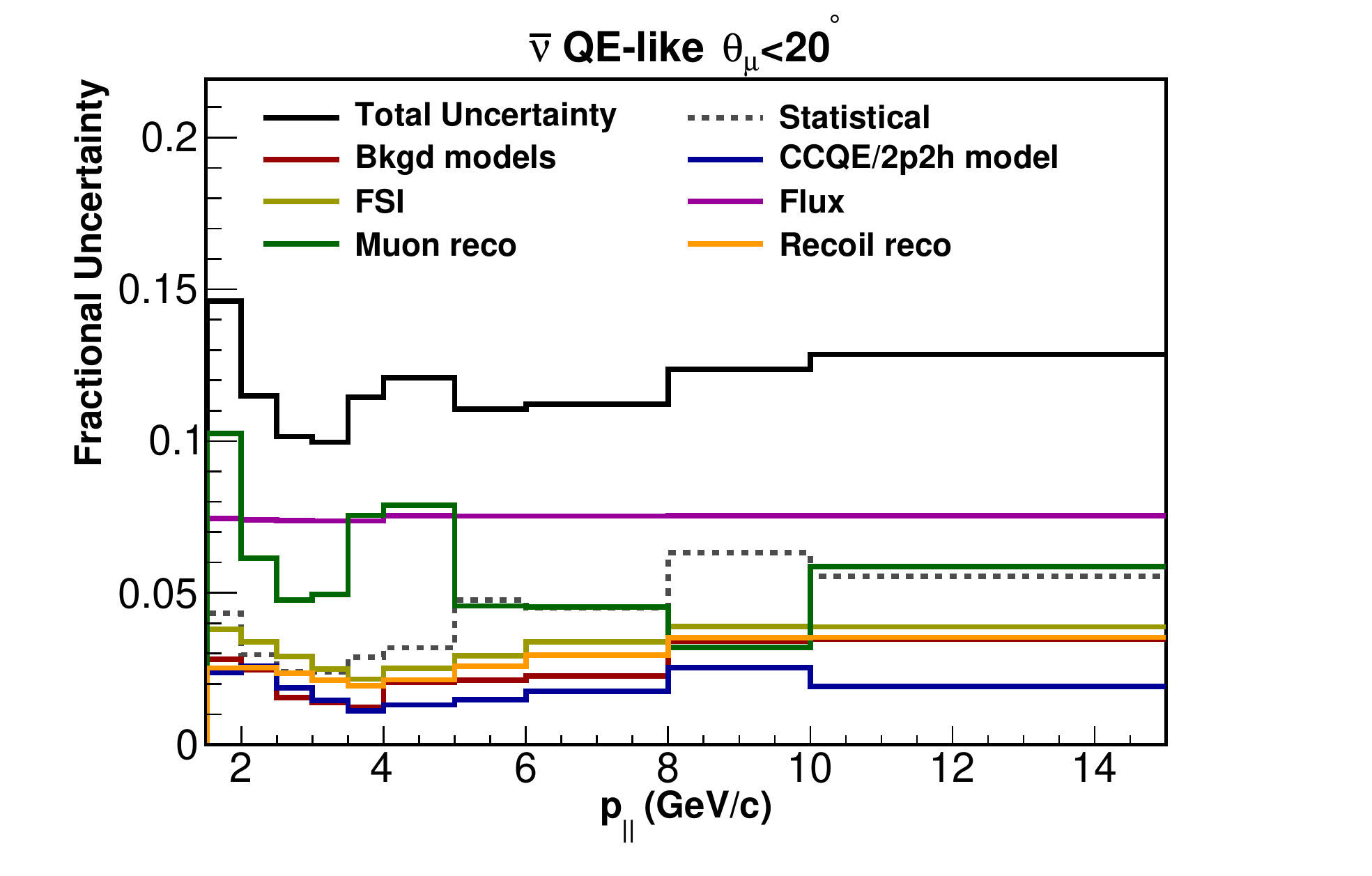}
                \includegraphics[width=0.49\textwidth]{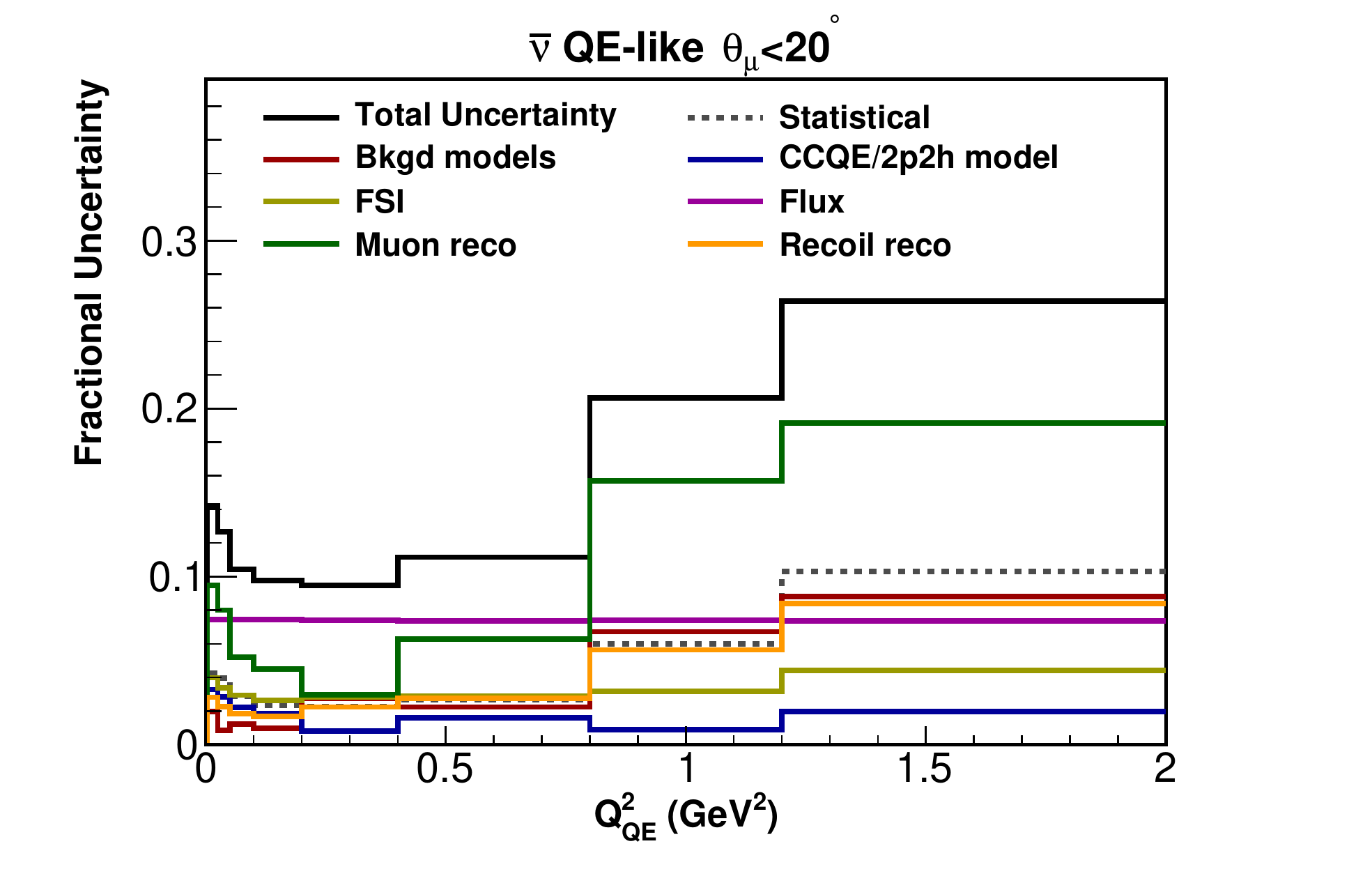}
                \includegraphics[width=0.49\textwidth]{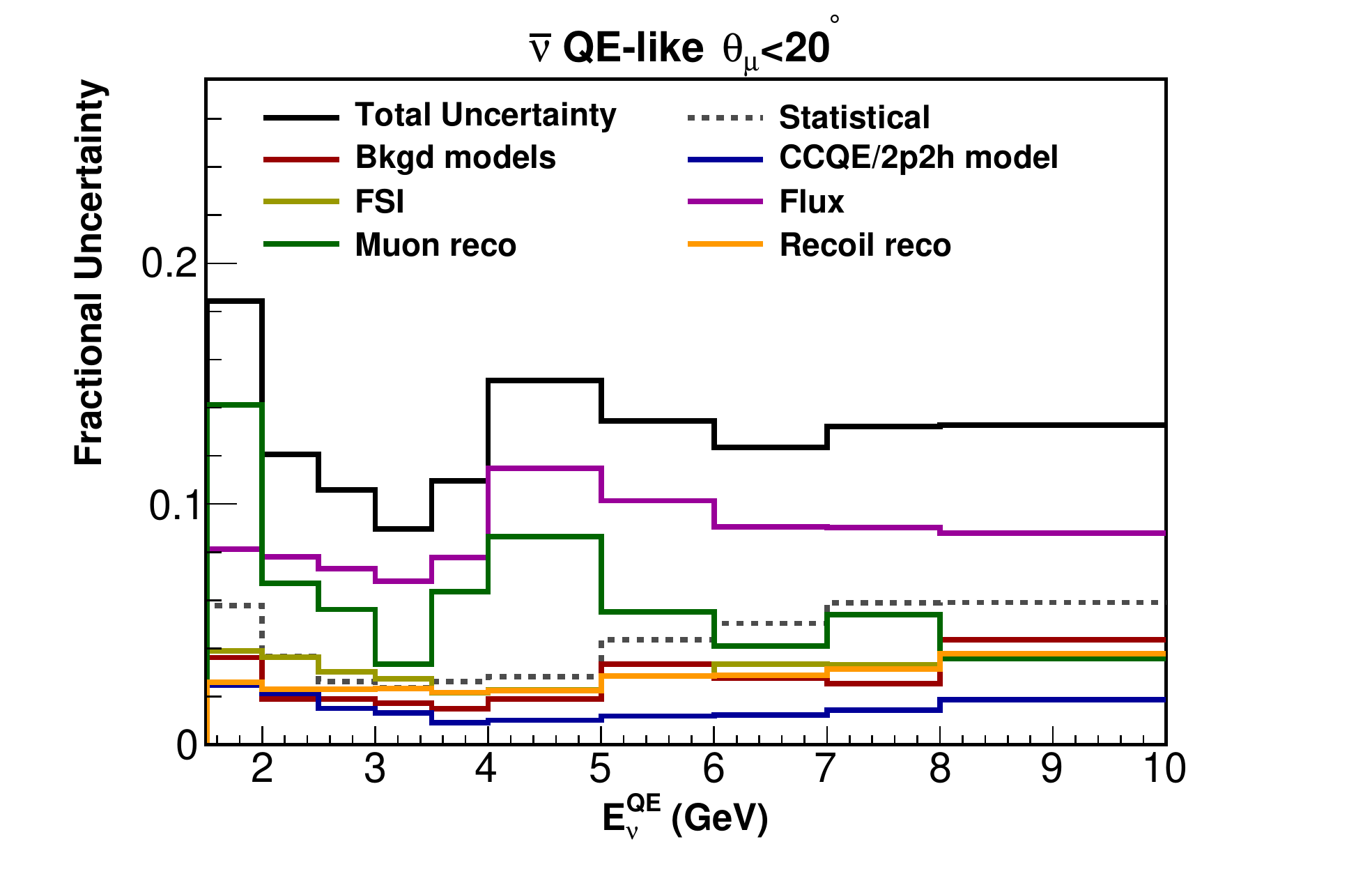}
                \caption{ \label{fig:xsec_systematics}{Summary of
                    fractional uncertainties on the
                    single-differential projections of the
                    double-differential QE-like cross section
                    measurements in data.  The cross sections
                    themselves are shown in Fig.~\ref{fig:cross_sections_qelike_models2}.}}        
    
\end{figure*}

\begin{figure*}\centering

                \includegraphics[width=0.9\textwidth]{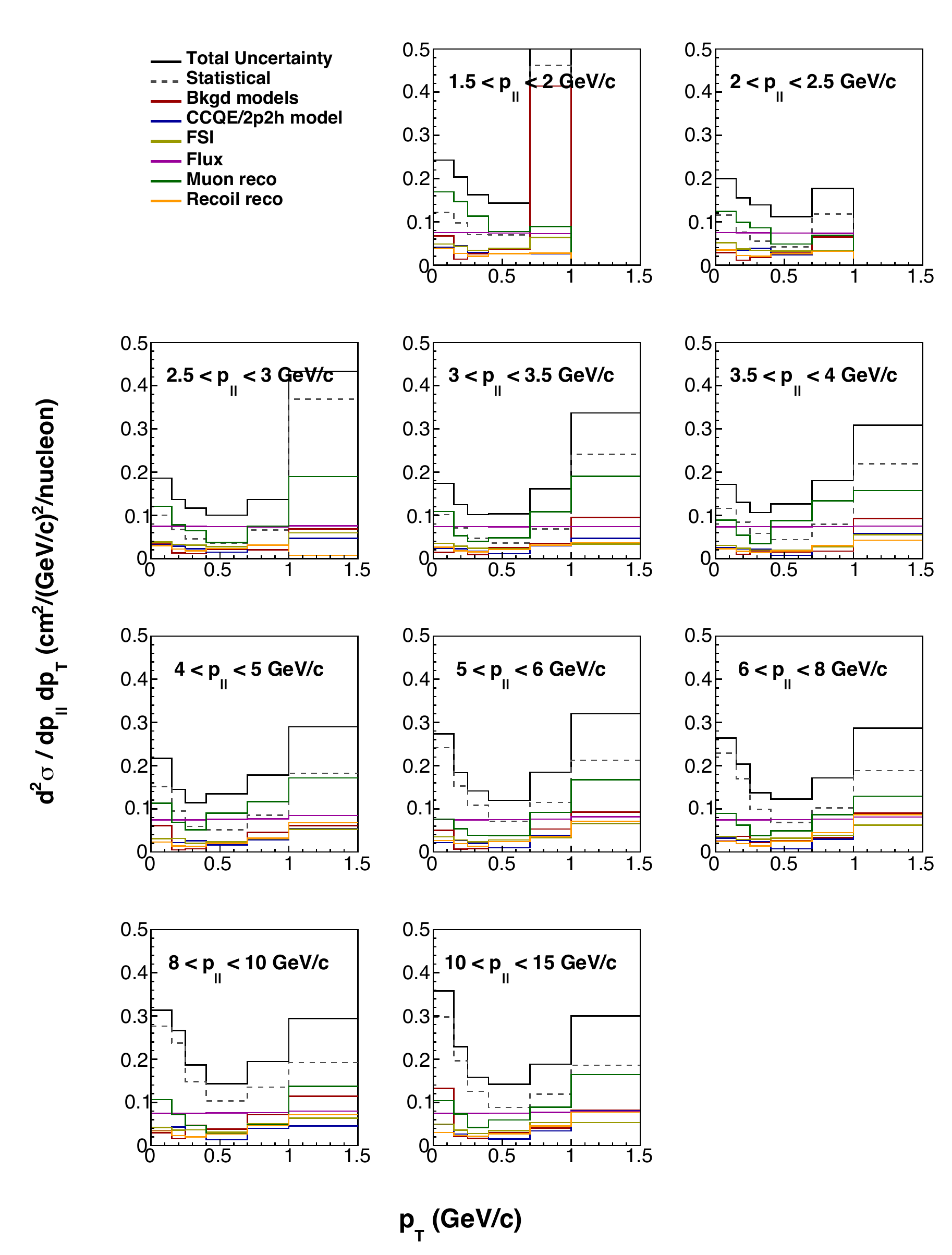}
        \caption{\label{fig:systs2ptpz}{Absolute fractional
            uncertainties on the double-differential QE-like cross section vs. muon transverse momentum, in bins of muon longitudinal momentum.}}        
\end{figure*}

\begin{figure*}\centering
                \includegraphics[width=0.9\textwidth]{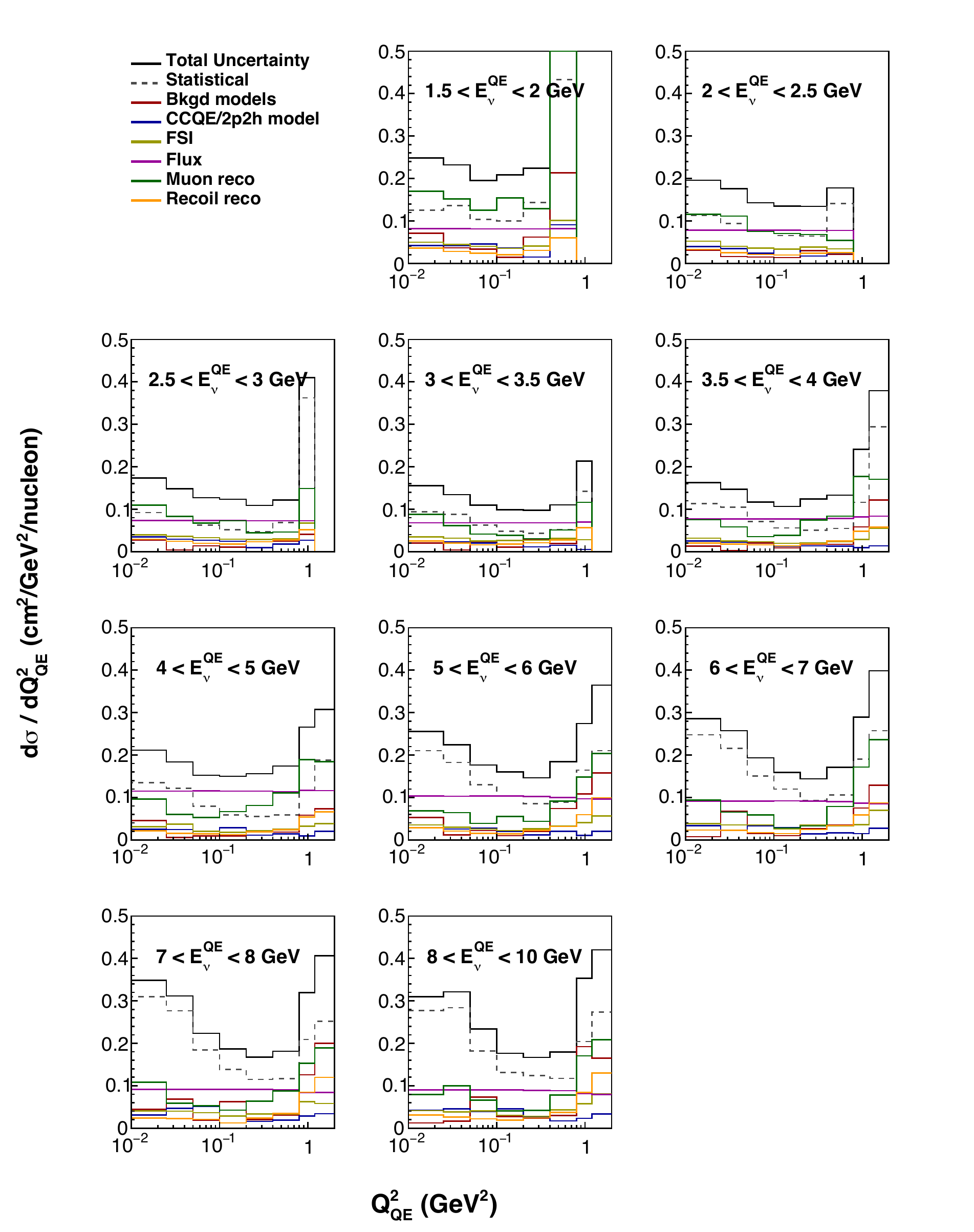}
        \caption{ \label{fig:systs2enuqsq}  {Absolute fractional uncertainties on the cross section vs. $Q^2$, in bins of \enu.}}

\end{figure*}

\begin{figure*}\centering
	 \includegraphics[width=0.9\columnwidth]{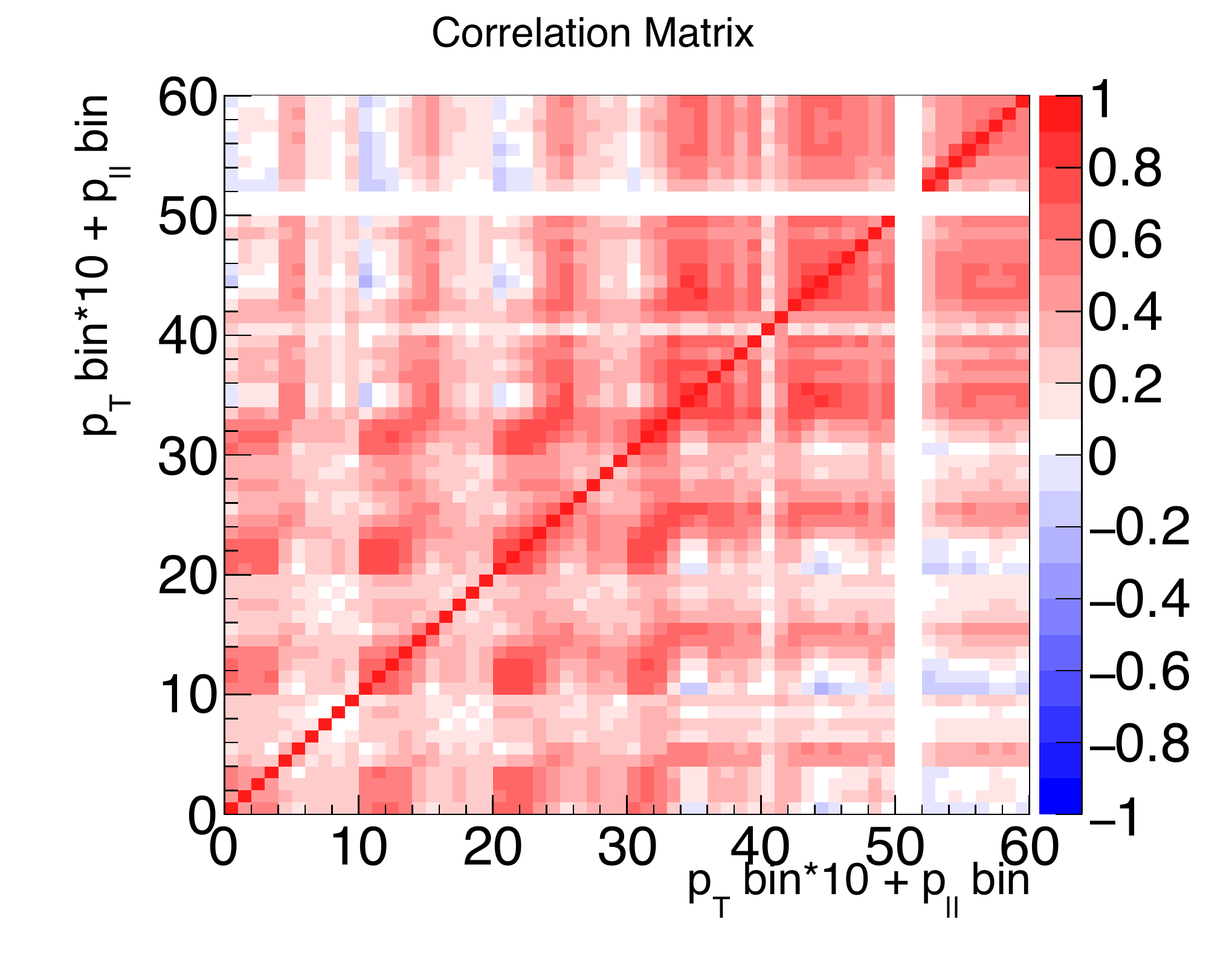}
	 \caption{\label{fix:corr}{Correlation matrix for the 6$\times$ 10 \pt and \pz bins.  As in the migration matrix, the larger blocks are \pt bins with the smaller scale bins \pz bins.}}
\end{figure*}

Figure \ref{fig:xsec_systematics} shows a summary of the fractional
systematic uncertainties from each of category of uncertainty
described in section~\ref{sec:systematics} for the one dimensional cross-sections.  The model uncertainties have been further subdivided into those primarily affecting the signal models (quasi-elastic and 2p2h), background models, and final state interactions.  Figures~\ref{fig:systs2ptpz} and~\ref{fig:systs2enuqsq} show fractional uncertainties on the double-differential cross sections.
Figure~\ref{fix:corr} shows the correlation matrix for all systematic uncertainties.

\clearpage

\clearpage
\section{Appendix: CCQE Cross-Sections}
\label{app:ccqe}\label{sec:ccqesig}
The main focus of the analysis was the calculation of CCQE-like double-differential cross sections shown above, which correspond to our measurement for the signal definition described in section \ref{sec:signal}. As an extension to the analysis, however, we also calculated a true CCQE cross section. Recall that, for the CCQE-like cross section, our  signal corresponded to interactions with a CCQE-like final state, even if that final state was generated by a resonant or DIS interaction followed by FSI. For the true CCQE definition, our signal corresponds only to events where the initial interaction was quasi-elastic, even if FSI created final-state particles such as pions that mimicked a non-quasi-elastic interaction. The signal also includes 2p2h events where a CCQE interaction takes place on a correlated pair. 

\begin{figure*}
\centering
 \includegraphics[width=.9\textwidth]{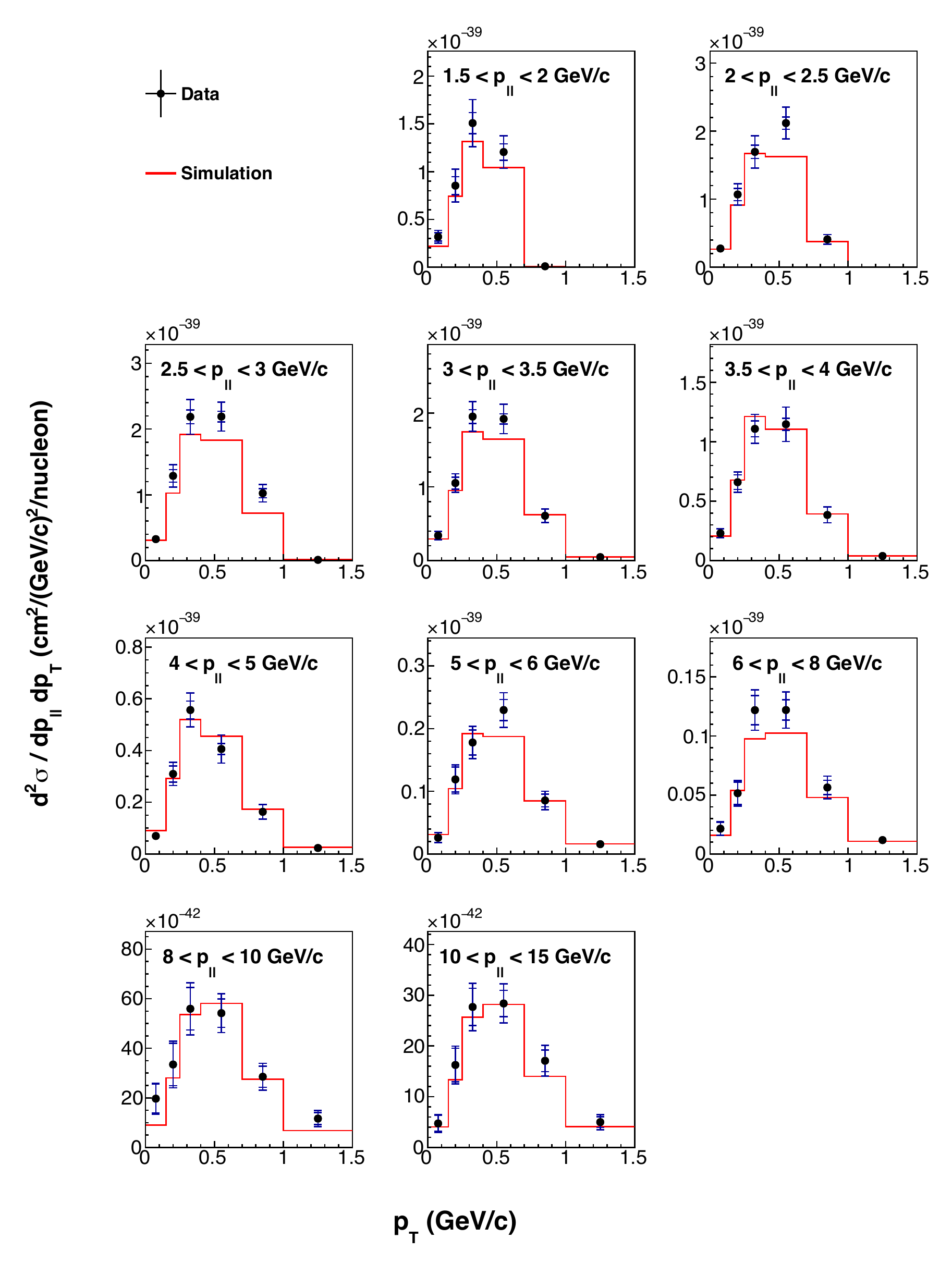}
\caption{{Double-differential flux-integrated true CCQE cross section $d^2\sigma/dp_Tdp_\parallel$ vs. muon transverse momentum, in bins of muon longitudinal momentum}\label{fig:ccqe_cross_sections_pt_binned}.  Inner error bars show statistical uncertainties; outer error bars show total (statistical and systematic) uncertainty. These results are tabulated in Tables \ref{tab:ccqe_xsec_pzpt}--\ref{tab:ccqe_sys_pzpt}.}        
\end{figure*}

\begin{figure*}
  \centering
   \includegraphics[width=.9\textwidth]{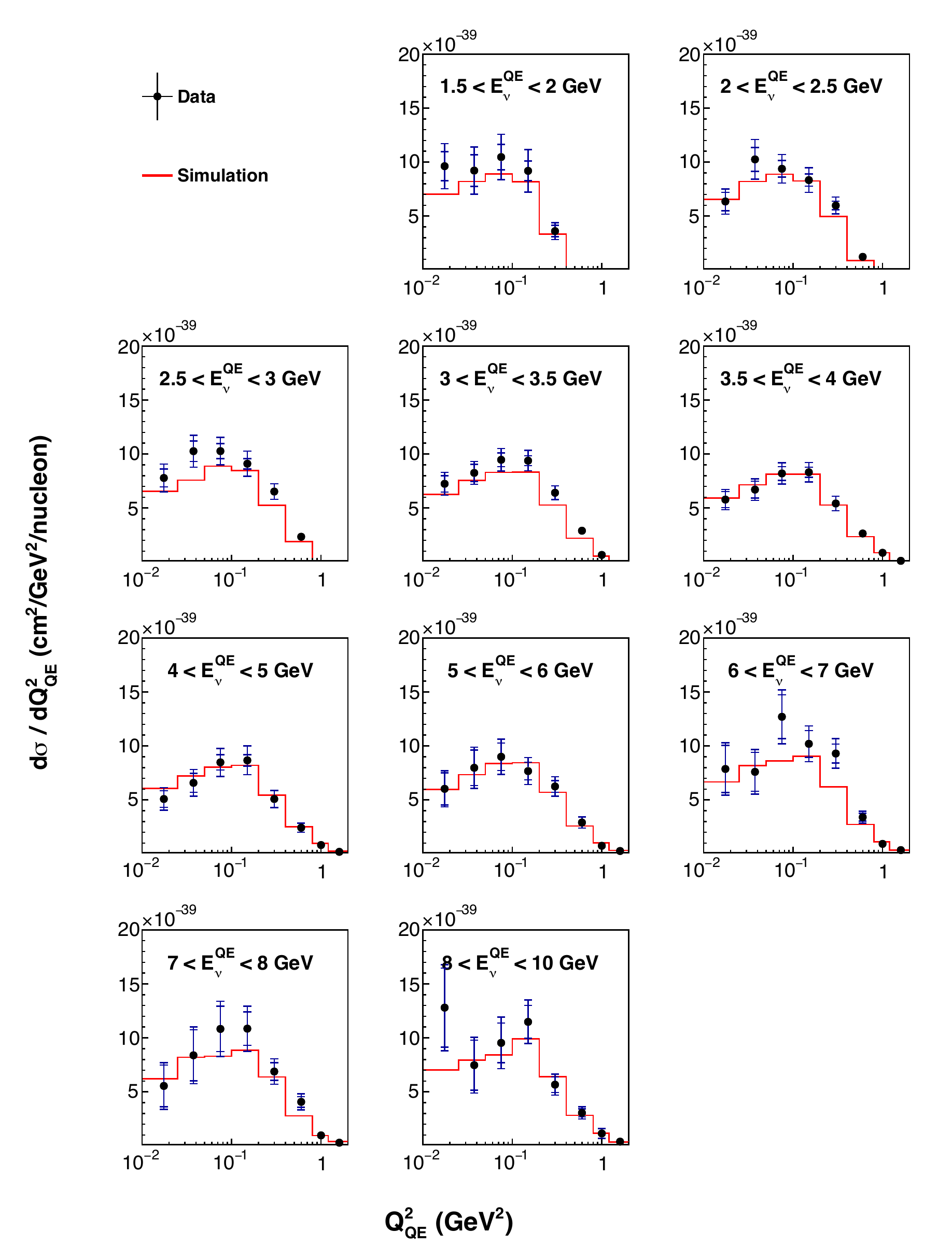}
  \caption{{True CCQE cross section  $d\sigma(E_\nu^{QE})/dQ^2_{QE}$, in bins of $E_\nu^{QE}$}\label{fig:ccqecross_sections_qsq_binned}.  Inner error bars show statistical uncertainties; outer error bars show total (statistical and systematic) uncertainty. These results are tabulated in Tables \ref{tab:ccqe_xsec_eqe}--\ref{tab:ccqe_sys_eqe}.}        
\end{figure*}

\begin{figure*}
     \includegraphics[width=0.49\textwidth]{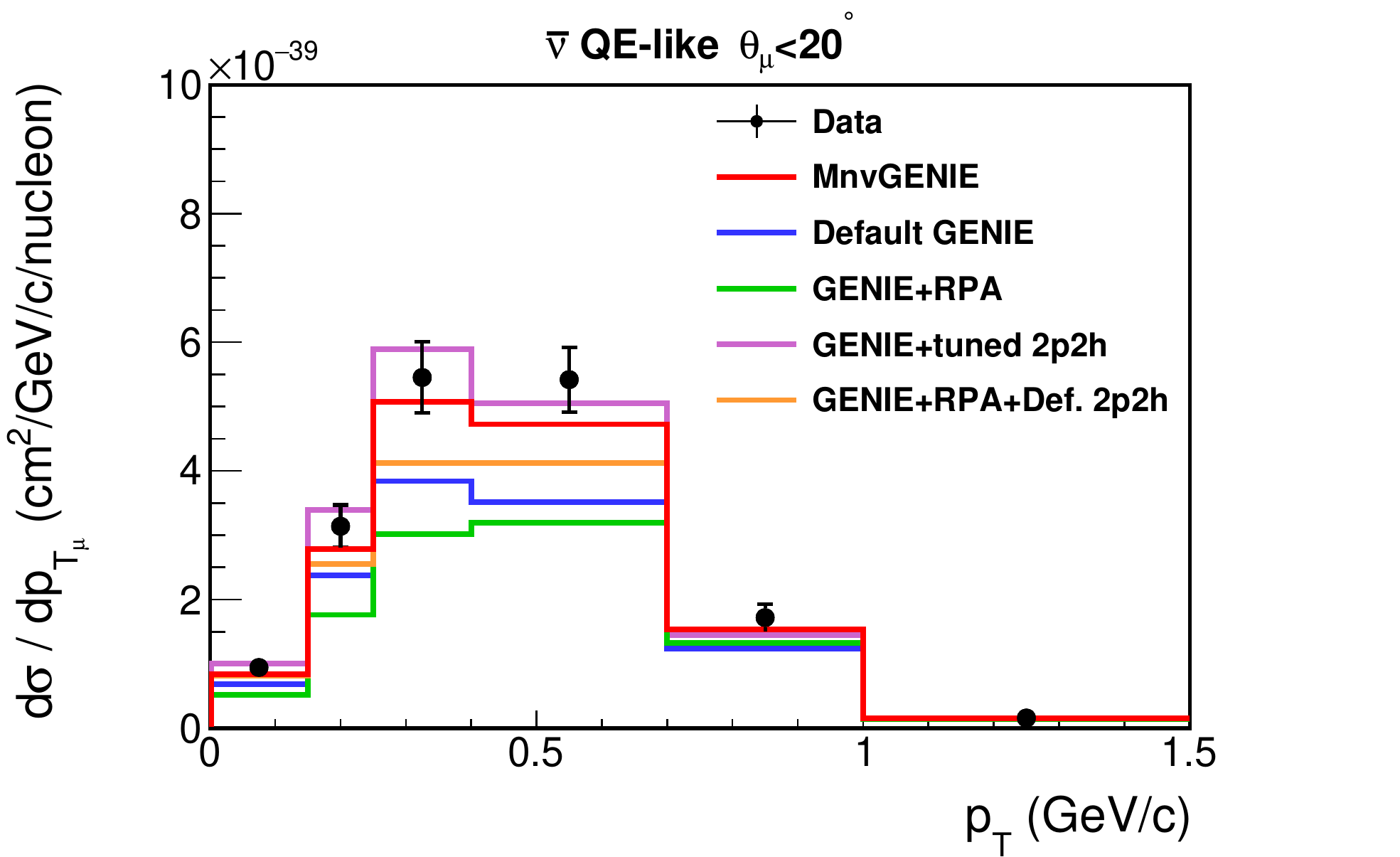}
     \includegraphics[width=0.49\textwidth]{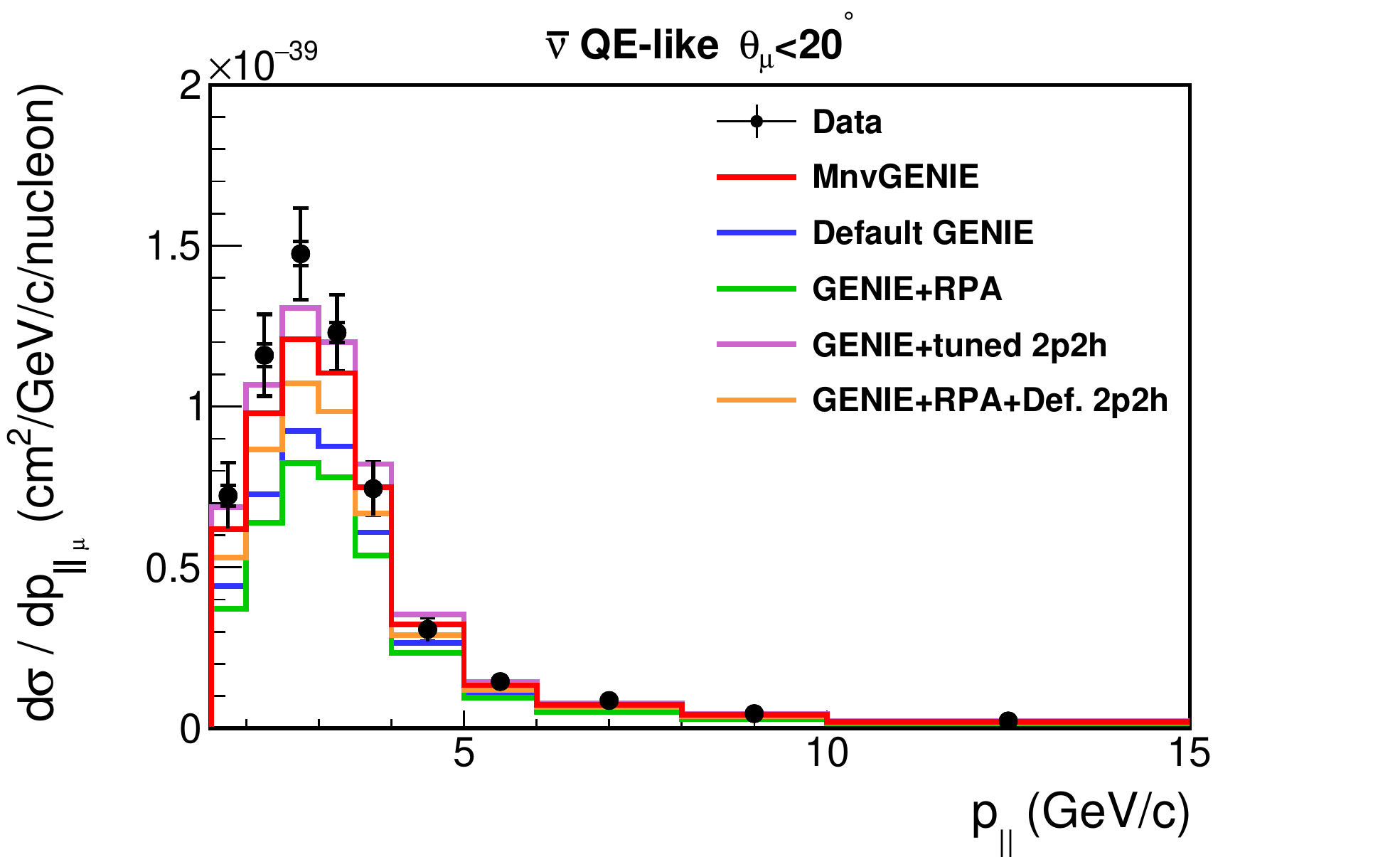}
     \includegraphics[width=0.49\textwidth]{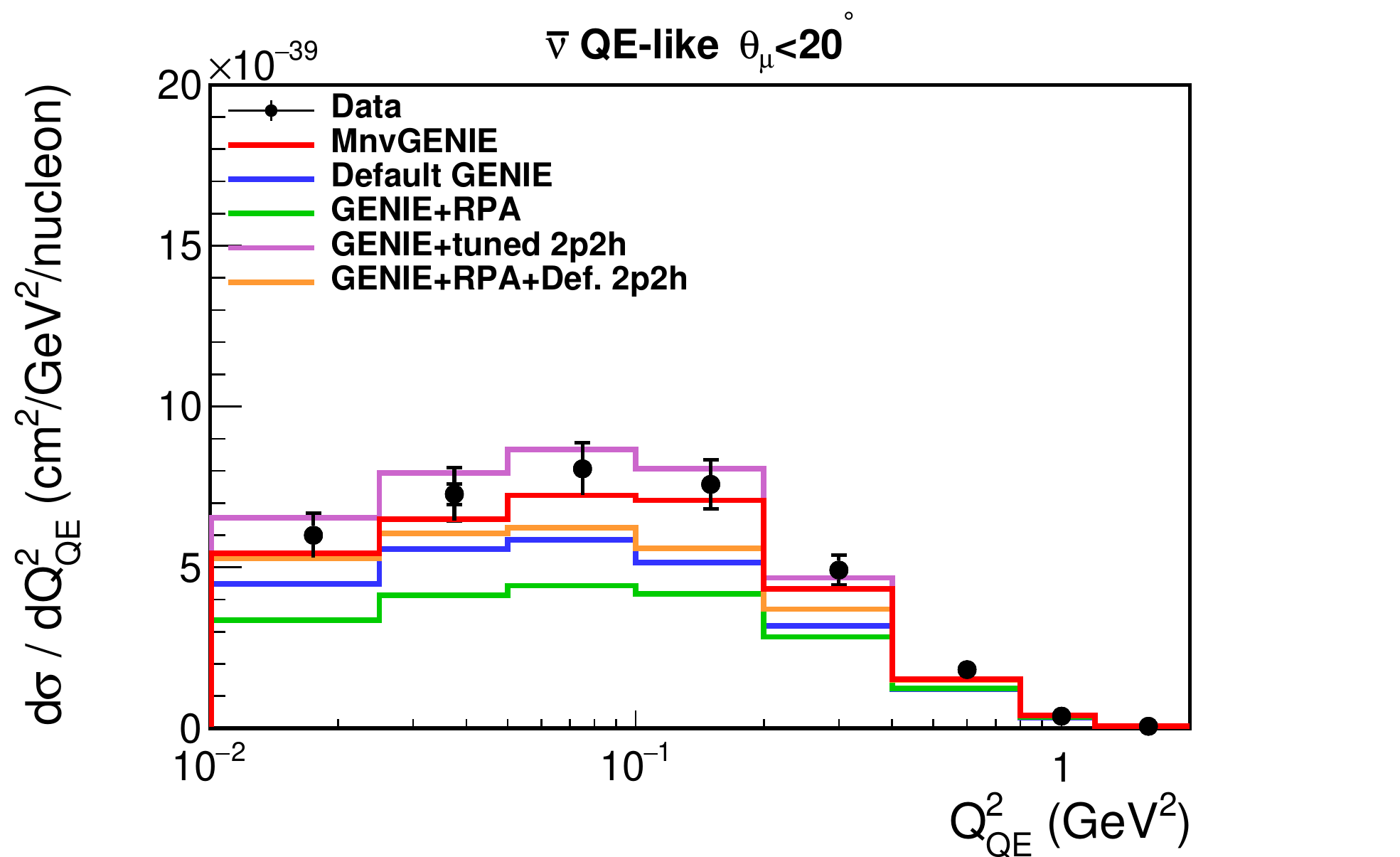}
     \includegraphics[width=0.49\textwidth]{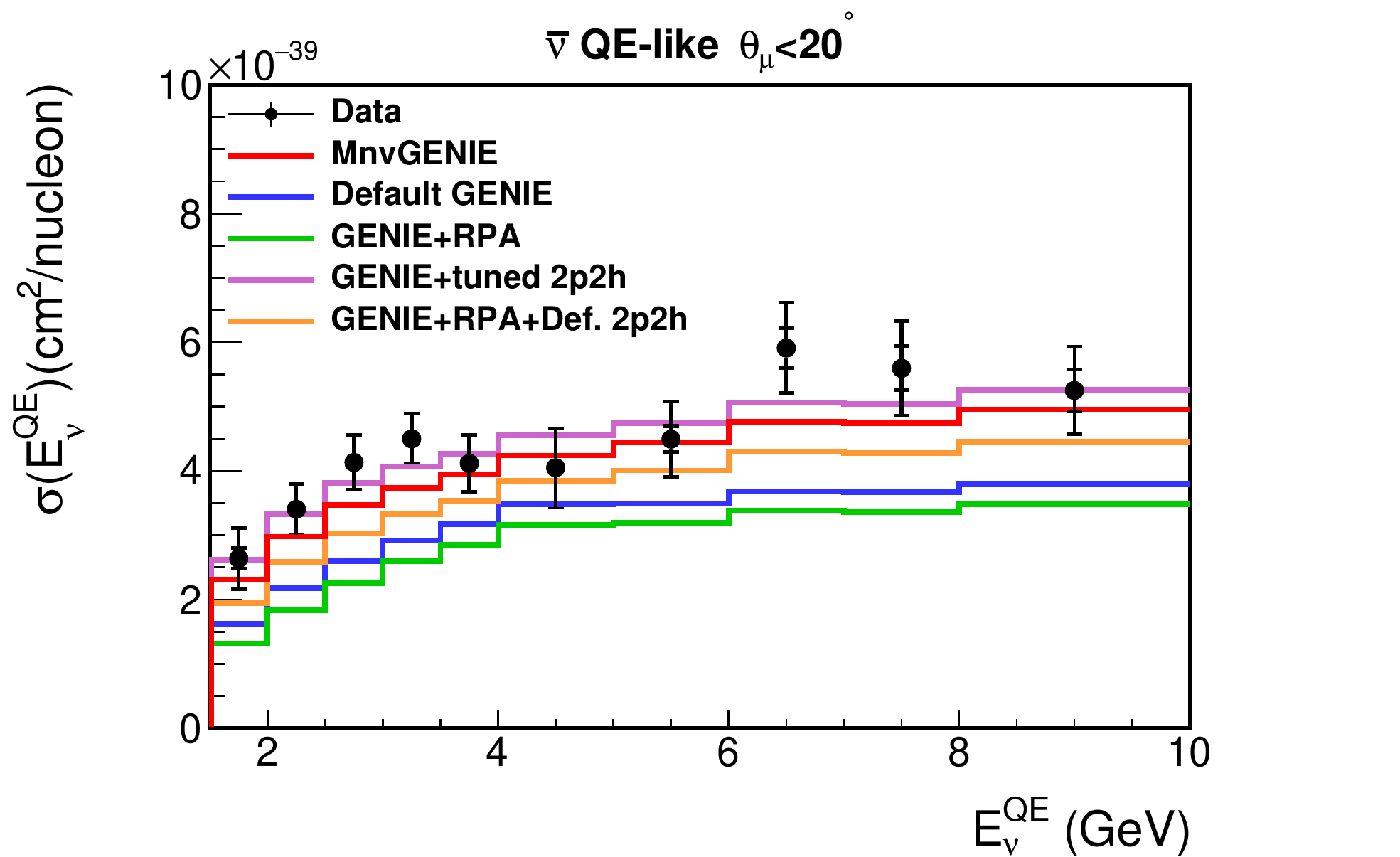}
\caption{Single-differential projections of the double-differential 
  true CCQE cross section measurements in data, compared to
  \minerva-tuned GENIE (red curve, includes RPA and MINERvA-tuned
  2p2h), GENIE without any modifications except the single
  non-resonant pion correction discussed in section
  \ref{sec:neutrinosim} (blue), GENIE with the RPA weight but no 2p2h
  component  (green), GENIE with MINERvA-tuned 2p2h but no RPA
  (violet), and GENIE with RPA and untuned 2p2h  (orange).  \label{fig:cross_sections_ccqe_projection}  Inner 
  error bars show statistical uncertainties; outer error bars show 
  total (statistical and systematic) uncertainty.  These results are tabulated in Tables \ref{tab:ccqe_xsec_pt}, \ref{tab:ccqe_xsec_pz}, \ref{tab:ccqe_xsec_enuqe_intqsq} and \ref{tab:ccqe_xsec_qsq_intenuqe}.}\end{figure*}

 The true CCQE double-differential cross sections are shown in Fig.~\ref{fig:ccqe_cross_sections_pt_binned} ($d^2\sigma/dp_Tdp_\parallel$) and \ref{fig:ccqecross_sections_qsq_binned} ($d\sigma(E_\nu^{QE})/dQ^2_{QE}$), while one-dimensional projections are shown in Fig.~\ref{fig:cross_sections_ccqe_projection}.  Also shown in these figures are the predictions in our default simulation, which includes both CCQE and 2p2h contributions.

\clearpage

\section{Appendix: Tables of cross section measurements}\label{app:xsec}

\subsection{CCQE-like cross sections}
The tables in this section list our cross section measurements for the
CCQE-like signal definition explained in section \ref{sec:signal}. For
all of the double-differential measurements, we show a table of values
corresponding to the cross section, followed by a table of statistical
uncertainties, then a table of systematic uncertainties. Units are explained in the captions to the tables.

\subsubsection{\pt and \pz}
The differential cross section with respect to muon parallel and
transverse momentum, $d^2\sigma/dp_Tdp_\parallel$ is shown
in Table \ref{tab:xsec_pzpt}. The statistical uncertainty on these
measurements are shown in Table \ref{tab:stat_pzpt}, and the systematic uncertainty in Table \ref{tab:sys_pzpt}.

Table \ref{tab:xsec_pt} shows the differential cross section
$d\sigma/dp_T$, generated by projecting the two-dimensional
measurement onto the $p_T$ axis. Table \ref{tab:xsec_pz} shows the
differential cross section $d\sigma/dp_\parallel$, generated by
projecting the two-dimensional measurement onto the $p_\parallel$
axis.   Both tables also include statistical and systematic
uncertainties.  

\begin{table*}[h]
	\makeatletter{}\begin{tabular}
{ l  r   r   r   r   r   r   r }\hline\hline
 & 0-0.15 & 0.15-0.25 & 0.25-0.4 & 0.4- 0.7 & 0.7-1 & 1-1.5\\
\hline
1.5 - 2 & 37.71 & 105.56 & 163.89 & 131.76 & 0.84 & 0.00\\
2 - 2.5 & 34.86 & 127.51 & 184.54 & 224.95 & 42.85 & 0.00\\
2.5 - 3 & 40.67 & 148.90 & 235.21 & 230.56 & 100.81 & 1.10\\
3 - 3.5 & 37.01 & 118.56 & 203.85 & 197.59 & 62.56 & 4.63\\
3.5 - 4 & 26.85 & 77.89 & 118.78 & 116.54 & 38.53 & 4.22\\
4 - 5 & 7.64 & 32.55 & 59.60 & 42.21 & 16.10 & 2.43\\
5 - 6 & 3.63 & 13.65 & 18.91 & 24.43 & 8.39 & 1.48\\
6 - 8 & 2.20 & 5.64 & 12.73 & 13.19 & 5.81 & 0.98\\
8 - 10 & 1.90 & 3.36 & 5.89 & 5.69 & 3.49 & 1.21\\
10 - 15 & 0.54 & 1.56 & 3.01 & 3.07 & 1.78 & 0.52\\
\hline\hline
\end{tabular} 
 	\caption{Measured double-differential CCQE-like cross section $d^2\sigma/dp_Tdp_\parallel$. Units are $10^{-41}$cm$^2$/GeV$^2$/nucleon. Columns represent bins of \pt (GeV), rows are bins of \pz (GeV).} 
	\label{tab:xsec_pzpt}
\end{table*}
\begin{table*}[h]
	\makeatletter{}\begin{tabular}
{ l   r   r   r   r   r   r   r } \hline\hline
 & 0-0.15 & 0.15-0.25 & 0.25-0.4 & 0.4- 0.7 & 0.7-1 & 1-1.5\\
\hline
1.5 - 2 & 4.58 & 10.27 & 11.49 & 9.18 & 0.39 & 0.00\\
2 - 2.5 & 4.02 & 9.62 & 10.17 & 9.43 & 5.05 & 0.00\\
2.5 - 3 & 4.09 & 9.97 & 10.77 & 8.27 & 6.62 & 0.40\\
3 - 3.5 & 3.76 & 8.43 & 9.50 & 7.02 & 4.30 & 1.12\\
3.5 - 4 & 3.12 & 6.58 & 6.89 & 5.09 & 3.06 & 0.93\\
4 - 5 & 1.16 & 3.09 & 3.54 & 2.16 & 1.37 & 0.44\\
5 - 6 & 0.88 & 2.09 & 2.05 & 1.73 & 0.96 & 0.32\\
6 - 8 & 0.51 & 0.96 & 1.26 & 0.90 & 0.60 & 0.18\\
8 - 10 & 0.53 & 0.80 & 0.87 & 0.59 & 0.47 & 0.23\\
10 - 15 & 0.16 & 0.31 & 0.38 & 0.27 & 0.21 & 0.10\\
\hline\hline
\end{tabular} 
 	\caption{Statistical uncertainty on the measured double-differential  CCQE-like cross section $d^2\sigma/dp_Tdp_\parallel$. Units are $10^{-41}$cm$^2$/GeV$^2$/nucleon. Columns represent bins of \pt (GeV), rows are bins of \pz (GeV).} 
	\label{tab:stat_pzpt}
\end{table*}
\begin{table*}[h]
	\makeatletter{}\begin{tabular}
{ l r   r   r   r   r   r   r }\hline\hline
 & 0-0.15 & 0.15-0.25 & 0.25-0.4 & 0.4- 0.7 & 0.7-1 & 1-1.5\\
\hline
1.5 - 2 & 6.32 & 17.38 & 23.03 & 15.69 & 0.36 & 0.00\\
2 - 2.5 & 4.17 & 15.24 & 22.51 & 22.07 & 5.01 & 0.00\\
2.5 - 3 & 4.70 & 15.92 & 23.89 & 20.42 & 10.60 & 0.24\\
3 - 3.5 & 3.95 & 10.85 & 17.21 & 18.32 & 8.45 & 1.04\\
3.5 - 4 & 2.61 & 6.72 & 10.14 & 13.55 & 5.96 & 0.89\\
4 - 5 & 0.97 & 3.28 & 5.60 & 5.26 & 2.49 & 0.53\\
5 - 6 & 0.39 & 1.21 & 1.64 & 2.36 & 1.21 & 0.35\\
6 - 8 & 0.22 & 0.55 & 1.14 & 1.35 & 0.80 & 0.21\\
8 - 10 & 0.22 & 0.34 & 0.63 & 0.56 & 0.49 & 0.27\\
10 - 15 & 0.09 & 0.15 & 0.27 & 0.34 & 0.26 & 0.12\\
\hline\hline
\end{tabular} 
 	\caption{Systematic uncertainty on the measured double-differential  CCQE-like cross section $d^2\sigma/dp_Tdp_\parallel$. Units are $10^{-41}$cm$^2$/GeV$^2$/nucleon. Columns represent bins of \pt (GeV), rows are bins of \pz (GeV).} 
	\label{tab:sys_pzpt}
\end{table*}
\begin{table*}[h]
	\makeatletter{}\begin{tabular}
{c  r r r}\hline\hline
Bin & Cross & Statistical  & Systematic \\
 & section &  uncertainty &  uncertainty\\
\hline
0 - 0.15 & 110.74 & 4.93 & 16.30\\
0.15 - 0.25 & 361.23 & 11.20 & 41.01\\
0.25 - 0.4 & 583.94 & 12.33 & 57.20\\
0.4 - 0.7 & 570.42 & 9.66 & 52.65\\
0.7 - 1 & 174.81 & 5.53 & 22.10\\
1 - 1.5 & 15.87 & 1.20 & 3.32\\
\hline\hline
\end{tabular} 
 	\caption{Differential  CCQE-like cross section $d\sigma/dp_T$, along with statistical and systematic uncertainties. Units are $10^{-41}$cm$^2$/GeV/nucleon. The $p_T$ bins are in GeV.} 
	\label{tab:xsec_pt}
\end{table*}
\begin{table*}[h]
	\makeatletter{}\begin{tabular}
{c  r r r}\hline\hline
Bin & Cross & Statistical  & Systematic \\
 & section &  uncertainty &  uncertainty\\
\hline
1.5 - 2 & 80.58 & 3.48 & 11.25\\
2 - 2.5 & 126.00 & 3.73 & 13.98\\
2.5 - 3 & 156.23 & 3.76 & 15.40\\
3 - 3.5 & 128.35 & 3.08 & 12.41\\
3.5 - 4 & 78.27 & 2.26 & 8.66\\
4 - 5 & 32.05 & 1.02 & 3.73\\
5 - 6 & 15.33 & 0.73 & 1.53\\
6 - 8 & 8.99 & 0.40 & 0.92\\
8 - 10 & 4.86 & 0.31 & 0.52\\
10 - 15 & 2.40 & 0.13 & 0.28\\
\hline\hline
\end{tabular} 
 	\caption{Differential  CCQE-like cross section $d\sigma/dp_\parallel$, along with statistical and systematic uncertainties. Units are $10^{-41}$cm$^2$/GeV/nucleon. The $p_\parallel$ bins are in GeV.} 
	\label{tab:xsec_pz}
\end{table*}

\subsubsection{$E_\nu^{QE}$ and $Q^2_{QE}$}
The reconstructed energy-dependent cross section vs $Q^2_{QE}$,
$d\sigma (E_\nu^{QE})/dQ^2_{QE}$ is shown in Table
\ref{tab:xsec_eqe}. The statistical uncertainty on these measurements
is in Table \ref{tab:stat_eqe}, and the systematic uncertainty in
Table \ref{tab:sys_eqe}. 

The total reconstructed energy-dependent
cross section $\sigma(E_\nu^{QE})$, generated by dividing the event
count $N_i(E_\nu^{QE})$ in each bin  by that energy bin's flux
$\Phi(E_\nu)$, is shown in Table \ref{tab:xsec_enuqe}. Note that
$E_\nu^{QE}$ is the neutrino energy reconstructed using muon
kinematics assuming a quasi-elastic hypothesis and is different than
true the true incoming neutrino energy \enutrue.  The differential
cross section with respect to reconstructed $Q^2_{QE}$,
$d\sigma/dQ^2_{QE}$, is shown in Table~\ref{tab:xsec_qsq_intenuqe}.
Tables~\ref{tab:xsec_enuqe} and~\ref{tab:xsec_qsq_intenuqe} also include statistical and systematic
uncertainties.  

\begin{table*}[h]
	\makeatletter{}\begin{tabular}
{ l   r   r   r   r   r   r   r   r   r }\hline\hline
 & 0.0-0.025 & 0.025-0.05 & 0.05-0.1 & 0.1-0.2 & 0.2-0.4 & 0.4-0.8 & 0.8-1.2 & 1.2-2\\
\hline
1.5 - 2 & 1118.91 & 1219.53 & 1224.29 & 968.17 & 404.04 & 1.08 & 0.00 & 0.00\\
2 - 2.5 & 823.22 & 1304.80 & 991.07 & 899.36 & 650.50 & 129.16 & 0.00 & 0.00\\
2.5 - 3 & 924.55 & 1219.48 & 1134.01 & 968.24 & 688.89 & 241.79 & 4.44 & 0.00\\
3 - 3.5 & 805.21 & 954.82 & 1009.61 & 976.76 & 670.47 & 296.84 & 63.60 & 0.00\\
3.5 - 4 & 682.27 & 810.94 & 944.45 & 851.43 & 553.80 & 270.56 & 81.75 & 8.34\\
4 - 5 & 562.66 & 688.42 & 884.91 & 914.05 & 524.28 & 244.33 & 81.80 & 18.68\\
5 - 6 & 746.09 & 964.90 & 1006.93 & 824.64 & 662.22 & 301.57 & 77.13 & 26.32\\
6 - 7  & 885.65 & 844.16 & 1366.60 & 1055.01 & 994.06 & 344.99 & 96.90 & 34.75\\
7 - 8  & 536.48 & 802.64 & 1164.94 & 1159.34 & 752.25 & 420.73 & 96.69 & 25.83\\
8 - 10 & 1151.00 & 736.27 & 1064.92 & 1161.83 & 605.77 & 351.78 & 132.23 & 39.22\\
\hline\hline
\end{tabular} 
 	\caption{Measured reconstructed  energy-dependent  CCQE-like cross section $d\sigma (E_\nu^{QE})/dQ^2_{QE}$. Units are $10^{-41}$cm$^2$/GeV$^2$/nucleon. Columns represent bins of \qsq (GeV$^2$), rows are bins of  $E_\nu^{QE}$ (GeV).} 
	\label{tab:xsec_eqe}
\end{table*}

\begin{table*}[h]
	\makeatletter{}\begin{tabular}
{ l   r   r   r   r   r   r   r   r   r }\hline\hline
 & 0.0-0.025 & 0.025-0.05 & 0.05-0.1 & 0.1-0.2 & 0.2-0.4 & 0.4-0.8 & 0.8-1.2 & 1.2-2\\
\hline
1.5 - 2 & 140.16 & 166.04 & 126.62 & 96.45 & 57.78 & 0.47 & 0.00 & 0.00\\
2 - 2.5 & 92.74 & 122.56 & 77.30 & 58.88 & 41.91 & 18.16 & 0.00 & 0.00\\
2.5 - 3 & 85.01 & 102.13 & 70.63 & 49.92 & 32.71 & 16.56 & 1.61 & 0.00\\
3 - 3.5 & 75.69 & 83.71 & 62.92 & 47.08 & 29.13 & 15.23 & 9.05 & 0.00\\
3.5 - 4 & 76.98 & 84.67 & 66.77 & 47.61 & 27.80 & 14.73 & 9.48 & 2.45\\
4 - 5 & 75.68 & 83.69 & 69.95 & 53.48 & 29.11 & 14.32 & 9.60 & 3.51\\
5 - 6 & 156.95 & 176.01 & 130.86 & 85.79 & 56.59 & 26.60 & 12.70 & 5.53\\
6 - 7  & 219.32 & 182.08 & 205.49 & 126.23 & 92.34 & 36.57 & 18.48 & 8.94\\
7 - 8  & 166.33 & 221.90 & 214.54 & 160.15 & 86.41 & 49.02 & 20.22 & 6.51\\
8 - 10 & 319.12 & 209.08 & 193.84 & 152.32 & 75.01 & 41.31 & 26.99 & 10.74\\
\hline\hline
\end{tabular} 
 	\caption{Statistical uncertainty on the measured reconstructed  energy-dependent CCQE-like cross section $d\sigma (E_\nu^{QE})/dQ^2_{QE}$. Units are $10^{-41}$cm$^2$/GeV$^2$/nucleon. Columns represent bins of \qsq (GeV$^2$), rows are bins of  $E_\nu^{QE}$ (GeV).} 
	\label{tab:stat_eqe}
\end{table*}

\begin{table*}[h]
	\makeatletter{}\begin{tabular}
{ l  r   r   r   r   r   r   r   r   r }\hline\hline
 & 0.0-0.025 & 0.025-0.05 & 0.05-0.1 & 0.1-0.2 & 0.2-0.4 & 0.4-0.8 & 0.8-1.2 & 1.2-2\\
\hline
1.5 - 2 & 210.28 & 206.37 & 196.50 & 173.09 & 68.41 & 2.11 & 0.00 & 0.00\\
2 - 2.5 & 106.15 & 171.52 & 110.83 & 100.71 & 74.64 & 12.81 & 0.00 & 0.00\\
2.5 - 3 & 112.17 & 128.79 & 118.34 & 103.32 & 65.44 & 22.03 & 0.83 & 0.00\\
3 - 3.5 & 78.84 & 82.57 & 86.28 & 78.44 & 56.73 & 27.09 & 8.78 & 0.00\\
3.5 - 4 & 66.43 & 72.48 & 83.54 & 74.37 & 61.94 & 31.70 & 15.99 & 1.86\\
4 - 5 & 82.56 & 88.82 & 113.47 & 124.45 & 75.79 & 39.39 & 18.12 & 4.12\\
5 - 6 & 96.66 & 111.76 & 117.26 & 98.77 & 77.03 & 46.78 & 14.84 & 6.92\\
6 - 7  & 104.65 & 105.73 & 155.69 & 107.91 & 106.41 & 43.98 & 18.84 & 9.34\\
7 - 8  & 79.68 & 106.44 & 142.15 & 143.67 & 89.35 & 55.58 & 20.11 & 7.04\\
8 - 10 & 140.22 & 94.41 & 148.86 & 132.28 & 65.51 & 44.85 & 34.39 & 10.59\\
\hline\hline
\end{tabular} 
 	\caption{Systematic uncertainty on the measured reconstructed  energy-dependent  CCQE-like cross section $d\sigma (E_\nu^{QE})/dQ^2_{QE}$. Units are $10^{-41}$cm$^2$/GeV$^2$/nucleon. Columns represent bins of \qsq (GeV$^2$), rows are bins of  $E_\nu^{QE}$ (GeV).} 
	\label{tab:sys_eqe}
\end{table*}

\begin{table*}[h]
	\makeatletter{}\begin{tabular}
{c  r r r}\hline\hline
Bin & Cross & Statistical  & Systematic \\
 & section &  uncertainty &  uncertainty\\
\hline
1.5 - 2 & 297.73 & 17.21 & 52.13\\
2 - 2.5 & 374.45 & 13.69 & 43.07\\
2.5 - 3 & 443.40 & 11.64 & 45.47\\
3 - 3.5 & 470.43 & 11.14 & 40.67\\
3.5 - 4 & 428.05 & 11.22 & 45.56\\
4 - 5 & 417.18 & 11.75 & 62.05\\
5 - 6 & 480.57 & 20.93 & 61.16\\
6 - 7  & 620.44 & 31.25 & 70.02\\
7 - 8  & 585.74 & 34.57 & 69.28\\
8 - 10 & 562.74 & 33.25 & 66.85\\
\hline\hline
\end{tabular} 
 	\caption{Reconstructed energy-dependent  quasi-elastic-like cross section $\sigma(E_\nu^{QE})$, along with statistical and systematic uncertainties. Units are $10^{-41}$cm$^2$/nucleon. The $E_\nu^{QE}$ bins are in GeV.} 
	\label{tab:xsec_enuqe}
\end{table*}

\begin{table*}[h]
	\makeatletter{}\begin{tabular}
{c  r r r}\hline\hline
Bin & Cross & Statistical  & Systematic \\
 & section &  uncertainty &  uncertainty\\
\hline
0.0 - 0.025 & 697.59 & 29.69 & 94.47\\
0.025 - 0.05 & 870.26 & 34.54 & 104.75\\
0.05 - 0.1 & 884.93 & 25.45 & 88.78\\
0.1 - 0.2 & 797.86 & 18.80 & 75.59\\
0.2 - 0.4 & 520.27 & 11.78 & 47.92\\
0.4 - 0.8 & 187.98 & 5.04 & 20.36\\
0.8 - 1.2 & 37.29 & 2.23 & 7.36\\
1.2 - 2.0& 5.63 & 0.58 & 1.37\\
\hline\hline
\end{tabular} 
 	\caption{Reconstructed $Q^2_{QE}$-dependent  quasi-elastic-like cross section $d\sigma/dQ^2_{QE}$, integrated over $E_\nu^{QE}$ along with statistical and systematic uncertainties. Units are $10^{-41}$cm$^2$/nucleon.  The $Q^2$ units are GeV$^2$} 
	\label{tab:xsec_qsq_intenuqe}
\end{table*}

\subsubsection{ \enutrue and $Q^2_{QE}$}

The energy-dependent cross section vs $Q^2_{QE}$,
$d\sigma (E_\nu^{true})/dQ^2_{QE}$ is shown in Table
\ref{tab:xsec_enu}. The statistical uncertainty on these measurements
is in Table \ref{tab:stat_enu}, and the systematic uncertainty in
Table \ref{tab:sys_enu}. The total energy-dependent
cross section $\sigma(E_\nu^{QE})$, generated by dividing the event
count $N_i(E_\nu^{true})$ in each bin  by that energy bin's flux
$\Phi(E_\nu)$, is shown in Table \ref{tab:xsec_enu_intqsq}. The differential
cross section with respect to reconstructed $Q^2_{QE}$ integrated over $E_\nu^{true}$,
$d\sigma/dQ^2_{QE}$, is shown in Table~\ref{tab:xsec_qsq_intenu}.
Tables~\ref{tab:xsec_enu_intqsq} and~\ref{tab:xsec_qsq_intenu} also include statistical and systematic
uncertainties.

\begin{table*}[h]
	\makeatletter{}\begin{tabular}
{ l   r   r   r   r   r   r   r   r   r }\hline\hline
 & 0.0-0.025 & 0.025-0.05 & 0.05-0.1 & 0.1-0.2 & 0.2-0.4 & 0.4-0.8 & 0.8-1.2 & 1.2-2\\
\hline
1.5 - 2 & 1056.88 & 1110.31 & 1100.17 & 815.95 & 286.66 & 3.36 & 0.00 & 0.00\\
2 - 2.5 & 787.14 & 1186.74 & 933.10 & 831.06 & 559.99 & 101.43 & 0.14 & 0.00\\
2.5 - 3 & 899.69 & 1221.31 & 1110.77 & 935.94 & 673.18 & 220.50 & 8.50 & 0.00\\
3 - 3.5 & 895.55 & 1070.57 & 1098.91 & 1051.38 & 705.41 & 298.99 & 60.06 & 0.73\\
3.5 - 4 & 810.59 & 962.01 & 1067.40 & 966.13 & 624.24 & 289.22 & 77.76 & 8.42\\
4 - 5 & 674.54 & 809.80 & 1042.99 & 1042.04 & 594.38 & 264.03 & 81.41 & 17.10\\
5 - 6 & 840.97 & 1057.84 & 1074.20 & 927.61 & 696.27 & 313.27 & 77.54 & 26.28\\
6 - 7  & 951.70 & 891.26 & 1486.97 & 1080.46 & 1027.38 & 358.63 & 103.35 & 34.62\\
7 - 8  & 568.51 & 859.62 & 1225.77 & 1227.45 & 822.94 & 436.67 & 97.58 & 24.18\\
8 - 10 & 1135.54 & 776.00 & 1115.78 & 1196.72 & 624.22 & 363.35 & 132.62 & 39.71\\
\hline\hline
\end{tabular} 
 	\caption{Measured reconstructed  energy-dependent  quasi-elastic-like cross section $d\sigma (E_\nu^{true})/dQ^2_{QE}$. Units are $10^{-41}$cm$^2$/GeV$^2$/nucleon. Columns represent bins of \qsq (GeV$^2$), rows are bins of  $E_\nu^{true}$ (GeV).} 
	\label{tab:xsec_enu}
\end{table*}

\begin{table*}[h]
	\makeatletter{}\begin{tabular}
{ l  r   r   r   r   r   r   r   r   r }\hline\hline
 & 0.0-0.025 & 0.025-0.05 & 0.05-0.1 & 0.1-0.2 & 0.2-0.4 & 0.4-0.8 & 0.8-1.2 & 1.2-2\\
\hline
1.5 - 2 & 152.11 & 163.37 & 127.60 & 92.36 & 44.66 & 1.15 & 0.00 & 0.00\\
2 - 2.5 & 87.13 & 115.65 & 73.41 & 55.57 & 37.84 & 14.58 & 0.16 & 0.00\\
2.5 - 3 & 82.33 & 98.70 & 68.37 & 47.89 & 31.89 & 14.74 & 2.01 & 0.00\\
3 - 3.5 & 78.40 & 87.50 & 65.14 & 48.76 & 29.46 & 14.55 & 7.92 & 0.26\\
3.5 - 4 & 85.62 & 92.55 & 71.05 & 51.79 & 29.66 & 14.77 & 7.95 & 2.30\\
4 - 5 & 85.13 & 93.14 & 77.97 & 58.61 & 31.72 & 15.07 & 9.45 & 3.12\\
5 - 6 & 167.12 & 189.80 & 135.44 & 91.96 & 58.20 & 27.33 & 12.61 & 5.35\\
6 - 7  & 224.38 & 193.56 & 217.48 & 128.48 & 94.48 & 37.57 & 19.22 & 8.72\\
7 - 8  & 175.62 & 221.82 & 218.54 & 166.11 & 93.00 & 50.85 & 20.11 & 5.90\\
8 - 10 & 315.55 & 219.43 & 201.38 & 154.21 & 77.07 & 42.35 & 26.56 & 10.53\\
\hline\hline
\end{tabular} 
 	\caption{Statistical uncertainty on the measured reconstructed  energy-dependent  quasi-elastic-like cross section $d\sigma (E_\nu^{true})/dQ^2_{QE}$. Units are $10^{-41}$cm$^2$/GeV$^2$/nucleon. Columns represent bins of \qsq (GeV$^2$), rows are bins of  $E_\nu^{true}$ (GeV).} 
	\label{tab:stat_enu}
\end{table*}

\begin{table*}[h]
	\makeatletter{}\begin{tabular}
{ l  r   r   r   r   r   r   r   r   r }\hline\hline
 & 0.0-0.025 & 0.025-0.05 & 0.05-0.1 & 0.1-0.2 & 0.2-0.4 & 0.4-0.8 & 0.8-1.2 & 1.2-2\\
\hline
1.5 - 2 & 219.23 & 207.33 & 186.49 & 164.18 & 53.03 & 3.74 & 0.00 & 0.00\\
2 - 2.5 & 98.35 & 161.44 & 107.43 & 91.54 & 62.44 & 9.33 & 0.15 & 0.00\\
2.5 - 3 & 110.37 & 134.90 & 120.03 & 101.43 & 65.45 & 19.16 & 1.45 & 0.00\\
3 - 3.5 & 93.09 & 99.51 & 98.99 & 90.46 & 63.23 & 27.62 & 8.26 & 0.27\\
3.5 - 4 & 85.21 & 92.10 & 100.69 & 92.56 & 71.16 & 35.37 & 15.09 & 2.41\\
4 - 5 & 105.66 & 109.92 & 137.20 & 142.26 & 88.41 & 43.85 & 18.36 & 3.74\\
5 - 6 & 112.54 & 128.37 & 129.79 & 113.05 & 82.31 & 48.33 & 15.52 & 7.24\\
6 - 7  & 115.33 & 101.43 & 179.81 & 109.58 & 110.78 & 48.34 & 20.41 & 9.64\\
7 - 8  & 76.97 & 103.14 & 162.17 & 160.27 & 102.16 & 58.86 & 19.67 & 7.37\\
8 - 10 & 135.35 & 105.94 & 142.31 & 132.12 & 68.13 & 47.87 & 34.70 & 10.40\\
\hline\hline
\end{tabular} 
 	\caption{Systematic uncertainty on the measured reconstructed  energy-dependent  quasi-elastic-like cross section $d\sigma (E_\nu^{true})/dQ^2_{QE}$. Units are $10^{-41}$cm$^2$/GeV$^2$/nucleon. Columns represent bins of \qsq (GeV$^2$), rows are bins of  $E_\nu^{true}$ (GeV).} 
	\label{tab:sys_enu}
\end{table*}

\begin{table*}[h]
	\makeatletter{}\begin{tabular}
{c  r r r}\hline\hline
Bin & Cross & Statistical  & Systematic \\
 & section &  uncertainty &  uncertainty\\
\hline
1.5 - 2 & 249.46 & 15.40 & 47.04\\
2 - 2.5 & 331.73 & 12.20 & 37.94\\
2.5 - 3 & 428.40 & 11.00 & 44.29\\
3 - 3.5 & 494.52 & 11.03 & 44.25\\
3.5 - 4 & 472.67 & 11.53 & 51.96\\
4 - 5 & 464.19 & 12.51 & 70.07\\
5 - 6 & 510.54 & 21.66 & 66.18\\
6 - 7  & 646.43 & 32.08 & 74.04\\
7 - 8  & 617.37 & 35.94 & 74.69\\
8 - 10 & 578.25 & 33.76 & 69.46\\
\hline\hline
\end{tabular} 
 	\caption{Reconstructed energy-dependent  quasi-elastic-like cross section $\sigma(E_\nu^{true})$, along with statistical and systematic uncertainties. Units are $10^{-41}$cm$^2$/nucleon. The $E_\nu^{true}$ bins are in GeV.} 
	\label{tab:xsec_enu_intqsq}
\end{table*}

\begin{table*}[h]
	\makeatletter{}\begin{tabular}
{c  r r r}\hline\hline
Bin & Cross & Statistical  & Systematic \\
 & section &  uncertainty &  uncertainty\\
\hline
0.0 - 0.025 & 728.99 & 30.66 & 101.92\\
0.025 - 0.05 & 899.15 & 34.70 & 110.94\\
0.05 - 0.1 & 913.31 & 25.80 & 94.13\\
0.1 - 0.2 & 813.52 & 18.79 & 78.43\\
0.2 - 0.4 & 519.00 & 11.12 & 48.32\\
0.4 - 0.8 & 187.09 & 4.70 & 20.55\\
0.8 - 1.2 & 37.03 & 2.04 & 7.31\\
1.2 - 2.0& 5.57 & 0.54 & 1.36\\
\hline\hline
\end{tabular} 
 	\caption{Reconstructed $Q^2_{QE}$-dependent  quasi-elastic-like differential cross section $d\sigma/dQ^2_{QE}$,  integrated over $E_\nu^{true}$, along with statistical and systematic uncertainties. Units are $10^{-41}$cm$^2$/nucleon.  The $Q^2$ units are GeV$^2$} 
	\label{tab:xsec_qsq_intenu}
\end{table*}

\subsection{CCQE cross sections}
The tables in this section list our cross section measurements for the true quasi-elastic signal definition explained in section \ref{sec:ccqesig}. In each case, we show a table of values corresponding to the cross section, followed by a table of statistical uncertainties, then one of systematic uncertainties. Units are explained in the captions to the tables.

\subsubsection{\pt and \pz}
The differential cross section with respect to muon parallel and transverse momentum, $d^2\sigma/dp_Tdp_\parallel$ is shown in Table \ref{tab:ccqe_xsec_pzpt}. The statistical uncertainty on these measurements is in Table \ref{tab:ccqe_stat_pzpt}, and the systematic uncertainty in Table \ref{tab:ccqe_sys_pzpt}. 

Table \ref{tab:ccqe_xsec_pt} shows the differential cross section $d\sigma/dp_T$, generated by projecting the two-dimensional measurement onto the $p_T$ axis. Table \ref{tab:ccqe_xsec_pz} shows the differential cross section $d\sigma/dp_\parallel$, generated by projecting the two-dimensional measurement onto the $p_\parallel$ axis. Both tables also include statistical and systematic
uncertainties.

\begin{table*}[h]
	\makeatletter{}\begin{tabular}
{ l  r   r   r   r   r   r   r } \hline\hline
 & 0-0.15 & 0.15-0.25 & 0.25-0.4 & 0.4- 0.7 & 0.7-1 & 1-1.5\\
\hline
1.5 - 2 & 31.73 & 85.28 & 150.76 & 120.45 & 0.85 & 0.00\\
2 - 2.5 & 27.31 & 106.82 & 169.45 & 211.73 & 40.76 & 0.00\\
2.5 - 3 & 32.60 & 128.78 & 218.57 & 218.99 & 102.36 & 1.08\\
3 - 3.5 & 33.87 & 105.03 & 195.17 & 191.95 & 60.66 & 4.66\\
3.5 - 4 & 22.96 & 65.93 & 110.79 & 114.65 & 38.36 & 3.85\\
4 - 5 & 6.97 & 30.94 & 55.64 & 40.58 & 16.31 & 2.33\\
5 - 6 & 2.60 & 11.90 & 17.78 & 22.97 & 8.54 & 1.57\\
6 - 8 & 2.14 & 5.15 & 12.19 & 12.20 & 5.64 & 1.18\\
8 - 10 & 1.97 & 3.35 & 5.60 & 5.42 & 2.85 & 1.17\\
10 - 15 & 0.47 & 1.63 & 2.77 & 2.84 & 1.71 & 0.50\\
\hline\hline
\end{tabular} 
 	\caption{Measured double-differential true CCQE cross section $d^2\sigma/dp_Tdp_\parallel$. Units are $10^{-41}$cm$^2$/GeV$^2$/nucleon. Columns represent bins of \pt (GeV), rows are bins of \pz (GeV).} 
	\label{tab:ccqe_xsec_pzpt}
\end{table*}
\begin{table*}[h]
	\makeatletter{}\begin{tabular}
{ l  r   r   r   r   r   r   r }\hline\hline
 & 0-0.15 & 0.15-0.25 & 0.25-0.4 & 0.4- 0.7 & 0.7-1 & 1-1.5\\
\hline
1.5 - 2 & 4.34 & 9.30 & 10.95 & 8.62 & 0.41 & 0.00\\
2 - 2.5 & 3.67 & 9.00 & 9.73 & 9.17 & 5.02 & 0.00\\
2.5 - 3 & 3.83 & 9.48 & 10.44 & 8.16 & 6.92 & 0.45\\
3 - 3.5 & 3.81 & 8.17 & 9.41 & 7.00 & 4.34 & 1.21\\
3.5 - 4 & 3.04 & 6.16 & 6.73 & 5.08 & 3.15 & 0.98\\
4 - 5 & 1.18 & 3.12 & 3.46 & 2.15 & 1.43 & 0.47\\
5 - 6 & 0.77 & 2.00 & 2.00 & 1.67 & 1.01 & 0.36\\
6 - 8 & 0.53 & 0.95 & 1.24 & 0.86 & 0.61 & 0.23\\
8 - 10 & 0.58 & 0.86 & 0.86 & 0.57 & 0.42 & 0.25\\
10 - 15 & 0.16 & 0.33 & 0.37 & 0.26 & 0.21 & 0.10\\
\hline\hline
\end{tabular} 
 	\caption{Statistical uncertainty on the measured double-differential true CCQE  cross section $d^2\sigma/dp_Tdp_\parallel$. Units are $10^{-41}$cm$^2$/GeV$^2$/nucleon. Columns represent bins of \pt (GeV), rows are bins of \pz (GeV).} 
	\label{tab:ccqe_stat_pzpt}
\end{table*}
\begin{table*}[h]
	\makeatletter{}\begin{tabular}
{ l  r   r   r   r   r   r   r }\hline\hline
 & 0-0.15 & 0.15-0.25 & 0.25-0.4 & 0.4- 0.7 & 0.7-1 & 1-1.5\\
\hline
1.5 - 2 & 4.69 & 13.97 & 20.89 & 14.00 & 0.30 & 0.00\\
2 - 2.5 & 2.87 & 12.26 & 20.43 & 20.46 & 4.58 & 0.00\\
2.5 - 3 & 3.68 & 13.30 & 22.88 & 19.54 & 10.65 & 0.31\\
3 - 3.5 & 3.43 & 8.95 & 16.30 & 18.15 & 7.80 & 1.10\\
3.5 - 4 & 2.00 & 5.45 & 9.64 & 13.38 & 5.77 & 0.85\\
4 - 5 & 0.83 & 3.07 & 5.31 & 5.01 & 2.43 & 0.47\\
5 - 6 & 0.26 & 1.06 & 1.52 & 2.17 & 1.05 & 0.31\\
6 - 8 & 0.21 & 0.47 & 1.08 & 1.27 & 0.74 & 0.21\\
8 - 10 & 0.22 & 0.37 & 0.55 & 0.53 & 0.33 & 0.21\\
10 - 15 & 0.07 & 0.15 & 0.27 & 0.28 & 0.21 & 0.11\\
\hline\hline
\end{tabular} 
 	\caption{Systematic uncertainty on the measured double-differential true CCQE  cross section $d^2\sigma/dp_Tdp_\parallel$. Units are $10^{-41}$cm$^2$/GeV$^2$/nucleon. Columns represent bins of \pt (GeV), rows are bins of \pz (GeV).} 
	\label{tab:ccqe_sys_pzpt}
\end{table*}

\begin{table*}[h]
	\makeatletter{}\begin{tabular}
{c  r r r}\hline\hline
Bin & Cross & Statistical  & Systematic \\
 & section &  uncertainty &  uncertainty\\
\hline
0 - 0.15 & 94.40 & 4.77 & 10.54\\
0.15 - 0.25 & 313.88 & 10.65 & 31.43\\
0.25 - 0.4 & 545.21 & 11.95 & 54.06\\
0.4 - 0.7 & 541.85 & 9.40 & 49.32\\
0.7 - 1 & 171.87 & 5.65 & 20.46\\
1 - 1.5 & 15.88 & 1.32 & 3.04\\
\hline\hline
\end{tabular} 
 	\caption{Differential   true CCQE cross section $d\sigma/dp_T$, along with statistical and systematic uncertainties. Units are $10^{-41}$cm$^2$/GeV/nucleon. The $p_T$ bins are in GeV.} 
	\label{tab:ccqe_xsec_pt}
\end{table*}
\begin{table*}[h]
	\makeatletter{}\begin{tabular}
{c  r r r}\hline\hline
Bin & Cross & Statistical  & Systematic \\
 & section &  uncertainty &  uncertainty\\
\hline
1.5 - 2 & 72.29 & 3.27 & 9.65\\
2 - 2.5 & 115.94 & 3.62 & 12.17\\
2.5 - 3 & 147.50 & 3.75 & 13.74\\
3 - 3.5 & 122.97 & 3.08 & 11.45\\
3.5 - 4 & 74.48 & 2.25 & 8.03\\
4 - 5 & 30.72 & 1.03 & 3.54\\
5 - 6 & 14.49 & 0.72 & 1.36\\
6 - 8 & 8.60 & 0.40 & 0.86\\
8 - 10 & 4.53 & 0.30 & 0.43\\
10 - 15 & 2.26 & 0.13 & 0.23\\
\hline\hline
\end{tabular} 
 	\caption{Differential   true CCQE cross section $d\sigma/dp_\parallel$, along with statistical and systematic uncertainties. Units are $10^{-41}$cm$^2$/GeV/nucleon. The $p_\parallel$ bins are in GeV.} 
	\label{tab:ccqe_xsec_pz}
\end{table*}

\subsubsection{$E_\nu^{QE}$ and $Q^2_{QE}$}
The reconstructed energy-dependent  true CCQE cross section vs
$Q^2_{QE}$, $d\sigma (E_\nu^{QE})/dQ^2_{QE}$ is shown in Table
\ref{tab:ccqe_xsec_eqe}. The statistical uncertainty on these
measurements is in Table \ref{tab:ccqe_stat_eqe}, and the systematic
uncertainty in Table \ref{tab:ccqe_sys_eqe}. 

 By projecting, we can get the total reconstructed energy-dependent cross section $\sigma(E_\nu^{QE})$, shown in Table \ref{tab:ccqe_xsec_enuqe_intqsq} and the single-differential cross section $d\sigma/dQ^2_{QE}$ shown in Table \ref{tab:ccqe_xsec_qsq_intenuqe}.

\begin{table*}[h]
	\makeatletter{}\begin{tabular}
{ l r   r   r   r   r   r   r   r   r }\hline\hline
 & 0.0-0.025 & 0.025-0.05 & 0.05-0.1 & 0.1-0.2 & 0.2-0.4 & 0.4-0.8 & 0.8-1.2 & 1.2-2\\
\hline
1.5 - 2 & 961.93 & 921.26 & 1046.39 & 918.71 & 360.39 & 0.99 & 0.00 & 0.00\\
2 - 2.5 & 634.90 & 1024.97 & 937.73 & 833.29 & 597.99 & 122.20 & 0.00 & 0.00\\
2.5 - 3 & 777.94 & 1026.35 & 1026.84 & 909.91 & 652.15 & 233.47 & 4.36 & 0.00\\
3 - 3.5 & 724.08 & 824.97 & 946.28 & 937.99 & 640.87 & 289.32 & 65.24 & 0.00\\
3.5 - 4 & 577.83 & 670.24 & 820.09 & 831.00 & 541.68 & 263.60 & 85.04 & 10.51\\
4 - 5 & 508.34 & 657.91 & 847.79 & 866.39 & 508.40 & 242.81 & 81.99 & 18.98\\
5 - 6 & 602.73 & 797.47 & 899.49 & 766.83 & 624.65 & 290.43 & 75.11 & 26.90\\
6 - 7  & 786.66 & 759.65 & 1269.84 & 1019.47 & 929.67 & 339.53 & 92.73 & 34.98\\
7 - 8  & 554.65 & 838.27 & 1082.64 & 1085.17 & 687.81 & 406.85 & 96.63 & 28.80\\
8 - 10 & 1279.92 & 747.12 & 953.36 & 1147.95 & 566.43 & 304.98 & 114.37 & 38.60\\
\hline\hline
\end{tabular} 
 	\caption{Measured reconstructed  energy-dependent  true CCQE cross section $d\sigma (E_\nu^{QE})/dQ^2_{QE}$. Units are $10^{-41}$cm$^2$/GeV$^2$/nucleon. Columns represent bins of \qsq (GeV$^2$), rows are bins of  $E_\nu^{QE}$ (GeV).} 
	\label{tab:ccqe_xsec_eqe}
\end{table*}

\begin{table*}[h]
	\makeatletter{}\begin{tabular}
{ l r   r   r   r   r   r   r   r   r }\hline\hline
 & 0.0-0.025 & 0.025-0.05 & 0.05-0.1 & 0.1-0.2 & 0.2-0.4 & 0.4-0.8 & 0.8-1.2 & 1.2-2\\
\hline
1.5 - 2 & 134.29 & 145.33 & 118.27 & 91.57 & 52.78 & 0.43 & 0.00 & 0.00\\
2 - 2.5 & 84.82 & 111.04 & 76.46 & 55.68 & 39.27 & 17.59 & 0.00 & 0.00\\
2.5 - 3 & 81.81 & 95.48 & 68.35 & 48.06 & 31.74 & 16.43 & 1.55 & 0.00\\
3 - 3.5 & 76.02 & 80.14 & 62.76 & 45.98 & 28.50 & 15.26 & 9.99 & 0.00\\
3.5 - 4 & 74.72 & 78.82 & 63.29 & 47.40 & 27.52 & 14.72 & 10.58 & 3.31\\
4 - 5 & 76.57 & 86.57 & 70.36 & 52.14 & 28.54 & 14.48 & 10.52 & 3.86\\
5 - 6 & 149.21 & 165.26 & 126.76 & 81.78 & 53.98 & 26.29 & 13.48 & 6.22\\
6 - 7  & 218.01 & 179.85 & 203.02 & 123.43 & 87.61 & 36.77 & 19.28 & 10.05\\
7 - 8  & 193.32 & 236.18 & 209.35 & 154.13 & 80.83 & 49.28 & 21.86 & 7.86\\
8 - 10 & 365.90 & 231.08 & 184.47 & 152.74 & 72.02 & 37.70 & 27.76 & 11.70\\
\hline\hline
\end{tabular} 
 	\caption{Statistical uncertainty on the measured reconstructed  energy-dependent true CCQE cross section $d\sigma (E_\nu^{QE})/dQ^2_{QE}$. Units are $10^{-41}$cm$^2$/GeV$^2$/nucleon. Columns represent bins of \qsq (GeV$^2$), rows are bins of  $E_\nu^{QE}$ (GeV).} 
	\label{tab:ccqe_stat_eqe}
\end{table*}
\begin{table*}[h]
	\makeatletter{}\begin{tabular}
{ l r   r   r   r   r   r   r   r   r }\hline\hline
 & 0.0-0.025 & 0.025-0.05 & 0.05-0.1 & 0.1-0.2 & 0.2-0.4 & 0.4-0.8 & 0.8-1.2 & 1.2-2\\
\hline
1.5 - 2 & 150.41 & 155.96 & 168.82 & 167.63 & 56.88 & 1.93 & 0.00 & 0.00\\
2 - 2.5 & 76.27 & 140.44 & 100.68 & 92.55 & 67.00 & 12.02 & 0.00 & 0.00\\
2.5 - 3 & 93.96 & 106.02 & 102.04 & 100.19 & 62.77 & 19.91 & 1.45 & 0.00\\
3 - 3.5 & 68.09 & 65.42 & 79.80 & 75.43 & 55.70 & 25.47 & 10.51 & 0.00\\
3.5 - 4 & 53.04 & 58.59 & 70.18 & 73.48 & 60.74 & 30.15 & 16.76 & 2.44\\
4 - 5 & 69.58 & 85.73 & 107.38 & 120.31 & 73.31 & 39.20 & 17.58 & 3.98\\
5 - 6 & 71.83 & 92.13 & 100.38 & 90.62 & 71.52 & 41.96 & 13.44 & 6.59\\
6 - 7  & 99.27 & 96.57 & 138.30 & 106.85 & 103.10 & 40.50 & 16.17 & 8.68\\
7 - 8  & 95.70 & 113.93 & 142.23 & 134.59 & 83.37 & 52.94 & 17.82 & 7.39\\
8 - 10 & 155.54 & 114.37 & 143.50 & 124.77 & 63.15 & 40.13 & 29.76 & 8.49\\
\hline\hline
\end{tabular} 
 	\caption{Systematic uncertainty on the measured reconstructed  energy-dependent true CCQE  cross section $d\sigma (E_\nu^{QE})/dQ^2_{QE}$. Units are $10^{-41}$cm$^2$/GeV$^2$/nucleon. Columns represent bins of \qsq (GeV$^2$), rows are bins of  $E_\nu^{QE}$ (GeV).} 
	\label{tab:ccqe_sys_eqe}
\end{table*}

%\onecolumngrid
\begin{table*}[h]
	\makeatletter{}\begin{tabular}
{c  r r r}\hline\hline
Bin & Cross & Statistical  & Systematic \\
 & section &  uncertainty &  uncertainty\\
\hline
1.5 - 2 & 263.74 & 15.96 & 44.39\\
2 - 2.5 & 340.19 & 13.00 & 36.97\\
2.5 - 3 & 413.00 & 11.34 & 40.76\\
3 - 3.5 & 449.84 & 11.15 & 37.21\\
3.5 - 4 & 411.51 & 11.39 & 43.12\\
4 - 5 & 404.97 & 11.89 & 59.61\\
5 - 6 & 449.33 & 20.47 & 55.18\\
6 - 7  & 590.92 & 30.87 & 63.22\\
7 - 8  & 559.46 & 34.23 & 65.00\\
8 - 10 & 525.04 & 32.88 & 59.24\\
\hline\hline
\end{tabular} 
 	\caption{Reconstructed energy-dependent   true CCQE cross section $\sigma(E_\nu^{QE})$, along with statistical and systematic uncertainties. Units are $10^{-41}$cm$^2$/nucleon. The $E_\nu^{QE}$ bins are in GeV.} 
	\label{tab:ccqe_xsec_enuqe_intqsq}
\end{table*}

\begin{table*}[h]
	\makeatletter{}\begin{tabular}
{c  r r r}\hline\hline
Bin & Cross & Statistical  & Systematic \\
 & section &  uncertainty &  uncertainty\\
\hline
0.0 - 0.025 & 599.77 & 28.92 & 62.26\\
0.025 - 0.05 & 727.63 & 32.10 & 75.72\\
0.05 - 0.1 & 806.35 & 24.66 & 77.79\\
0.1 - 0.2 & 757.95 & 18.11 & 74.43\\
0.2 - 0.4 & 491.60 & 11.22 & 44.70\\
0.4 - 0.8 & 182.10 & 5.00 & 18.65\\
0.8 - 1.2 & 37.48 & 2.45 & 7.82\\
1.2 - 2.0& 5.98 & 0.68 & 1.41\\
\hline\hline
\end{tabular} 
 	\caption{Reconstructed $Q^2_{QE}$-dependent  true CCQE differential  cross section $d\sigma/dQ^2_{QE}$ integrated over $E_\nu^{QE}$, along with statistical and systematic uncertainties. Units are $10^{-41}$cm$^2$/nucleon.  The $Q^2$ units are GeV$^2$} 
	\label{tab:ccqe_xsec_qsq_intenuqe}
\end{table*}

\subsubsection{\enutrue and $Q^2_{QE}$}
The energy-dependent true CCQE cross section vs $Q^2_{QE}$, $d\sigma (E_\nu)/dQ^2_{QE}$ is shown in Table \ref{tab:ccqe_xsec_enuq2}. The statistical uncertainty on these measurements is in Table \ref{tab:ccqe_stat_enuq2}, and the systematic uncertainty in Table \ref{tab:ccqe_sys_enuq2}. Table \ref{tab:ccqe_xsec_enu_intqsq} shows the energy-dependent cross section $\sigma(E_\nu^{true})$ and \ref{tab:ccqe_xsec_qsq_intenu} shows the single differential cross section $d\sigma/dQ^2_{QE}$ integrated over $E_\nu^{true}$.

\begin{table*}[h]
	\makeatletter{}\begin{tabular}
{ l r   r   r   r   r   r   r   r   r }\hline\hline
 & 0.0-0.025 & 0.025-0.05 & 0.05-0.1 & 0.1-0.2 & 0.2-0.4 & 0.4-0.8 & 0.8-1.2 & 1.2-2\\
\hline
1.5 - 2 & 890.86 & 846.55 & 961.45 & 815.88 & 285.51 & 3.45 & 0.00 & 0.00\\
2 - 2.5 & 617.73 & 948.99 & 890.74 & 806.55 & 548.30 & 107.21 & 0.14 & 0.00\\
2.5 - 3 & 760.88 & 1040.54 & 1009.35 & 877.58 & 640.49 & 227.43 & 9.22 & 0.00\\
3 - 3.5 & 780.77 & 886.12 & 1010.46 & 989.15 & 660.75 & 291.88 & 66.50 & 0.82\\
3.5 - 4 & 648.05 & 756.86 & 902.74 & 901.89 & 586.61 & 270.99 & 80.57 & 10.23\\
4 - 5 & 584.32 & 736.70 & 946.70 & 959.07 & 552.33 & 249.86 & 77.97 & 18.00\\
5 - 6 & 648.49 & 859.86 & 957.90 & 829.65 & 647.02 & 292.48 & 73.34 & 26.18\\
6 - 7  & 826.04 & 786.51 & 1351.17 & 1050.72 & 955.91 & 342.25 & 94.24 & 34.43\\
7 - 8  & 579.24 & 878.83 & 1121.80 & 1126.58 & 718.53 & 415.85 & 95.83 & 26.61\\
8 - 10 & 1267.73 & 773.88 & 985.76 & 1166.76 & 578.91 & 309.48 & 114.34 & 38.68\\
\hline\hline
\end{tabular} 
 	\caption{Measured energy-dependent true CCQE  cross section $d\sigma (E_\nu)/dQ^2_{QE}$. Units are $10^{-41}$cm$^2$/GeV$^2$/nucleon. Columns represent bins of \qsq (GeV$^2$), rows are bins of  $E_\nu$ (GeV).} 
	\label{tab:ccqe_xsec_enuq2}
\end{table*}
\begin{table*}[h]
	\makeatletter{}\begin{tabular}
{ l r   r   r   r   r   r   r   r   r }\hline\hline
 & 0.0-0.025 & 0.025-0.05 & 0.05-0.1 & 0.1-0.2 & 0.2-0.4 & 0.4-0.8 & 0.8-1.2 & 1.2-2\\
\hline
1.5 - 2 & 133.08 & 140.17 & 117.28 & 91.03 & 44.74 & 1.19 & 0.00 & 0.00\\
2 - 2.5 & 82.43 & 105.63 & 73.89 & 54.48 & 37.40 & 15.38 & 0.16 & 0.00\\
2.5 - 3 & 81.00 & 95.30 & 66.85 & 46.34 & 31.01 & 15.44 & 2.19 & 0.00\\
3 - 3.5 & 79.73 & 83.44 & 65.00 & 47.14 & 28.53 & 14.73 & 9.34 & 0.30\\
3.5 - 4 & 81.51 & 84.85 & 66.87 & 50.01 & 28.69 & 14.41 & 9.10 & 2.86\\
4 - 5 & 85.00 & 94.36 & 76.25 & 55.94 & 30.22 & 14.74 & 10.06 & 3.54\\
5 - 6 & 153.43 & 176.05 & 132.63 & 86.12 & 55.17 & 26.38 & 13.09 & 5.94\\
6 - 7  & 224.19 & 184.42 & 212.42 & 126.15 & 89.90 & 37.02 & 19.39 & 9.90\\
7 - 8  & 197.24 & 244.05 & 213.82 & 159.30 & 83.43 & 49.87 & 21.40 & 7.07\\
8 - 10 & 362.75 & 238.19 & 192.43 & 153.43 & 73.44 & 38.04 & 27.29 & 11.34\\
\hline\hline
\end{tabular} 
 	\caption{Statistical uncertainty on the measured energy-dependent  true CCQE cross section $d\sigma (E_\nu)/dQ^2_{QE}$. Units are $10^{-41}$cm$^2$/GeV$^2$/nucleon. Columns represent bins of \qsq (GeV$^2$), rows are bins of  $E_\nu$ (GeV).} 
	\label{tab:ccqe_stat_enuq2}
\end{table*}
\begin{table*}[h]
	\makeatletter{}\begin{tabular}
{ l r   r   r   r   r   r   r   r   r }\hline\hline
 & 0.0-0.025 & 0.025-0.05 & 0.05-0.1 & 0.1-0.2 & 0.2-0.4 & 0.4-0.8 & 0.8-1.2 & 1.2-2\\
\hline
1.5 - 2 & 145.69 & 146.91 & 163.02 & 170.41 & 49.20 & 3.84 & 0.00 & 0.00\\
2 - 2.5 & 72.08 & 131.50 & 95.52 & 88.68 & 62.40 & 10.38 & 0.12 & 0.00\\
2.5 - 3 & 88.35 & 111.16 & 100.91 & 96.18 & 63.43 & 19.84 & 2.78 & 0.00\\
3 - 3.5 & 76.03 & 72.91 & 85.99 & 83.45 & 58.50 & 25.79 & 10.53 & 0.25\\
3.5 - 4 & 61.43 & 69.74 & 81.06 & 82.99 & 65.75 & 31.49 & 15.67 & 2.77\\
4 - 5 & 82.69 & 98.86 & 121.50 & 132.02 & 80.61 & 40.49 & 16.68 & 3.65\\
5 - 6 & 73.23 & 99.01 & 109.59 & 101.45 & 75.13 & 41.53 & 13.68 & 6.01\\
6 - 7  & 109.31 & 96.58 & 147.95 & 106.63 & 106.94 & 41.04 & 16.37 & 8.75\\
7 - 8  & 100.24 & 111.66 & 142.86 & 154.27 & 85.81 & 54.34 & 16.30 & 6.98\\
8 - 10 & 144.27 & 124.00 & 143.31 & 120.49 & 64.28 & 40.38 & 29.63 & 8.33\\
\hline\hline
\end{tabular} 
 	\caption{Systematic uncertainty on the measured energy-dependent  true CCQE cross section $d\sigma (E_\nu)/dQ^2_{QE}$. Units are $10^{-41}$cm$^2$/GeV$^2$/nucleon. Columns represent bins of \qsq (GeV$^2$), rows are bins of  $E_\nu$ (GeV).} 
	\label{tab:ccqe_sys_enuq2}
\end{table*}

\begin{table*}[h]
	\makeatletter{}\begin{tabular}
{c  r r r}\hline\hline
Bin & Cross & Statistical  & Systematic \\
 & section &  uncertainty &  uncertainty\\
\hline
1.5 - 2 & 231.58 & 14.86 & 42.54\\
2 - 2.5 & 316.96 & 12.18 & 34.83\\
2.5 - 3 & 406.02 & 10.94 & 40.24\\
3 - 3.5 & 467.26 & 11.06 & 39.20\\
3.5 - 4 & 436.57 & 11.38 & 46.17\\
4 - 5 & 432.26 & 12.31 & 63.48\\
5 - 6 & 465.24 & 20.86 & 57.31\\
6 - 7  & 606.27 & 31.47 & 64.60\\
7 - 8  & 574.87 & 34.80 & 67.11\\
8 - 10 & 533.25 & 33.08 & 58.93\\
\hline\hline
\end{tabular} 
 	\caption{Energy-dependent   true CCQE cross section $\sigma(E_\nu^{true})$, along with statistical and systematic uncertainties. Units are $10^{-41}$cm$^2$/nucleon. The $E_\nu$ bins are in GeV.} 
	\label{tab:ccqe_xsec_enu_intqsq}
\end{table*}

\begin{table*}[h]
	\makeatletter{}\begin{tabular}
{c  r r r}\hline\hline
Bin & Cross & Statistical  & Systematic \\
 & section &  uncertainty &  uncertainty\\
\hline
0.0 - 0.025 & 615.63 & 29.49 & 64.50\\
0.025 - 0.05 & 744.62 & 32.34 & 77.84\\
0.05 - 0.1 & 824.27 & 24.95 & 80.20\\
0.1 - 0.2 & 769.26 & 18.26 & 76.57\\
0.2 - 0.4 & 491.04 & 10.83 & 45.13\\
0.4 - 0.8 & 181.83 & 4.75 & 18.69\\
0.8 - 1.2 & 37.44 & 2.28 & 7.80\\
1.2 - 2.0& 5.90 & 0.62 & 1.40\\
\hline\hline
\end{tabular} 
 	\caption{True CCQE differential cross section $d\sigma/dQ^2_{QE}$ integrated over $E_\nu^{true}$, along with statistical and systematic uncertainties. Units are $10^{-41}$cm$^2$/GeV$^2$/nucleon. The $Q^2_{QE}$ bins are in GeV$^2$.} 
	\label{tab:ccqe_xsec_qsq_intenu}
\end{table*}

\

\renewcommand{\textfraction}{0.05}
\renewcommand{\topfraction}{0.95}
\renewcommand{\bottomfraction}{0.95}
\renewcommand{\floatpagefraction}{0.95}
\renewcommand{\dblfloatpagefraction}{0.95}
\renewcommand{\dbltopfraction}{0.95}
\setcounter{totalnumber}{5}
\setcounter{bottomnumber}{3}
\setcounter{topnumber}{3}
\setcounter{dbltopnumber}{3}

\end{document}